\documentclass[11pt,a4paper]{article}

\usepackage[utf8]{inputenc}
\usepackage[T1]{fontenc}
\usepackage{amsmath,amssymb,amsthm}
\usepackage{mathtools}
\usepackage{bm}
\usepackage{graphicx}
\usepackage{xcolor}
\usepackage[colorlinks=true]{hyperref}
\usepackage[noabbrev,capitalize,nameinlink]{cleveref}
\usepackage[numbers,sort&compress]{natbib}
\usepackage{geometry}
\usepackage{booktabs}
\usepackage{algorithm}
\usepackage{algpseudocode}
\usepackage[numbers]{natbib} 
\usepackage{subcaption}
\usepackage{todonotes}
\usepackage{nicefrac}
\usepackage{enumitem}
\usepackage{tikz}
\usetikzlibrary{decorations.pathreplacing, arrows.meta, shapes.geometric}

\newtheorem{theorem}{Theorem}[section]
\newtheorem{lemma}[theorem]{Lemma}
\newtheorem{definition}[theorem]{Definition}
\newtheorem{remark}[theorem]{Remark}

\newcommand{\R}{\mathbb{R}}
\newcommand{\T}{\mathbb{T}}

\newcommand{\df}{\delta\!f}
\newcommand{\finit}{f_{\mathrm{init}}}
\newcommand{\fbulk}{f_0}
\newcommand{\dw}{\delta w}

\newcommand{\flow}{\Phi}
\newcommand{\Np}{N_p}
\newcommand{\dt}{\Delta t}

\newcommand{\abs}[1]{\left|#1\right|}
\newcommand{\bi}{{\boldsymbol i}}
\newcommand{\bj}{{\boldsymbol j}}

\title{%
    A neural delta-$f$ particle-in-cell method: evolving the background with learned backward flows%
}

\author{%
  Victor Fournet$^1$
  \and Martin Campos Pinto$^1$
  \and Emmanuel Franck$^2$
  \and Victor Michel-Dansac$^2$
  \\[1ex]
  \small $^1$ Max Planck Institute for Plasma Physics, Garching, Germany\\
  \small $^2$ Université de Strasbourg, CNRS, Inria, IRMA, F-67000, Strasbourg, France
}

\date{\today}

\begin{document}

\maketitle

\begin{abstract}

    We present a novel $\df$ particle-in-cell (PIC) method for the kinetic
    simulation of electrostatic plasmas in which the bulk density, acting as a
    control variate, is evolved using symplectic neural
    networks (SympNets). The SympNets are used as an approximation of the backward flow and trained using the particle trajectories. We introduce a
    periodic variant of the SympNet architecture that encodes the spatial
    periodicity of the problem into the network itself. We validate the approach with numerical results in 1D1V and 3D3V for the Vlasov-Poisson system.
\end{abstract}

\tableofcontents
\bigskip

\section{Introduction}
\label{sec:intro}

High-fidelity plasma simulations often require solving kinetic equations, which are nonlinear transport equations posed in phase spaces of moderate to high dimension (up to six in the case of three-dimensional space and velocity).

Grid-based methods such as
semi-Lagrangian~\cite{Sonnendrucker1999} or Eulerian~\cite{arber2002critical} schemes can compute
the distribution function on a grid, but in $d$ dimensions their storage cost scales as
$\mathcal{O}(n^d)$, where $n$ is the number of grid points per direction, making fine-resolution
simulations prohibitively expensive. This phenomenon is known as the \emph{curse of dimensionality}.
An additional fundamental difficulty is \emph{filamentation}: distribution function often develop
fine-scale structures in phase space as time evolves, requiring a fine enough grid
to be accurately resolved.

The Particle-In-Cell (PIC) method \cite{Birdsall1991,bottino2015monte} is a popular method to avoid a full discretisation of the phase space, representing the distribution as a collection of $\Np$
numerical markers or macro-particles. The computational cost of the method grows only linearly with $\Np$, but
it suffers from statistical noise that decreases in \smash{$\Np^{-\nicefrac 1 2}$} which comes from the Monte Carlo approximation of the moments of the distribution function.

A classical strategy to reduce this noise is the \emph{$\df$ approach}
\cite{Kotschenreuther1988,Dimits1993,Parker1993,parker1993fully,hu1994generalized}, in which the distribution is
decomposed as $f = \fbulk + \df$, where $\fbulk$ is an analytically known bulk
density, while the term $\df$ is discretised by numerical particles carrying time-dependent weights
so that the underlying transport equation is the same.
The moments are split in the same way: those of the bulk are evaluated analytically and only the $\df$ part is evaluated by a Monte Carlo approximation.
If $\fbulk$ is a good approximation to $f$, the weights are small, reducing the
statistical error.
For this reason the $\df$ method may be interpreted as a \emph{control variate method}~\cite{Aydemir1994}, a well-known noise reduction technique in Monte Carlo methods.

In many practical problems, $\fbulk$ is a steady state of the system, which is a valid choice as long as the
plasma stays close to that equilibrium, as it is often the case for magnetic fusion simulations~\cite{lanti2020}.
However, in regimes where the distribution diverges significantly from its initial state, such as in simulations close to the edge of a tokamak plasma, which involves steep gradients and low density levels~\cite{murugappan2022gyrokinetic,murugappan2024},
a static bulk is no longer adequate and the particle weights become large.
This eventually cancels the noise-reduction benefit.

A natural idea to recover a good noise reduction is then to also evolve the bulk in time.
Several approaches have been developed in this direction~\cite{Allfrey2003,CamposPinto2023,murugappan2024}. A common strategy is to assume that the bulk density $\fbulk$ takes the form of a Maxwellian distribution, and to evolve its moments in time using fluid equations~\cite{murugappan2024,brunner1999collisional}. This has the advantage that the moments of the distribution function are easy to compute. Another way is to directly represent the bulk on a phase space grid using B-splines~\cite{CamposPinto2023,Allfrey2003}. For instance in~\cite{CamposPinto2023}, the flow is approximated using a collection of auxiliary particles that are initialized on a grid and then pushed forward by the PIC scheme. The approximated backward flow is then computed using linear or quadratic Taylor expansions around these particles, and the density is transported by this flow and projected on a coarse spline grid. This method works well in 1D1V; however, its isotropic nature limits its performance in higher dimensions where flows may exhibit anisotropic smoothness~\cite{CamposPinto2023}.

Recently, there has been a growing interest in incorporating neural networks in classical methods \cite{ray_deep_2024},
in particular for high dimensional problems where they can be used to mitigate the curse of
dimensionality~\cite{mfo_learning_2026}. One example is Physics-Informed Neural Networks (PINNs)~\cite{raissi2019physics,de2022error,beltran2022physics,hu2024tackling,zhang2023physics} where the PDE residual is directly incorporated in the loss function, and a space-time approximation is used. Another possibility is to use a time-discrete approach, where the network only approximates the solution in space, and its parameters evolve at each time step. Such examples are discrete PINNs~\cite{biesek2023burgers,stiasny2023physics}, Neural Galerkin methods~\cite{bruna2024neural}, and neural semi-Lagrangian methods~\cite{franck2026neural}. All these works show that neural networks can be a promising approach to approximate functions living in a high dimensional space, while circumventing the curse of dimensionality.

In this work, focusing on the Vlasov-Poisson system as a proof of concept,
we propose an extension of the method described in~\cite{CamposPinto2023}:
again the bulk density $\fbulk$ is represented on a coarse grid with B-splines, but it is no longer
updated with a backward flow computed from isotropic Taylor expansions. Instead,
we use a neural network to approximate the backward flow associated with particles pushed forward by a given PIC scheme.
Composing this ``neural flow'' with the initial density provides us with a fine representation of the full solution $f$,
which is then projected on a coarse grid of B-splines at a relatively small computational cost to obtain the bulk density $\fbulk$.
Note that the fine representation is easy to evaluate at any point in phase space, but its moments are difficult to compute due to the phase space filamentations. The spline bulk density does not suffer from this issue, as both its point values and its moments are straightforward to compute. Nevertheless, as it is always updated from the fine representation, it remains close to the full solution which allows us to keep the weights small and reduce the noise.

In long-time simulations, one would eventually need to remap the fine representation, that is, to re-approximate it so as to reset the growing composition of neural flows, at the price of some loss of information, once this composition becomes too costly to evaluate. Performed at every time step in a standard semi-Lagrangian scheme, this remapping is here required only over very long times, which makes our method only \emph{weakly} semi-Lagrangian, in the spirit of the Characteristic Mapping Method~\cite{krah2024characteristic}. In all the simulations reported in this work, it was in fact never necessary.

The main advantages of this new method are twofold. First, it strongly mitigates the curse of dimensionality, as neural networks have the ability to efficiently approximate functions with anisotropic smoothness in high dimensions~\cite{franck2026neural,mfo_learning_2026}. 
Second, it allows to get rid of the auxiliary particles used in~\cite{CamposPinto2023} to approximate the flow,
which simplifies its integration with an existing PIC scheme.
Indeed, to train the flow networks at any desired point in time, we use the current and previous particle coordinates computed with the PIC scheme as training data, without requiring them to be on a structured grid.

Since the Vlasov-Poisson system is Hamiltonian, its flow is symplectic.
We choose to approximate it using the well-known paradigm of symplectic neural networks (SympNets, see~\cite{Jin2020}). Such neural networks are designed to directly incorporate symplecticity into their architecture, bypassing the need to add an additional penalization term in the loss function.
Since we are dealing with periodic boundary conditions, as is natural in plasma physics simulations, we furthermore introduce a new \emph{periodic} SympNet architecture that natively encodes the spatial periodicity into the network, once again avoiding an additional penalization term.

The resulting method, which we call the \emph{Neural $\df$-PIC scheme}, has the
following features:
\begin{itemize}[nosep]
    \item The bulk density is updated entirely from particle data, without requiring
          auxiliary markers or fine grid semi-Lagrangian steps.

    \item The symplectic structure of the Vlasov flow is preserved at the level of
          the fine (lagrangian) representation.

    \item The method extends naturally to high-dimensional phase spaces, as neural network architectures scale well to high dimensional problems and the training data are directly provided by the existing PIC particles.
\end{itemize}

We stress that the goal of this work is a proof of concept: the Vlasov-Poisson system is used as a controlled and well-understood test bed to assess whether symplectic neural networks coupled can effectively denoise the density by dynamically evolving the bulk on a coarse spline grid. In this study, the objective is not to obtain a method that would be more efficient than a standard $\df$-PIC scheme in every possible regime. Our motivation lies in regimes where a static, or even a Maxwellian, bulk is inadequate and where evolving it is genuinely necessary, such as edge gyrokinetic simulations~\cite{murugappan2024,murugappan2022gyrokinetic} where prohibitive numbers of particles would be required without a dynamic control variate.

The remainder of the paper is organised as follows. \Cref{sec:vlasov}
recalls the Vlasov-Poisson system and the classical $\df$-PIC framework.
\Cref{sec:sympnet} describes the SympNet architecture and its periodic
extension. \Cref{sec:algorithm} presents the Neural $\df$-PIC algorithm.
\Cref{sec:numerics} reports numerical results in 1D1V and 3D3V.
\Cref{sec:conclusion} concludes with a discussion and perspectives.

\section{The \texorpdfstring{$\df$}{delta-f}-PIC framework}
\label{sec:vlasov}

\subsection{The Vlasov-Poisson system and its flow map}

We consider the Vlasov-Poisson system (in normalised units) for a electron distribution function
$f(t,x,v)\geq 0$ in a domain $[0,T] \times L\T^d\times\R^d$,
where the dimension is $d \in \mathbb{N}$,
and where $L\T^d := (\R/L\mathbb{Z})^d$ denotes the periodic torus of side length $L$.
It is governed by the system of equations
\begin{equation}
    \begin{aligned}
        \partial_t f + v\cdot\nabla_x f - \nabla_x\phi\cdot\nabla_v f & = 0,                                      \\
        -\Delta\phi                                                   & = \rho \coloneqq \int f\,\mathrm{d}v - 1, \\
        f(t=0)                                                        & = \finit,
    \end{aligned}
    \label{eq:vlasov}
\end{equation}
where the second equation is a normalised Poisson equation for the electric
potential $\phi$, with a constant background ion density set to 1 so that the total charge
(assuming $\iint f\,\mathrm{d}x\mathrm{d}v = L^d$) is zero.

The characteristic curves associated with \eqref{eq:vlasov} are the solutions
to the system of nonlinear differential equations,
\begin{equation}
    \begin{dcases}
        \dot{X}(t) = V(t), \\
        \dot{V}(t) = -\nabla_x\phi(t,X(t)),
    \end{dcases}
    \label{eq:characteristics}
\end{equation}
defined for $t \geq s \geq 0$ and with initial condition $(X,V)|_{t=s} = (x,v)$. Assuming that $\phi$ is sufficiently smooth, the system~\eqref{eq:characteristics} has a unique solution and one can define
the \emph{forward flow}
\begin{equation*}
    \flow_{s,t} \colon (x,v) = (X(s),V(s)) \mapsto (X(t),V(t)),
\end{equation*}
which maps phase-space positions at time $s$ to
positions at time $t$. We recall that, for all times $t \geq s \geq 0$, $\flow_{s,t}$ is a diffeomorphism. The backward flow is denoted by $\Psi_{s,t} \coloneqq \Phi_{s,t}^{-1}$.
This backward flow gives a compact expression of the solution to \eqref{eq:vlasov}, as evidenced by the following lemma.
\begin{lemma}[\cite{Sonnendrucker1999}]
    The following two statements hold.
    \begin{itemize}
        \item Let $0 = t_0 < t_1 <\dots < t_n = t$ be a partition of the time interval $[0, t]$. Then, the inverse flow map between times $0$ and $t$ is equal to the composition of the inverse flow maps over each subinterval:
              \begin{equation*}
                  \Psi_{0,t} =   \Psi_{t_0,t_1} \circ \dots \circ  \Psi_{t_{n-1},t_n}.
              \end{equation*}
        \item
              The density $f$ at time $t$ can be expressed in terms of the initial density $\finit$ and the flow
              map as:
              \begin{equation}
                  f(t,x,v) = \finit\!\bigl(\Psi_{0,t}(x,v)\bigr).
                  \label{eq:lagrangian_sol}
              \end{equation}
    \end{itemize}
\end{lemma}

\subsection{Particle-in-cell approximation}
In a \emph{full-f} PIC method, the distribution function is approximated by a
weighted sum of shape functions centered on $\Np$ numerical markers
$z_k^n = (x_k^n, v_k^n) \in \R^{2d}$:
\begin{equation}
    f^n(z) = \sum_{k=1}^{\Np} w_k\,\varphi_\varepsilon(x - x_k^n) \delta(v - v_k^n)
    \approx f(t^n,z),
    \label{eq:pic_approx}
\end{equation}
where $z=(x,v)$, and $\varphi_\varepsilon$ is a smooth shape function of integral one and width
$\varepsilon > 0$. For solving Vlasov-Poisson equations a classical choice is to use spline functions scaled with
$\varepsilon = \Delta x$ the step size of the
grid used for the Poisson solver. Specifically, we set
\begin{equation}
    \varphi_{\varepsilon}(x) :=  \frac{1}{\varepsilon^d}  \varphi \left( \frac{x}{\varepsilon} \right),
    \label{eq:shape_function}
\end{equation}
with a reference shape function $\varphi$  defined as a centered cardinal B-spline of degree $p$:
\begin{equation}
    \varphi(x) = \prod\limits_{i = 1}^d B_p(x_i), \quad \mbox{ with support } \left[ -\frac{p+1}{2}, \frac{p+1}{2} \right]^d,
    \label{eq:spline}
\end{equation}
involving standard univariate B-splines defined recursively by
\begin{equation*}
    B_0(x) := \mathbf{1}_{\left[ - \frac{1}{2}, \frac{1}{2} \right]}(x) \quad \mbox{and } B_p(x) = \int_{x - \frac{1}{2}}^{x+ \frac{1}{2}} B_{p-1}(y) \, \mathrm{d}y \mbox{ for } p \geq 1.
\end{equation*}
The weights
\begin{equation}
    w_k = \frac{\finit(z_k^0)}{\Np\,g(z_k^0)}
    \label{eq:weight_full_f}
\end{equation}
are computed at $t=0$, with $g$ the sampling distribution of the initial markers, to be determined in the numerical experiments.
Each time step consists of
computing the velocity integral of $f^n$ needed to evaluate the potential $\phi$ and the right-hand
side of~\eqref{eq:characteristics}, and pushing the markers by a discretisation of the characteristic
equations~\eqref{eq:characteristics}. If we consider an integral of the form
\begin{equation}
    I(\alpha)(t^n) = \int_{\mathbb{R}^{2d}} \alpha(z)\,f(t^n, z)\,\mathrm{d}z
    \label{eq:velocity_integral}
\end{equation}
and substitute $f(t^n)$ by its particle approximation $f^n$ (in the $\varepsilon \to 0$ limit), we find
\begin{equation}
    I^n(\alpha)
    = \int_{\mathbb{R}^{2d}} \alpha(z)\,f^n(z)\,\mathrm{d}z
    =
    \sum_{k=1}^{\Np} w_k
    \,\alpha(z_k^n)
    \label{eq:mc_estimator}
\end{equation}
Now, denote by $g^n$ the probability distribution of the markers $z_k^n$, $1 \le k \le \Np$, which is transported by the same characteristic flow as $f$: we have
$$
    w_k \approx \frac{f(t^n,z_k^n)}{\Np\,g^n(z_k^n)}
$$
hence
$$
    I^n(a)(x) \approx \frac 1 {\Np} \sum_{k=1}^{\Np} \frac{f(t^n,z_k^n)}{g^n(z_k^n)}\,\alpha(z_k^n) . 
$$
The latter term can be seen as a Monte Carlo estimate: indeed
it is the empirical mean of the random variable
$$
    X^n(z) = \frac{f(t^n,z)}{g^n(z)}\,\alpha(z)
$$
whose expectation is the desired integral $I^n(\alpha)(t^n,x)$.
The error scales as $(\sigma^2[X^n]/\Np)^{\nicefrac 1 2}$, with
the variance \cite{Aydemir1994,hatzky_reduction_2019}
\begin{equation}
    \sigma^2[X^n] = \mathbb{E}[(X^n)^2] - \mathbb{E}[X^n]^2
    = \int_{\mathbb{R}^{2d}} \frac{f(t^n,z)^2}{g^n(z)^2}\,\alpha(z)^2\,g^n(z)\,\mathrm{d}z - I(\alpha)(t^n)^2.
\end{equation}
This statistical error, or ``noise'', can be reduced by
increasing $\Np$ or by decreasing $\sigma^2[X^n]$. In some applications,
notably gyrokinetic simulations \cite{garbet2010gyrokinetic}, the required
accuracy demands a prohibitively large number $N_p$ of particles.
Hence, decreasing the variance is a better way to achieve a lower error in this case.
The $\df$ method, seen as a \emph{control variate} technique in a Monte Carlo perspective
\cite{Aydemir1994}, directly reduces $\sigma^2[X^n]$ without increasing $\Np$.
Observe that for computing $\rho(t^n)$ at a given point $x$, one would take
$\alpha(z^n) = \delta(x - x^n)$ in \eqref{eq:velocity_integral}, or a smoothed version thereof to
avoid singularities in the above discussion: this motivates the use of the shape function $\varphi_\varepsilon$ in \eqref{eq:pic_approx}.
We refer to \cite{sonnendrucker2015split,hatzky_reduction_2019} for a detailed account of the
connection between PIC methods and Monte Carlo estimation.

For completeness, we summarise here the main steps of the standard full-$f$ PIC algorithm,
which will serve as a reference for the $\df$ variant introduced in the next section.
At each time step $n$, the algorithm proceeds as follows:
\begin{enumerate}[nosep]
    \item \textbf{Half-step in position.} The positions of the markers are advanced by half a time step with periodicity:
          \begin{equation*}
              x_k^{n+1/2} = \Bigl(x_k^n + \frac{\dt}{2}\,v_k^n\Bigr) \bmod L.
          \end{equation*}
    \item \textbf{Charge deposition.} The charge density is assembled by depositing
          the markers onto a grid of size $\Delta x$:
          \begin{equation*}
              \rho^{n+1/2}(x) = \sum_{k=1}^{\Np} w_k\,\varphi_{\Delta x} (x - x_k^{n+1/2}),
          \end{equation*}
          where $\varphi_{\Delta x}$ is the shape function $\varphi_{\varepsilon}$ from \eqref{eq:shape_function} with $\varepsilon = \Delta x$.
    \item \textbf{Electric field computation.} The Poisson equation
          $-\Delta \phi^{n+1/2} = \rho^{n+1/2}$ is solved spectrally using a
          Fast Fourier Transform (FFT)-based solver on a space grid of size $\Delta x$. The electric field
          $E^{n+1/2} = -\nabla\phi^{n+1/2}$ is then interpolated back to
          the particle positions using the same shape function $\varphi_{\Delta x}$.
    \item \textbf{Full step in velocity.} The velocities are updated using the electric field:
          \begin{equation*}
              v_k^{n+1} = v_k^n + \dt\, E^{n+1/2}(x_k^{n+1/2}).
          \end{equation*}
    \item \textbf{Half-step in position.} The positions are advanced by half a time step with periodicity:
          \begin{equation*}
              x_k^{n+1} = \Bigl(x_k^{n+1/2} + \frac{\dt}{2}\,v_k^{n+1}\Bigr) \bmod L.
          \end{equation*}
\end{enumerate}
The main differences with the $\df$ method, described in the next section, are that the weights $w_k$ are fixed throughout the simulation, and that the full velocity integral of $f^n$ is evaluated by a Monte Carlo approximation.

\subsection{The $\df$ method}
\label{sec:the_df_method}

The basic idea of the $\df$ method is to reduce the amplitude of the Monte Carlo estimate~\eqref{eq:mc_estimator} in the evaluation of the velocity integrals.
To achieve this goal 
the numerical distribution function is decomposed in two parts:
\begin{equation}
    f(t^n) = \fbulk + \df(t^n)
    \qquad \text{ with } \quad
    \df(t^n) := f(t^n) - \fbulk
    \label{eq:df_ansatz}
\end{equation}
where the velocity integrals of $\fbulk$ are easy to compute, and only $\df(t^n)$ is approximated by numerical particles.
For simplicity, we consider in this section the case where $\fbulk$ is constant in time.
A typical choice for such $\fbulk$ is a local Maxwellian, or an equilibrium of the system~\cite{lanti2020}.

The integrals $I(t^n,\alpha)$ are then split accordingly, 
\begin{equation}
    I(t^n,\alpha)
    = \underbrace{\int_{\mathbb{R}^{2d}} \alpha(z)\fbulk(z)\,\mathrm{d}z}_{\text{noiseless}}
    + \underbrace{\int_{\mathbb{R}^{2d}} \alpha(z) \df(t^n,z)\,\mathrm{d}z}_{\text{Monte Carlo}}.
    \label{eq:charge_split}
\end{equation}
Here the first term can be computed exactly, and the remainder $\df(t^n)$ is
approximated by particles
\begin{equation}
    \df(t^n,z) \approx \df^n(z) \coloneqq \sum_{k=1}^{\Np} \dw_k^{n}\,\varphi_\varepsilon(x-x_k^n)\delta(v-v_k^n),
    \label{eq:df_particles}
\end{equation}
with weights given by
\begin{equation}
    \dw_k^{n} \coloneqq \frac{\finit(z_k^0) - \fbulk(z_k^n)}{N_p g^0(z_k^0)}
    \approx \frac{f(t^n,z_k^n) - \fbulk(z_k^n)}{N_p g^n(z_k^n)} = \frac{\df(t^n,z_k^n)}{N_p g^n(z_k^n)}
    \label{eq:df_weights}
\end{equation}
Substituting \eqref{eq:df_particles} and \eqref{eq:df_weights} into the second term of
\eqref{eq:charge_split} gives, analogously to \eqref{eq:mc_estimator},
\begin{equation}
    \int_{\mathbb{R}^{2d}} \alpha(z) \df(t^n,z)\,\mathrm{d}z
    \approx
    \int_{\mathbb{R}^{2d}} \alpha(z) \df^n(z)\,\mathrm{d}z
    = \frac 1 {\Np}  \sum_{k=1}^{\Np}
    \frac{\finit(z_k^0) - \fbulk(z_k^n)}{g(z_k^0)}
    \,\alpha(z_k^n). 
    \label{eq:df_mc}
\end{equation}
This sum has the same structure as \eqref{eq:mc_estimator}, but with
$\finit(z_k^0)$ replaced by $\finit(z_k^0) - \fbulk(z_k^n)$. Its error still
scales as \smash{$\Np^{-\nicefrac 1 2}$}, but with a prefactor proportional to the magnitude of the
weights $\dw_k^n$. As long as $\fbulk$ closely tracks $f^n$, these weights remain
small and the noise is significantly reduced compared to the full-$f$ estimate
\eqref{eq:mc_estimator}.
This approach comes at the cost of updating the bulk and the weights, which should be smaller than that of
increasing the number of particles to achieve the same noise reduction.

We summarise here the main steps of the static $\df$-PIC algorithm,
and highlight the differences with respect to the full-$f$ PIC method.
This will serve as a direct reference for the Neural $\df$-PIC method introduced in \cref{sec:algorithm}.
At each time step $n$, the algorithm proceeds as follows:
\begin{enumerate}[nosep]
    \item \textbf{Half-step in position.} Same as in the full-$f$ case:
          \begin{equation*}
              x_k^{n+1/2} = \Bigl(x_k^n + \frac{\dt}{2}\,v_k^n\Bigr) \bmod L.
          \end{equation*}
    \item \textbf{Weight update.} The weights are updated
          \begin{equation*}
              \dw_k^{n+1/2} = \frac{\finit(x_k^0, v_k^0) - \fbulk(x_k^{n+1/2}, v_k^{n})}{\Np\, g(x_k^0, v_k^0)}.
          \end{equation*}
    \item \textbf{Charge deposition.} The charge density is split into a noiseless bulk
          contribution and a noisy $\df$ contribution,
          \begin{equation}
              \rho^{n+1/2}(x) = \rho_0(x)
              + \sum_{k=1}^{\Np} \dw_k^{n+1/2}\,\varphi_\varepsilon(x - x_k^{n+1/2}),
              \label{eq:df_charge_deposition}
          \end{equation}
          where $\rho_0(x) = \int \fbulk(x,v)\,\mathrm{d}v$ may be computed once and stored.

    \item \textbf{Electric field computation.} Same FFT-based spectral solver as in the full-$f$ case, but $E^{n+1/2}$ is computed using the charge density split in the two terms \eqref{eq:df_charge_deposition}.
    \item \textbf{Full step in velocity.} Same as in the full-$f$ case:
          \begin{equation*}
              v_k^{n+1} = v_k^n + \dt\, E^{n+1/2}(x_k^{n+1/2}).
          \end{equation*}
    \item \textbf{Half-step in position.} Same as in the full-$f$ case:
          \begin{equation*}
              x_k^{n+1} = \Bigl(x_k^{n+1/2} + \frac{\dt}{2}\,v_k^{n+1}\Bigr) \bmod L.
          \end{equation*}

\end{enumerate}

Compared to the full-$f$ algorithm, the main differences are: (i) the charge density
is split into a precomputed bulk part and a particle $\df$ part,
and (ii) the particle weights are no longer fixed but evolve in time to track the
deviation of $f^n$ from the static bulk $\fbulk$.

\subsection{Static versus dynamic bulk densities}

In the classical $\df$ approach presented above, termed the static one, the bulk density $\fbulk$ is fixed throughout the simulation.
While simple and effective for near-equilibrium problems, this becomes inadequate when the distribution significantly evolves:
in this case, the weights $\dw_k^{n}$ grow 
and the statistical noise becomes comparable to that of a full-$f$ PIC scheme.

A natural idea is then to also evolve the bulk density $\fbulk$, albeit on a slower time-scale than that of $\df$. This approach has been explored in related works (see for instance \cite{brunner1999collisional,Allfrey2003,ku2016new} and the references therein). A natural choice, especially when collisions are involved~\cite{brunner1999collisional}, is to assume that the bulk $\fbulk$ has the form of a Maxwellian. The parameters (density, average velocity
and thermal velocity) of the background $\fbulk$ are then evolved according to fluid equations, with a closure computed from the $\df$ part. The simultaneous evolution of $\df$ and the fluid equations thus defines a self-consistent
hybrid fluid-kinetic procedure.

Another approach, presented in~\cite{CamposPinto2023}, is based on a
Forward-Backward Lagrangian (FBL) reconstruction of the bulk density.
At each remapping step, a collection of passive auxiliary markers $\tilde{z}_j$ are reset on a
Cartesian grid of spacing $h_*$, i.e., $\tilde{z}_j := jh_*$ for $j \in \mathbb{Z}^d$,
and then pushed forward by the PIC flow alongside the standard markers, in order
to track the forward characteristic flow.
The bulk density is then updated via a semi-Lagrangian step of the form
\begin{equation*}
    \fbulk^n := A_* \,\mathcal{T}_{\mathrm{fbl}}[\tilde{\mathbf{z}}^n]\, \fbulk^m,
\end{equation*}
where $A_*$ is a spline interpolation (or quasi-interpolation) operator on the $h_*$ grid, and
$\mathcal{T}_{\mathrm{fbl}}[\tilde{\mathbf{z}}^n]$ is a transport operator
that approximates the exact backward flow using a local quadratic inversion
of the auxiliary marker trajectories.
In low dimensions this approach has shown promising results,
however its cost becomes quickly expensive in high dimensions because of the isotropic nature of the backward flow reconstruction.

In this work, we propose to improve the foregoing FBL-$\df$ method through two main modifications:
\begin{itemize}
    \item First, we replace the local polynomial approximation of the backward flow by a neural network, which is trained on the numerical markers pushed by the underlying PIC code.
    \item Second, we distinguish between two ``bulk densities'':
          \begin{itemize}
              \item[(i)] a \emph{fine} representation of the full solution $f$ obtained by composing the initial density $\finit$ with the backward flow approximated by a neural network, using again the Lagrangian formula \eqref{eq:lagrangian_sol}, that is $f(t,x,v) = \finit(\Psi_{0,t}(x,v))$, and
              \item[(ii)] a \emph{coarse} representation, obtained by interpolating the fine one on a coarse grid of B-splines.
                    As the former one is easy to evaluate at any point in phase space but difficult to integrate in velocity due to the phase space filamentations, we use it to update the latter one, which is then used as a bulk density in the $\df$ PIC steps.
          \end{itemize}
\end{itemize}

The neural network that we use to approximate the backward flow $\Psi_{0,t}$ will be detailed in the following section:
it will be trained on the numerical markers computed by the PIC code.
This allows us to compute an appropriate form for $\fbulk$ without any assumption on the general shape of the bulk
and without introducing any auxiliary markers as in~\cite{CamposPinto2023}.
Furthermore, the approximation of the backward flow by a neural network can a priori handle very anisotropic flows.

\section{Symplectic neural networks}
\label{sec:sympnet}
In this section, we explain the architecture of the neural network used to
approximate the backward flow.  The characteristic flow of the Vlasov-Poisson
system is a symplectic map: its Jacobian matrix $J_{\Phi}$ satisfies $J_{\Phi}^\top \Omega J_{\Phi} =
    \Omega$, where $\Omega = \bigl(\begin{smallmatrix}0&I\\-I&0\end{smallmatrix}\bigr)$
is the standard symplectic matrix.
It is therefore natural to approximate
it by a neural network that is symplectic by construction. We begin, in \cref{sec:SympNet}, by recalling the
standard SympNet architecture \cite{Jin2020}.
Then, we describe in \cref{sec:periodic_SympNet} the periodic variant used in our experiments.

\subsection{SympNet architecture}
\label{sec:SympNet}

Symplectic neural networks were introduced in~\cite{Jin2020} as
networks that are symplectic by construction. Their architecture is motivated by the classical result that a symplectic map $\Psi:\R^{2d}\to\R^{2d}$ can be approximated by a compositions of multiple \emph{shear maps}
\begin{equation}
    \psi_{\mathrm{up}}(x,v)
    = \begin{pmatrix} x + \nabla_v T(v) \\ v \end{pmatrix},
    \qquad
    \psi_{\mathrm{down}}(x,v)
    = \begin{pmatrix} x \\ v + \nabla_x V(x) \end{pmatrix},
    \label{eq:shear_maps_exact}
\end{equation}
for scalar potentials $T:\R^d\to\R$ and $V:\R^d\to\R$.

These maps have a direct physical interpretation: $\psi_{\mathrm{up}}$ is the
exact unit-time flow of the purely kinetic Hamiltonian $H_T(x,v) = T(v)$, and
$\psi_{\mathrm{down}}$ is the exact unit-time flow of the purely potential
Hamiltonian $H_V(x,v) = V(x)$. In particular, the composition
$\psi_{\mathrm{up}} \circ \psi_{\mathrm{down}}$ with $\nabla_v T(v) = \dt\,v$
and $\nabla_x V(x) = \dt\,\nabla_x\phi(x)$ is precisely the symplectic Euler
scheme for the Vlasov-Poisson Hamiltonian $H = |v|^2/2 + \phi(x)$.

In the \emph{gradient-based} (SympNet) variant, the potentials $T$ and $V$
are parametrized by shallow neural networks.
Such a neural network is defined by
\begin{equation*}
    \begin{aligned}
        \mathcal{N}_{K,b} \colon \R^d & \to \R^d,                            \\
        x                             & \mapsto \mathbf{1}^\top\Sigma(Kx+b),
    \end{aligned}
\end{equation*}
and it is parameterized by its weights and biases $K\in\R^{w\times d}$ and $b\in\R^w$. The function $\Sigma(s)$ is of the form $\int_0^s \sigma $, with $\sigma \colon \R \to \R$ an activation function (such as a sigmoid or tanh). Classically, we also denote by $ \Sigma \colon \R^w \to \R^w$ and $\sigma \colon \R^w \to \R^w$ their elementwise extensions. The vector $\mathbf{1} \in \R^{w \times 1}$ is the vector whose components are all equal to $1$.
Then, the potentials are defined by shallow networks as
\begin{equation}
    V_{K,b}(x) = \mathcal{N}_{K,b}(x)
    \text{\quad and \quad}
    T_{K,b}(v) = \mathcal{N}_{K,b}(v),
    \label{eq:gradient_module}
\end{equation}
where we emphasize that $V_{K,b}$ and $T_{K,b}$ have different weights and biases $K$ and $b$ in practice.
Then, to compute $\psi_{\mathrm{up}}$ and $\psi_{\mathrm{down}}$,
we compute the gradients of $V_{K,b}$ and $T_{K,b}$ as
\begin{equation*}
    \nabla_x V_{K,b}(x) = K^\top \sigma(Kx + b) \eqqcolon \hat{\sigma}_{K,b}(x)
    \text{\quad and \quad}
    \nabla_v T_{K,b}(v) = K^\top \sigma(Kv + b) \eqqcolon \hat{\sigma}_{K,b}(v).
\end{equation*}
Thus, $\psi_{\mathrm{up}}$ and $\psi_{\mathrm{down}}$ read
\begin{equation*}
    \psi_{\mathrm{up}}(x,v)
    = \begin{pmatrix} x + \hat{\sigma}_{K,b}(v) \\ v \end{pmatrix}
    \text{\quad and \quad}
    \psi_{\mathrm{down}}(x,v)
    = \begin{pmatrix} x \\ v + \hat{\sigma}_{K,b}(x) \end{pmatrix}.
\end{equation*}
A SympNet $\Psi_\theta $ of depth $2 \ell$ and width $w$ is finally defined as the composition
\begin{equation}
    \Psi_\theta = \psi_{\mathrm{up}}^\ell \circ \psi_{\mathrm{down}}^\ell
    \circ \cdots \circ \psi_{\mathrm{up}}^1 \circ \psi_{\mathrm{down}}^1,
    \label{eq:sympnet}
\end{equation}
with $\theta = (K^j, b^j)_{1 \leq j \leq 2\ell}$ the trainable parameters of the network. Being a composition of
symplectic maps, $\Psi_\theta$ is symplectic for any set $\theta$ of trainable parameters.

\begin{theorem}[Universal approximation, {\cite[Theorem.~3.2]{Jin2020}}]
    Any symplectic map $\Psi:\R^{2d}\to\R^{2d}$ can be approximated arbitrarily
    well in the $C^0$ norm by a SympNet of the form \eqref{eq:sympnet}, given
    sufficient depth and width.
\end{theorem}

\begin{remark}[SympNet as a learned symplectic integrator]
    A SympNet of depth $\ell$ can be understood as a symplectic splitting
    integrator with $\ell$ steps, where the potentials $(T^j, V^j)$ at each step
    are learned from data rather than prescribed by the physics. The Störmer-Verlet
    scheme for $H = |v|^2/2 + \phi(x)$ is recovered as the special case where $\ell=2$ and
    $
        \Psi_\theta =
        \psi_{\mathrm{down}}^2 \circ \psi_{\mathrm{up}}^2
        \circ \psi_{\mathrm{up}}^1 \circ \psi_{\mathrm{down}}^1,
    $
    where the potentials are given as the physical ones instead of neural networks.
\end{remark}

\subsection{Periodic SympNet}
\label{sec:periodic_SympNet}

For the Vlasov-Poisson problem with spatial periodicity $x\in L\T^d$, the
characteristic flow satisfies $\flow_{s,t}(x+Le_j, v) = \flow_{s,t}(x,v)$ for each canonical basis vector $e_j$. We encode this constraint into the
SympNet architecture by replacing the standard gradient module in
$\psi_{\mathrm{down}}$ by a \emph{periodic gradient module}, and by introducing a modulo operator in $\psi_{\mathrm{up}}$.

To parametrise $L$-periodic functions, a natural choice (see e.g.~\cite{MilSriTanBarRamNg2021}) is to use a
one-hidden-layer network with a fixed trigonometric embedding as input features:
\begin{equation}
    x \in \R^d \;\longmapsto\;
    \begin{pmatrix}\cos(2\pi x/L)\\\sin(2\pi x/L)\end{pmatrix} \in \R^{2d}.
    \label{eq:trig_embedding}
\end{equation}
This is a standard approach for learning periodic functions with neural
networks: the trigonometric embedding encodes the
periodicity directly into the input features, so any network built on top
of it is automatically $L$-periodic.
A natural building block to parameterize a $L$-periodic gradient module (to be used in a velocity shear map) is then
\begin{equation}
    \tilde{V}_{K_1,K_2,b_1,b_2}(x) =
    \mathbf{1}^\top \Sigma\left(K_1\cos\left(\frac{2\pi x}{L}\right)+b_1\right)
    + \mathbf{1}^\top \Sigma\left(K_2\sin\left(\frac{2\pi x}{L}\right)+b_2\right),
    \label{eq:periodic_potential}
\end{equation}
where the weights and biases are $K_1,K_2\in\R^{w\times d}$ and $b_1,b_2\in\R^w$,
and where $\Sigma(s) = \int_0^s\sigma$ remains the antiderivative of an activation function $\sigma$.
Since the embedding \eqref{eq:trig_embedding} is $L$-periodic, so is \smash{$\tilde{V}_{K_1,K_2,b_1,b_2}$},
for any values of the parameters and the activation function.

\begin{definition}[Periodic gradient module]
    \label{def:periodic_module}
    A periodic gradient module is defined by
    \begin{equation}
        \tilde{\sigma}_{K_1,K_2,b_1,b_2} =
        \nabla_x\tilde{V}_{K_1,K_2,b_1,b_2}.
    \end{equation}
    In other words, for all $x \in \R^d$,
    \begin{equation}
        \begin{aligned}
            \tilde{\sigma}_{K_1,K_2,b_1,b_2}(x) =
             & -\frac{2\pi}{L}\,K_1^\top\!\left[\sigma\left(K_1\cos\left(\frac{2\pi x}{L}\right)+b_1\right)
            \odot \sin\left(\frac{2\pi x}{L}\right)\right]                                                  \\
             & +\frac{2\pi}{L}\,K_2^\top\!\left[\sigma\left(K_2\sin\left(\frac{2\pi x}{L}\right)+b_2\right)
                \odot \cos\left(\frac{2\pi x}{L}\right)\right],
            \label{eq:periodic_module}
        \end{aligned}
    \end{equation}
    where $\odot$ denotes elementwise multiplication. Since $\tilde{V}$ is
    $L$-periodic, so is $\tilde{\sigma}$.
\end{definition}
The symplecticity of the resulting shear map follows from the same argument as
in the non-periodic case.
Dropping the subscripts for clarity, since $\tilde{\sigma} = \nabla_x\tilde{V}$,
the Jacobian matrix of
\begin{equation}
    (x,v) \mapsto \tilde{\psi}_{\mathrm{down}}(x,v) = \bigl(x,\; v + \tilde{\sigma}(x)\bigr)
    \label{eq:periodic_down}
\end{equation}
is lower triangular with unit diagonal.
Therefore, $\tilde{\psi}_{\mathrm{down}}$ is a symplectic map;
this result holds for all weights and biases $(K_1,K_2,b_1,b_2)$.

To make sure that the position shear map is $L$-periodic in $x$, we introduce a modulo operation, and we define
\begin{equation}
    \tilde{\psi}_{\mathrm{up}}(x,v)
    = \bigl((x + \hat{\sigma}(v))\bmod L,\; v\bigr).
    \label{eq:periodic_up}
\end{equation}
where $\hat{\sigma}$ is the gradient module introduced in the non-periodic case.
\begin{definition}[Periodic SympNet]
    A periodic SympNet of depth $2\ell$ is the composition
    \begin{equation}
        \Psi_\theta =
        \tilde{\psi}_{\mathrm{up}}^\ell \circ \tilde{\psi}_{\mathrm{down}}^\ell
        \circ \cdots \circ
        \tilde{\psi}_{\mathrm{up}}^1 \circ \tilde{\psi}_{\mathrm{down}}^1,
        \label{eq:periodic_sympnet}
    \end{equation}
    where each \smash{$\tilde{\psi}_{\mathrm{down}}^j$} and \smash{$\tilde{\psi}_{\mathrm{up}}^j$}
    uses their own parameters
    $(K_1^j, K_2^j, b_1^j, b_2^j)$ and $(K^j, b^j)$, respectively.

\end{definition}
The trainable parameters of the periodic SympNet are finally grouped in the vector
\begin{equation*}
    \theta = (K_1^j, K_2^j, b_1^j, b_2^j, K^j, b^j)_{1 \leq j \leq \ell}.
\end{equation*}
Note that this vector is larger than the one corresponding to the traditional SympNets. The total number of parameters in this case is $3 w \ell (d+1)$.
Each factor is symplectic, so $\Psi_\theta$ is symplectic for every $\theta$,
and $\Psi_\theta:L\T^d\times\R^d\to L\T^d\times\R^d$ is smooth.

\section{The Neural \texorpdfstring{$\df$}{delta-f}-PIC algorithm}
\label{sec:algorithm}
We finally describe our proposed Neural $\df$-PIC scheme, combining periodic SympNets from \cref{sec:sympnet} with the $\df$-PIC scheme from \cref{sec:vlasov}.

\subsection{Algorithm overview}
We recall that the bulk density is approximated by projecting a fine representation (involving a neural network flow) on a coarse spline grid.
This bulk density is updated every $N_{\Psi}$ time steps.
The goal is then to compute, at each time step $n$, a density of the form
\begin{equation*}
    f^n = \fbulk^m + \df^n,
    \text{\quad where \quad}
    m = \left\lfloor\frac{n}{N_{\Psi}} \right\rfloor.
\end{equation*}
\begin{itemize}
    \item The $\df$ part is updated at each time step. It involves computing the markers' positions and velocities $(x_k^n,v_k^n)$, and their weights $\dw_k^n$. It takes the form of~\eqref{eq:df_particles}.

    \item The bulk density $\fbulk^m$ is updated every $N_{\Psi}$ time steps, i.e., when $n=m  N_{\Psi}$. It is updated by first computing a neural network $\Psi_{\theta^m}$ that is trained on data corresponding to particle pairs $\{ (x_k^n, v_k^n), (x_k^{n-N_{\Psi}}, v_k^{n-N_{\Psi}}) \}$ at times $t^n$ and $t^{n-N_{\Psi}}$. The fine representation is next expressed by composing the approximate backward flows and the initial distribution:
          \begin{equation}
              \widetilde{\fbulk^m} = \finit(\Psi_{\theta^1} \circ \dots \circ \Psi_{\theta^m}),
              \label{eq:fbulk_fine}
          \end{equation}
          which is then approximated on a coarse B-spline grid,
          \begin{equation}
              \fbulk^m := A_{*} \widetilde{\fbulk^m}
              \label{eq:fbulk_coarse}
          \end{equation}
          in order to update the velocity integrals and the weights in the $\df$ particle scheme.
          Here, $A_{*}$ is a spline interpolation (or quasi-interpolation) operator: a simple choice is to take
          \begin{equation}
              A_* \widetilde{\fbulk^m}(x,v) = \sum_{\bi \in \{1, \dots, N_x\}^d} \sum_{\bj \in \{1, \dots, N_v\}^d}
              \widetilde{\fbulk^m}(x_\bi,v_\bj)
              \varphi_{\Delta x} (x -x_\bi) \varphi_{\Delta v} (v -v_\bj),
              \label{eq:spline_interpolation}
          \end{equation}
          where $x_\bi$, $v_\bj$ are the nodes of a cartesian spline grid with $N_x$ (respectively $N_v$) points per
          spatial (respectively velocity) dimension, and $\varphi_{\Delta x}$, $\varphi_{\Delta v}$ are B-splines on these respective grids. Note that \eqref{eq:spline_interpolation} corresponds to a smoothing of the piecewise interpolation of $\widetilde{\fbulk^m}$ on the coarse grid. In practice we have used cubic splines to ensure a smooth bulk density.
\end{itemize}

\begin{figure}[!ht]
    \begin{tikzpicture}[
            >=Stealth,
            sketchGreen/.style={color=teal!80!black},
            sketchOrange/.style={color=orange!70!black},
            sketchBlue/.style={color=blue!80},
            nodeBox/.style={draw, minimum width=1.4cm, minimum height=1.0cm,
                    very thick, font=\small, align=center, rounded corners=2pt},
            nodeCircle/.style={draw, circle, inner sep=1pt, minimum size=0.52cm, very thick}
        ]

        \draw[very thick, ->] (0,0) -- (14.5,0) node[below left] {$t$};

        \foreach \x in {0.7, 1.4, 2.1, 2.8, 3.5, 4.2, 4.9, 5.6,
                7.0, 7.7, 8.4, 9.1, 9.8, 10.5, 11.2, 11.9, 12.6, 13.3} {
                \draw[very thick] (\x, 0.07) -- (\x, -0.07);
            }

        \node[above, font=\small] at (0.35, 0.06) {$\Delta t$};

        \node[nodeCircle, sketchGreen] (c0) at (0,0) {};
        \fill[sketchGreen] (0,0) circle (1.5pt);

        \node[nodeCircle, sketchGreen] (c1) at (2.8,0) {\tiny $+$};
        \node[nodeCircle, sketchGreen] (c2) at (5.6,0) {\tiny $+$};


        \node[nodeCircle, sketchGreen]  (c4) at (9.1,0) {\tiny $+$};
        \node[nodeCircle, sketchGreen]  (c4) at (11.9,0) {\tiny $+$};


        \def\boxHeight{1.5}

        \node[nodeBox, sketchGreen]  (box0) at (0,\boxHeight)    {bulk\\($\fbulk$)};
        \node[nodeBox, sketchGreen]  (box1) at (2.8,\boxHeight)  {new\\bulk};
        \node[nodeBox, sketchGreen]  (box2) at (5.6,\boxHeight)  {new\\bulk};

        \node[nodeBox, sketchGreen]  (box3) at (9.1,\boxHeight) {new\\bulk};
        \node[nodeBox, sketchGreen]  (box4) at (11.9,\boxHeight) {new\\bulk};

        \node[font=\normalsize] at (13.5, \boxHeight) {$\cdots$};

        \draw[dashed, sketchGreen, very thick] (box0.east) -- (box1.west);
        \draw[dashed, sketchGreen, very thick] (box1.east) -- (box2.west);
        \draw[dashed, sketchGreen, very thick] (box3.east) -- (box4.west);
        \node[font=\normalsize] at (7.5, \boxHeight) {$\cdots$};
        \def\picX{0.0}
        \def\picY{3.8}

        \node[sketchBlue, font=\bfseries\large, anchor=west]
        at (\picX-0.5, \picY) {PIC};

        \fill[sketchBlue] (\picX-0.35, \picY-0.35) circle (2pt);
        \fill[sketchBlue] (\picX-0.15, \picY-0.45) circle (2pt);
        \fill[sketchBlue] (\picX-0.25, \picY-0.6)  circle (2pt);
        \fill[sketchBlue] (\picX-0.5,  \picY-0.5)  circle (2pt);

        \def\arcY{3.2}
        \def\arcH{0.4}
        \def\gap{0.07}

        \foreach \xstart/\xend in {
                0.0/0.7, 0.7/1.4, 1.4/2.1, 2.1/2.8, 2.8/3.5, 3.5/4.2, 4.2/4.9, 4.9/5.6} {
                \draw[sketchBlue, very thick, ->]
                (\xstart+\gap, \arcY)
                .. controls (\xstart+\gap+0.1, \arcY+\arcH) and (\xend-\gap-0.1, \arcY+\arcH) ..
                (\xend-\gap, \arcY);
            }

        \node[sketchBlue, font=\normalsize] at (6.7, \arcY+0.1) {$\cdots$};

        \foreach \xstart/\xend in {
                7.7/8.4, 8.4/9.1, 9.1/9.8, 9.8/10.5, 10.5/11.2, 11.2/11.9, 11.9/12.6} {
                \draw[sketchBlue, very thick, ->]
                (\xstart+\gap, \arcY)
                .. controls (\xstart+\gap+0.1, \arcY+\arcH) and (\xend-\gap-0.1, \arcY+\arcH) ..
                (\xend-\gap, \arcY);
            }

        \node[sketchBlue, font=\normalsize] at (13.4, \arcY+0.1) {$\cdots$};

        \draw[decorate, decoration={brace, amplitude=7pt, mirror}, sketchGreen, very thick]
        (0,-0.5) -- (2.8,-0.5)
        node[sketchGreen, midway, yshift=-0.55cm, font=\normalsize] {$N_\Psi \Delta t$};

        \draw[decorate, decoration={brace, amplitude=7pt, mirror}, sketchGreen, very thick]
        (2.8,-0.5) -- (5.6,-0.5)
        node[sketchGreen, midway, yshift=-0.55cm, font=\normalsize] {$N_\Psi \Delta t$};

        \draw[decorate, decoration={brace, amplitude=7pt, mirror}, sketchGreen, very thick]
        (9.1,-0.5) -- (11.9,-0.5)
        node[sketchGreen, midway, yshift=-0.55cm, font=\normalsize] {$N_\Psi \Delta t$};

    \end{tikzpicture}
    \caption{The Neural $\df$ algorithm. Particles are pushed by a PIC scheme
        (blue arrows), and after every $N_\Psi$ time steps, the bulk density $\fbulk$is updated by
        first computing a new fine representation $\widetilde{\fbulk}$ which composes $\finit$ with a neural backward flow (a SympNet trained on the particles' positions at two given times), and approximating this fine representation on a coarse spline grid
        as described in \eqref{eq:fbulk_fine}--\eqref{eq:fbulk_coarse}.
        Note that when an incremental training strategy is used, a given flow network may be reused over different time intervals of increasing size (not pictured here).}
    \label{fig:sketch_df_PIC}

\end{figure}
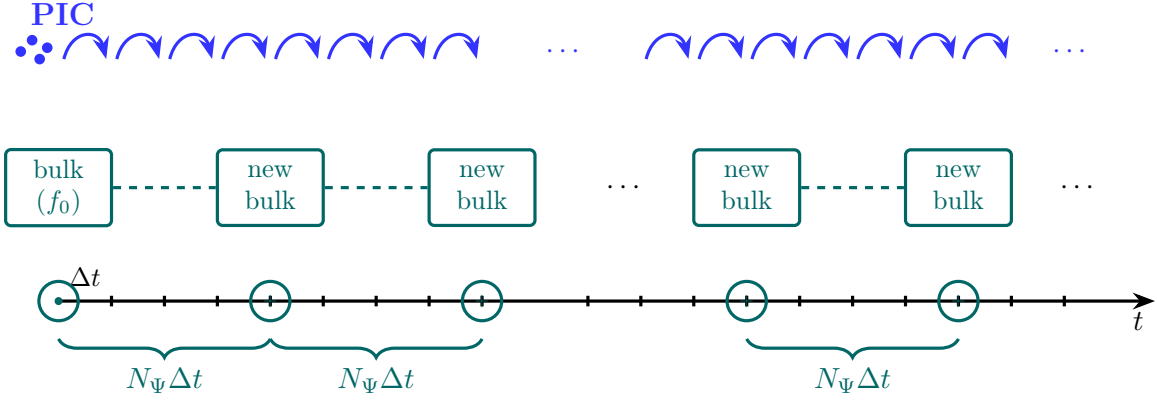

The particle pusher and the
Poisson solver remain standard components~\cite{Birdsall1991}: a leap-frog
(Strang splitting) scheme advances the markers at the fine timescale $\dt$,
and a spectral FFT-based solver computes the self-consistent electric field.
The novel contribution of this work is the dynamic update of the bulk density
$\fbulk^m$ using symplectic neural networks, which are trained on the particle trajectories to
approximate the backward characteristic flow.
The proposed algorithm is illustrated on \cref{fig:sketch_df_PIC}.

\subsection{Incremental training of the neural flows.}
\label{sec:inc_train}

Above we have considered for simplicity that each flow of the form $\Psi_{[m\dt_\Psi, (m+1)\dt_\Psi]}$, with $\dt_{\Psi} = N_{\Psi} \Delta t$, was approximated by a distinct neural network $\Psi_{\theta^{m+1}}$ trained on the particle pairs
$\{(x_k^{m N_{\Psi}}, v_k^{m N_{\Psi}}), (x_k^{(m+1) N_{\Psi}}, v_k^{(m+1) N_{\Psi}})\}_{k=1}^{\Np}$.
In practice however, it often happen that the previous network, say $\Psi_{\theta^m}$, which has been trained for the flow over the time interval $[(m-1)\dt_\Psi, m\dt_\Psi]$, is also able to accurately approximate the flow over the larger time interval
$[(m-1)\dt_\Psi, (m+1)\dt_\Psi]$.
In such a case, a natural strategy is to reuse this network $\Psi_{\theta^m}$ with its current weights $\theta^m$ as a warm start, and train it on the new data corresponding to the extended time interval, that is on the particle pairs
$\{(x_k^{(m-1) N_{\Psi}}, v_k^{(m-1) N_{\Psi}}), (x_k^{(m+1) N_{\Psi}}, v_k^{(m+1) N_{\Psi}})\}_{k=1}^{\Np}$.
We still denote the resulting weights by $\theta^{m+1}$, but the corresponding network
$\Psi_{\theta^{m+1}}$ now approximates the flow over a larger time interval.
In terms of network flows, this approach corresponds to replacing
\begin{equation}
    \Psi_{\theta^{m}} \circ \Psi_{\theta^{m+1}} \rightsquigarrow \Psi_{\theta^{m+1}}
    \label{eq:inc_train}
\end{equation}
and discarding the previous parameters $\theta^{m}$ since they are no longer needed.
Compared to training a new network for a $\dt_\Psi$ time interval, this approach saves computational resources and avoids composing several network flows (which may result in a loss of accuracy) where a single one can be used.
We call this strategy an \emph{incremental training} of the neural flows.
In terms of non-linear approximation of complex flows, this approach is similar to a preconditioning. It may also be seen as a form of \emph{curriculum learning}~\cite{bengio2009curriculum,wang2021survey}, where the training data is presented in a sequence of increasing complexity.

In practice we apply this strategy with a prescribed tolerance $\varepsilon_{\mathrm{tol}}$. At each bulk update step $(m+1)$, we first try to extend the time interval covered by the current network.
If the approximation loss reaches the tolerance then this updated network is used for the larger time interval, otherwise the
previous network (with its weights $\theta^m$) is kept and a new network is trained (from scratch) for the last sub-interval, i.e., $[m \dt_{\Psi}, (m+1) \dt_{\Psi}]$.
This adaptive strategy balances the competing goals of minimising the number of stored networks while ensuring that each network remains accurately trained.

\subsection{Initialisation of the markers and the bulk density.}

The initialisation consists in drawing $\Np$ numerical markers in phase space,
\begin{equation}
    (x_k^0, v_k^0) \in L\T^d \times \R^d, \quad k = 1, \dots, \Np,
    \label{eq:initial_markers}
\end{equation}
from a sampling distribution $g$. In all our experiments we use the Gaussian density
\begin{equation}
    g(x,v) = \frac{1}{V_{\mathrm{tot}}}
    \frac{1}{(2\pi v_{\mathrm{th}}^2)^{d/2}}\,
    e^{-|v|^2 / (2v_{\mathrm{th}}^2)},
    \label{eq:sampling_dist}
\end{equation}
where $V_{\mathrm{tot}} = L^d$ is the volume of the spatial domain and
$v_{\mathrm{th}} > 0$ is a thermal velocity chosen to cover the support of
$\finit$ in velocity space.
Setting then $\widetilde{\fbulk^0} = \finit$, the initial bulk density is computed on the coarse spline grid as
\begin{equation}
    \fbulk^0 = A_* \finit,
    \label{eq:fbulk_init}
\end{equation}
where $A_*$ is the spline quasi-interpolation operator defined in \eqref{eq:spline_interpolation}.


\subsection{$\df$-PIC phase}
\label{sec:pic_step}

Each PIC step with index $n \in \mathbb{N}$ advances the markers from time $t^n$ to $t^{n+1} = t^n + \dt$,
using a standard leap-frog (Strang splitting) scheme. As above we let $N_{\Psi} > 0$ be the number of PIC steps between two updates of the neural network representing the backward flow, and we denote by \smash{$m = \lfloor\frac{n}{N_{\Psi}} \rfloor $} the index of the last bulk update.
At the beginning of the PIC step $n$ we thus assume that the corresponding bulk density
$\fbulk^m = A_* \widetilde{\fbulk^m}$ (which is frozen until the next bulk update)
has been computed and stored on the coarse spline grid.
Accordingly, its charge density
\begin{equation}
    \rho_0^m(x) = \int_{\mathbb{R}^d} \fbulk^m(x,v)\,\mathrm{d}v
    = \sum_{\bi \in \{1, \dots, N_x\}^d} \widetilde{\fbulk^m}(x_\bi) \varphi_{\Delta x} (x -x_\bi),
    \label{eq:bulk_charge_density}
\end{equation}
may be computed exactly and stored as well.

\begin{enumerate} \item
          \textbf{Half-step in position.}
          We start with a predictive half-step, updating the positions of the particles:
          \begin{equation}
              x_k^{n+1/2} = \Bigl(x_k^n + \frac{\dt}{2}\,v_k^n\Bigr) \bmod L,
              \label{eq:half_step_x}
          \end{equation}
          where the modulo operation enforces the periodic boundary condition
          $x \in L\T^d$.
    \item \textbf{Weight update.}
          The weights of the $\df$ part are updated as
          \begin{equation}
              \dw_k^{n+1/2}
              = \frac{\finit(x_k^0, v_k^0) - \fbulk^m(x_k^{n+1/2}, v_k^{n})}
              {\Np\, g(x_k^0, v_k^0)}.
              \label{eq:weight_update}
          \end{equation}
    \item
          \textbf{Electric field computation.}
          The total charge density at the half-step is the sum of the exact contribution \eqref{eq:bulk_charge_density}
          of the spline bulk and the Monte Carlo evaluation of the $\df$ part, corresponding to \eqref{eq:df_mc}
          with $\alpha(z_k) = \varphi_{\Delta x}(x-x_k)$, that is:
          \begin{equation}
              \rho^{n+1/2}(x) = \rho_0^m(x)
              + \sum_{k=1}^{\Np} \dw_k^{n+1/2}\,\varphi_{\Delta x} \!\Bigl(x - x_k^{n+1/2}\Bigr).
              \label{eq:charge_assembly}
          \end{equation}
          The Poisson equation $-\Delta \phi^{n+1/2} = \rho^{n+1/2}$ is then solved with a spectral, FFT-based scheme
          and the electric field $E^{n+1/2} \approx -\nabla \phi^{n+1/2}$ is interpolated 
          from the resulting point values, using the same shape function $\varphi_{\Delta x}$.
          We write this entire procedure as
          \begin{equation*}
              E^{n+1/2} = \mathcal{F}((x_k^{n+1/2}, \dw_k^{n+1/2})_{1 \le k \le \Np}, \rho_0^m).
          \end{equation*}

    \item
          \textbf{Full step in velocity, half-step in position.}
          Equipped with the electric field and positions at the half-step, we complete the update of the velocity and position as follows:
          \begin{equation*}
              v_k^{n+1} = v_k^n + \dt\, E^{n+1/2}\Bigl(x_k^{n+1/2}\Bigr),
              \qquad \qquad
              x_k^{n+1}  = \Bigl(x_k^{n+1/2} + \frac{\dt}{2}\,v_k^{n+1}\Bigr) \bmod L.
          \end{equation*}

\end{enumerate}

\subsection{Update of the neural flow and the bulk density}

\paragraph{Training of the backward flow.}
\label{sec:training}

Recall that $N_{\Psi} > 0$ is the number of PIC substeps between two updates of the neural flow network.
According to the incremental training strategy described in Section~\ref{sec:inc_train}, at each update step
$m+1$ we first consider the last network $\Psi_{\theta^{m}}$ which has been trained to approximate the backward flow over a time interval of the form $[r \dt_\psi, m \dt_\psi]$ with $r < m$, and train it for the flow corresponding to the
larger time interval  $[r \dt_\psi, (m+1) \dt_\psi]$.

For this we use the particle pairs $\{ (x_k^{rN_{\Psi}}, v_k^{rN_{\Psi}}), (x_k^{(m+1)N_{\Psi}}, v_k^{(m+1)N_{\Psi}}) \}_{k=1}^{\Np}$ as training data, and a loss function defined as
\begin{equation}
    \mathcal{L}(\theta) 
    = \frac{1}{\Np}\sum_{k=1}^{\Np}
    \bigl\|\Psi_\theta(x_k^{(m+1)N_{\Psi}},v_k^{(m+1)N_{\Psi}}) - (x_k^{rN_{\Psi}},v_k^{rN_{\Psi}})\bigr\|^2,
    \label{eq:training_loss}
\end{equation}
so that the resulting network $\Psi_{\theta^{m+1}}$ is a good approximation of the associated backward flow.
If the optimisation fails to reach the prescribed tolerance $\varepsilon_{\mathrm{tol}}$, then we train a new network
for the last time interval only, corresponding to the same loss function but with $r = m$.
The minimisation of \eqref{eq:training_loss} is performed in two steps: the first step use the classical optimiser Adam~\cite{kingma2015adam}, then we switch to the
\emph{natural gradient} method \cite{Amari1998}, which preconditions the
gradient by the Fisher information matrix of the network. Compared to
standard gradient descent, the natural gradient accounts for the Riemannian
geometry of the parameter space and typically achieves better convergence in
practice \cite{muller2023}. Here, since the loss function~$\eqref{eq:training_loss}$
corresponds to a $L^2$ minimisation, the natural gradient method is equivalent to the Gauss-Newton algorithm.
Using this algorithm enables much more precise training,
but at the cost of solving an ill-conditioned linear system with a full matrix
at each iteration of the optimisation process.


\paragraph{Bulk density update.}

Given a composed backward flow at time $t^{mN_\Psi}$, of the form
\begin{equation}
    \Psi^m = \Psi_{\theta^{r}}\circ\cdots\circ\Psi_{\theta^m}
    \label{eq:composed_flow}
\end{equation}
(the precise number of networks in the composition depending on the convergence of the sucessive trainings in the adaptive incremental strategy,
see Section~\ref{sec:inc_train}), the bulk density at update step $m$ is computed in two steps:
First, the fine representation is defined by the Lagrangian transport formula
\begin{equation}
    \widetilde{\fbulk^m}(x,v) = \finit\!\bigl(\Psi^m(x,v)\bigr),
    \label{eq:bulk_eval}
\end{equation}
and it is then approximated on the coarse B-spline grid through
\begin{equation*}
    \fbulk^m = A_{*} \widetilde{\fbulk^m}
\end{equation*}
as described in \eqref{eq:spline_interpolation}. The resulting bulk density is then used in the following $N_{\Psi}$ PIC steps, until the next update of the neural flow.

We note that some structural properties readily follow from our approach:
\begin{itemize}
    \item \emph{Positivity}: we have
          $$
              \widetilde{\fbulk^{m}}(x,v) \geq 0
              \qquad \text{ and } \qquad \fbulk^{m}(x,v) \geq 0
          $$
          as a result of $\finit \geq 0$ (for the first property) and the spline nodal formula \eqref{eq:spline_interpolation} (for the second one).

    \item \emph{Mass conservation}: using a change of variables and the fact that
          a symplectic map preserves the volume in phase space, we have
          \begin{equation}
              \int_{\mathbb{R}^d} \int_{L\T^d} \widetilde{\fbulk^{m}}(x,v)\,\mathrm{d}x\,\mathrm{d}v
              = \int_{\mathbb{R}^d} \int_{L\T^d} \finit(x,v)\,\mathrm{d}x\,\mathrm{d}v.
              \label{eq:mass_conservation}
          \end{equation}
\end{itemize}
We note that the mass conservation property does not a priori hold for the spline bulk $\fbulk^m$, but since the latter
is periodically recomputed from $\widetilde{\fbulk^{m}}$, we do not expect large deviations over long time ranges.
Moreover a rescaling of the spline coefficients in \eqref{eq:spline_interpolation} is always possible if that should be an issue.
Finally we emphasize that the above structural properties hold for \emph{any} values of the parameters
$\theta^r,\ldots,\theta^{m}$, regardless of training accuracy.

\begin{remark}
    In long-time simulations, the number of flow networks involved in the composition \eqref{eq:composed_flow}
    can increase significantly, together with the evaluation cost. One mitigation strategy is to periodically project the fine density
    on a space of neural networks: every $K < m$ bulk updates, train a MLP $f_{\mu}$
    to approximate the full composed map $\widetilde{\fbulk^m}$ from a sampling of the phase space, and reset the list of learned backward flows. This resets the evaluation
    cost back to $\mathcal{O}(\ell w)$ every $K$ bulk updates. In the numerical experiments
    reported in this manuscript, the number of bulk updates is small enough that
    the cost growth is not a bottleneck, as the cost of training the different networks remains by far the main computational task.
    Conceptually, such a reset plays the role of a remapping in the semi-Lagrangian sense: a re-approximation of the solution that discards the accumulated flow and thus loses some information. In this respect, our method is only \emph{weakly} semi-Lagrangian: in contrast with standard semi-Lagrangian schemes, which remap at every time step, this operation would only be needed in long time simulations. The Characteristic Mapping Method~\cite{krah2024characteristic} is weakly semi-Lagrangian in the same spirit, the flow map being built over long time intervals rather than remapped at every step.
\end{remark}

\begin{algorithm}[H]
    \caption{Summary of \texttt{full-f}, \texttt{delta-f} and \texttt{neural-delta-f} schemes}
    \label{alg:neural_df_pic}
    \begin{algorithmic}[1]
        \State \textbf{Initialisation:}
        Draw markers $(x_k^0, v_k^0)$ from $g$;

        \textbf{if} \texttt{full-f}: set $\fbulk^0 = 0$,
        \textbf{else}: $\fbulk^0 = \finit$ or a spline approximation of it.

        \textbf{if} \texttt{full-f}: set $\dw_k = \frac{\finit(x_k^0, v_k^0)}{\Np\, g(x_k^0, v_k^0)}$

        Compute $\rho_0^0 = \int \fbulk^0 dv$.

        \For{$n = 0, 1, 2, \ldots$}
        \State \textbf{// Particles update}
        \State $x_k^{n+1/2} = (x_k^n + \frac{\dt}{2} v_k^n) \bmod L$
        \State Assemble field:
        \State $\quad$
        \textbf{if} \texttt{full-f}: Set $\dw_k^{n+1/2} = \dw_k$:
        \textbf{else}: $\dw_k^{n+1/2} = \frac{\finit(x_k^0, v_k^0) - \fbulk^m(x_k^{n+1/2}, v_k^{n})}{\Np\, g(x_k^0, v_k^0)}$;
        \State $\quad$ Compute $E^{n+1/2} = \mathcal{F}(x^{n+1/2}, \dw^{n+1/2}, \rho_0^m)$ via~\eqref{eq:charge_assembly}
        \State $v_k^{n+1} = v_k^n + \dt\, E^{n+1/2}(x_k^{n+1/2})$
        \State $x_k^{n+1} = (x_k^{n+1/2} + \frac{\dt}{2} v_k^{n+1}) \bmod L$
        \If{\texttt{neural-delta-f} and $ n+1 = (m+1) {N_\Psi}$}
        \State \textbf{// Bulk update }

        \State Train $\Psi_{\theta^{m+1}}$ by minimising \eqref{eq:training_loss}
        using natural gradient descent
        \State Update the fine representation of the density: $\widetilde{\fbulk^{m+1}} = \widetilde{\fbulk^{m}} \circ \Psi_{\theta^{m+1}}$
        \State Approximate it on the coarse spline grid: $\fbulk^{m+1} = A_* \widetilde{\fbulk^{m+1}}$
        \State Compute $\rho_0^{m+1} = \int \fbulk^{m+1} dv$
        and store on spatial grid
        \Else
        \State Set $\fbulk^{m+1} = \fbulk^{0}$ and $\rho_0^{m+1} = \rho_0^{0}$
        \EndIf
        \EndFor
    \end{algorithmic}
\end{algorithm}

\section{Numerical results}
\label{sec:numerics}



In this section we perform several experiments to assess the accuracy of our neural $\df$-PIC solver.
To this end we run several Vlasov-Poisson test cases in 1D1V and 3D3V, and compare the results with
a standard $\df$-PIC scheme which relies on a static bulk density as described above.

These test-cases have been chosen because they are standard, however one shoud keep in mind that their solutions do not deviate very much from the initial distribution, so that a standard $\df$-PIC scheme with static bulk density is expected to perform well already. The goal of our experiments is thus to show that the neural $\df$-PIC scheme performs at least as well as a standard $\df$-PIC scheme, while being able to adapt to more complex situations where the bulk density evolves significantly in time.
In all experiments, the particle pusher uses a Strang splitting scheme and a spectral Poisson solver with 32 grid cells per direction,
and in the neural $\df$ scheme we represent our bulk densities with cubic B-splines on a grid of 32 cells per direction.
We also compare our results with backward semi-Lagrangian (BSL) schemes which compute accurate solutions at the price of meshing the phase space with fine grids: In 1D1V, we use a standard BSL scheme with directional splitting on a fine $1024\times 1024$ phase-space grid, and in 3D3V we use the BSL6D code presented in~\cite{schild2024performance}.

The hyperparameters of the networks are chosen as follows. All SympNets we use here have the same architecture, with $\ell = 10$ layers of width $w = 8$ with $\tanh$ as activation functions. The total number of parameters per network is then 480 in 1D1V and 960 in 3D3V (we recall that for the periodic Sympnet proposed in this paper, the total number of parameters is $3w\ell(d+1)$).
To train each network, we use the Adam optimizer for 200 epochs, followed by 500 steps of natural gradient descent in 1D1V and 1000 steps of natural gradient in 3D3V. Furthermore, we fixed a tolerance of $\varepsilon_{\mathrm{tol}} = 10^{-5}$ for the curriculum learning of the networks.

In our experiments, the results turned out to be fairly robust with respect to the network architecture and to the training hyperparameters, as long as the networks are expressive enough and trained for sufficiently many epochs. The parameter that genuinely influences the results is $N_\Psi$, the number of time steps between two consecutive trainings. It should be large enough to keep the number of trainings, and hence the computational cost, moderate, yet small enough so that the marker displacement between two trainings remains not too large to be accurately learned by the network.

We emphasize that the phase-space density we show in all the numerical experiments is the fine bulk density $\widetilde{f_0^m}$.
\subsection{Numerical study of flow learning}
\label{sec:numerical_study_flow_learning}

Before investigating the performance of our neural $\df$-PIC scheme, we investigate the ability of periodic SympNets to learn a given characteristic flow of the Vlasov-Poisson system.

In this section we consider a flow associated with the 1D1V two-stream instability described in \cref{sec:tsi} below,
between two times $T_0=30$ and $T_1=35$, chosen somewhat arbitrarily. Concretely, we first compute a discrete solution
$f(T_0)$ at time $T_0$ by using a grid-based BSL scheme, and next use this solution as the initial condition of a standard PIC scheme (by drawing particles with a Maxwellian probability and weighting them according to $f(T_0)$) that we run between times $T_0$ and
$T_0+t$, for $t \in \{1, 2, \ldots, 5\}$.
The resulting sets of particle coordinates for the times $T_0$ and $t$ are then decomposed in two sets of equal size: the first half is used as training data for periodic SympNets of various depths and widths to approximate the backward flow $\Psi_{[T_0,T_1]}$ as described above, and the second half is used as a test set to evaluate the performance of the trained networks.

\begin{figure}[!ht]
    \centering
    \begin{subfigure}[b]{0.32\textwidth}
        \centering
        \includegraphics[width=\textwidth]{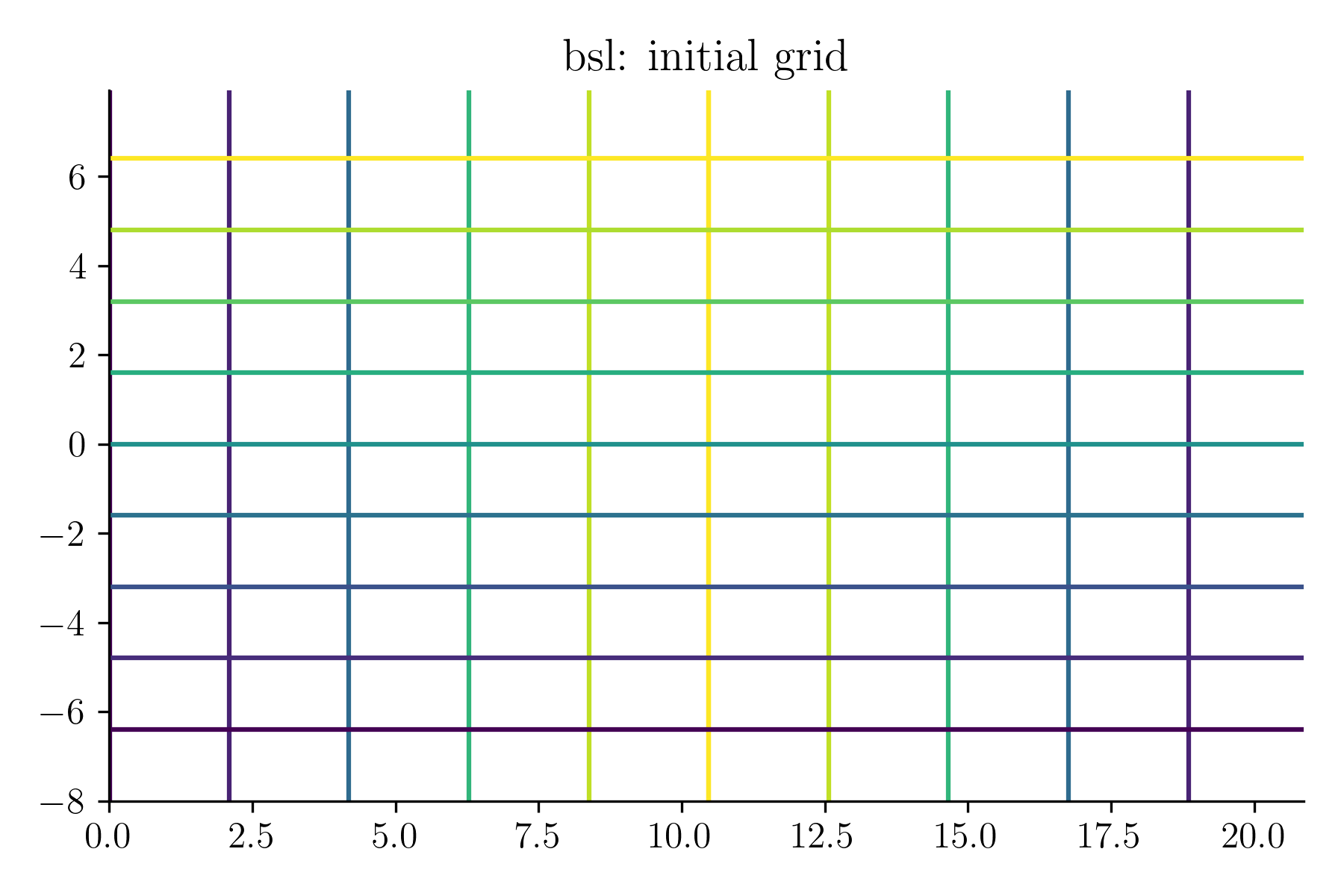}
    \end{subfigure}
    \hfill
    \begin{subfigure}[b]{0.32\textwidth}
        \centering
        \includegraphics[width=\textwidth]{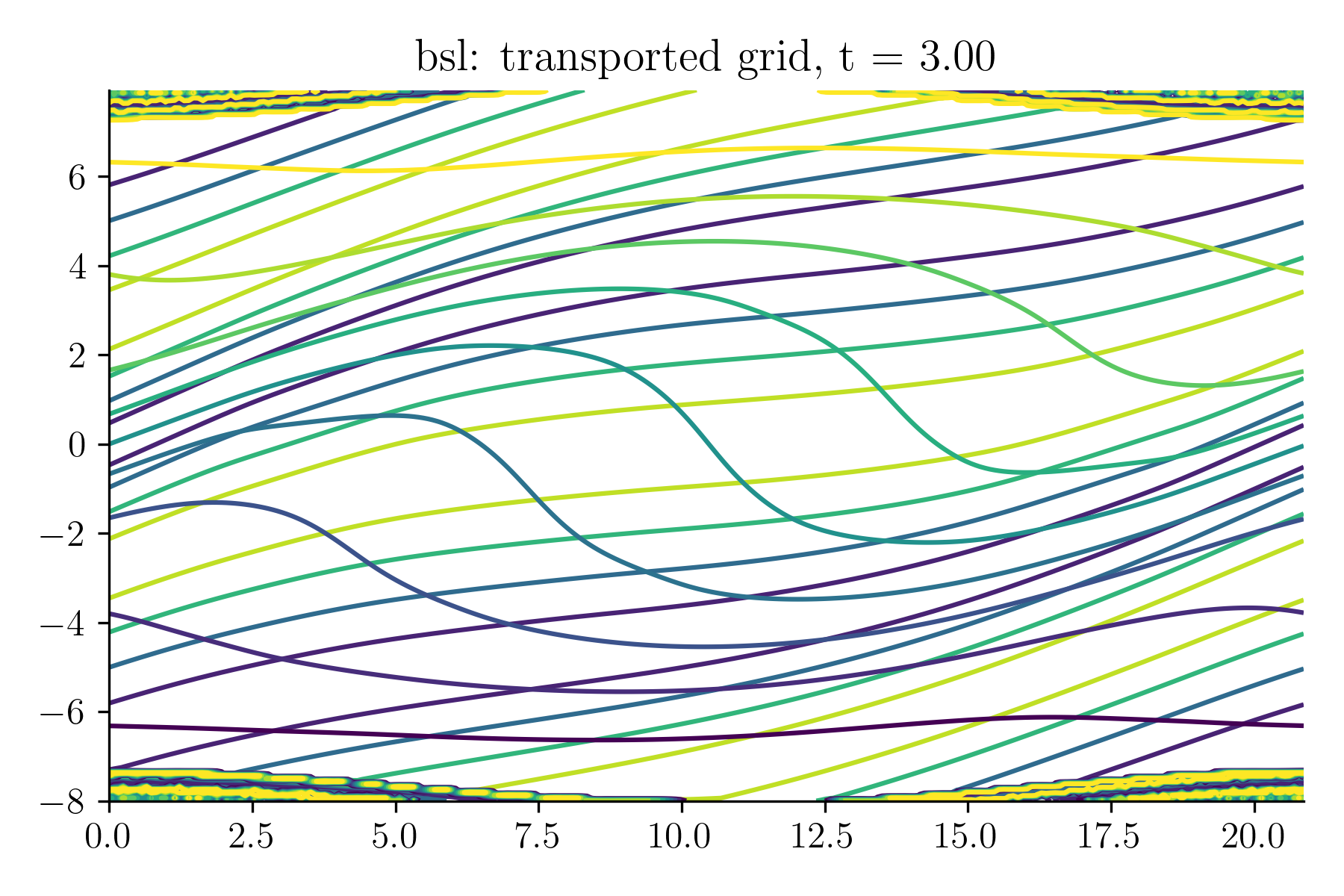}
    \end{subfigure}
    \hfill
    \begin{subfigure}[b]{0.32\textwidth}
        \centering
        \includegraphics[width=\textwidth]{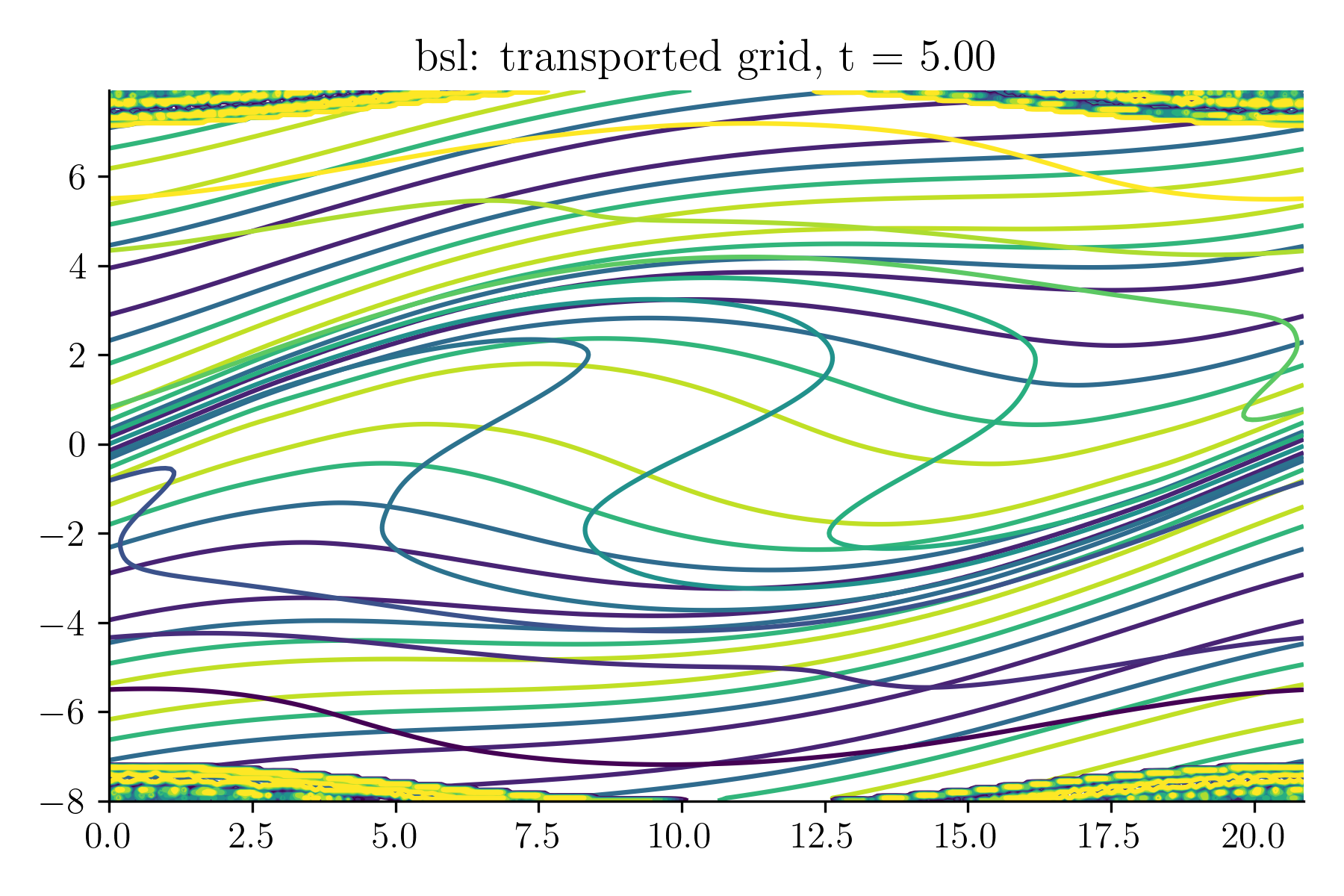}
    \end{subfigure}
    \\
    \begin{subfigure}[b]{0.32\textwidth}
        \centering
        \includegraphics[width=\textwidth]{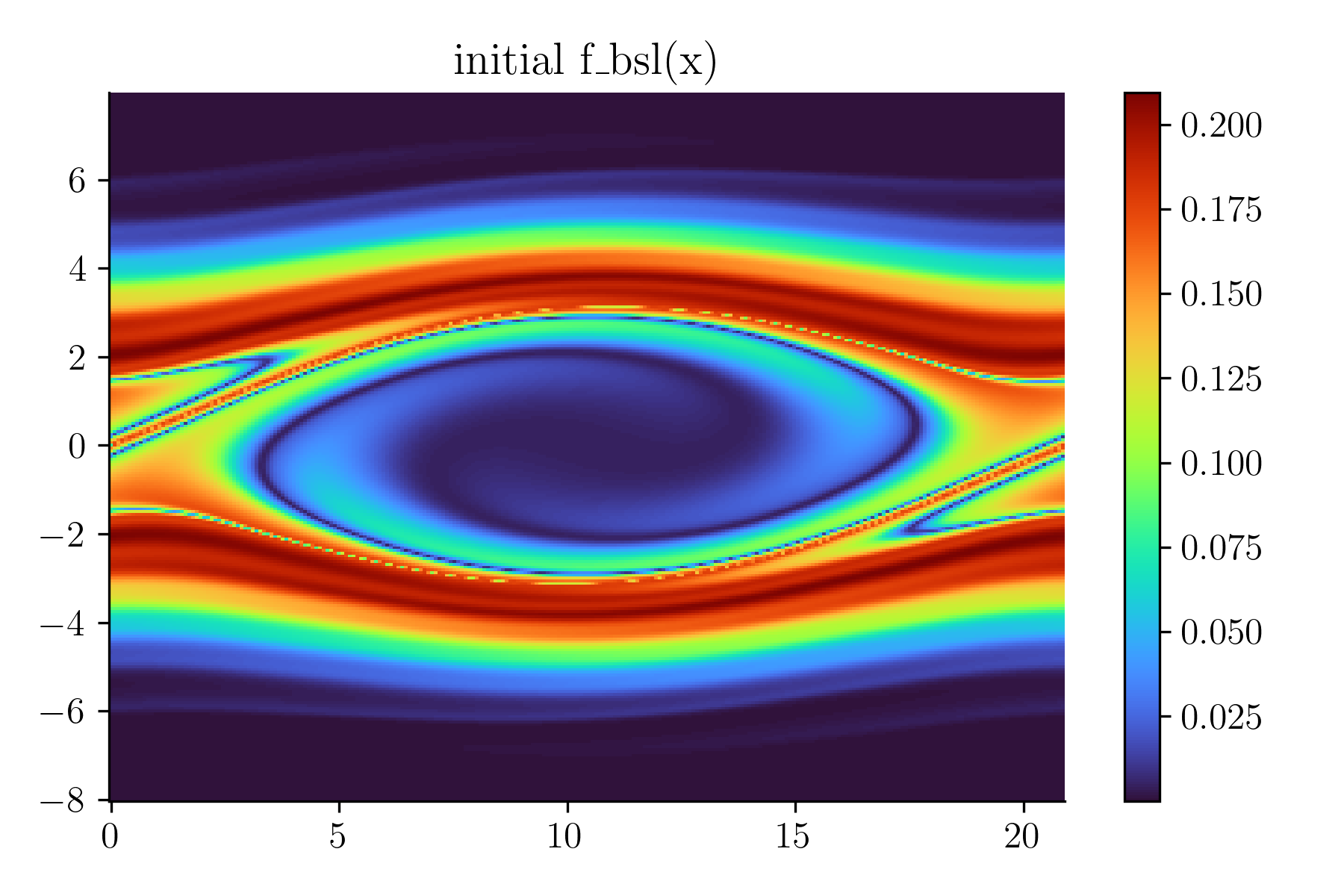}
    \end{subfigure}
    \hfill
    \begin{subfigure}[b]{0.32\textwidth}
        \centering
        \includegraphics[width=\textwidth]{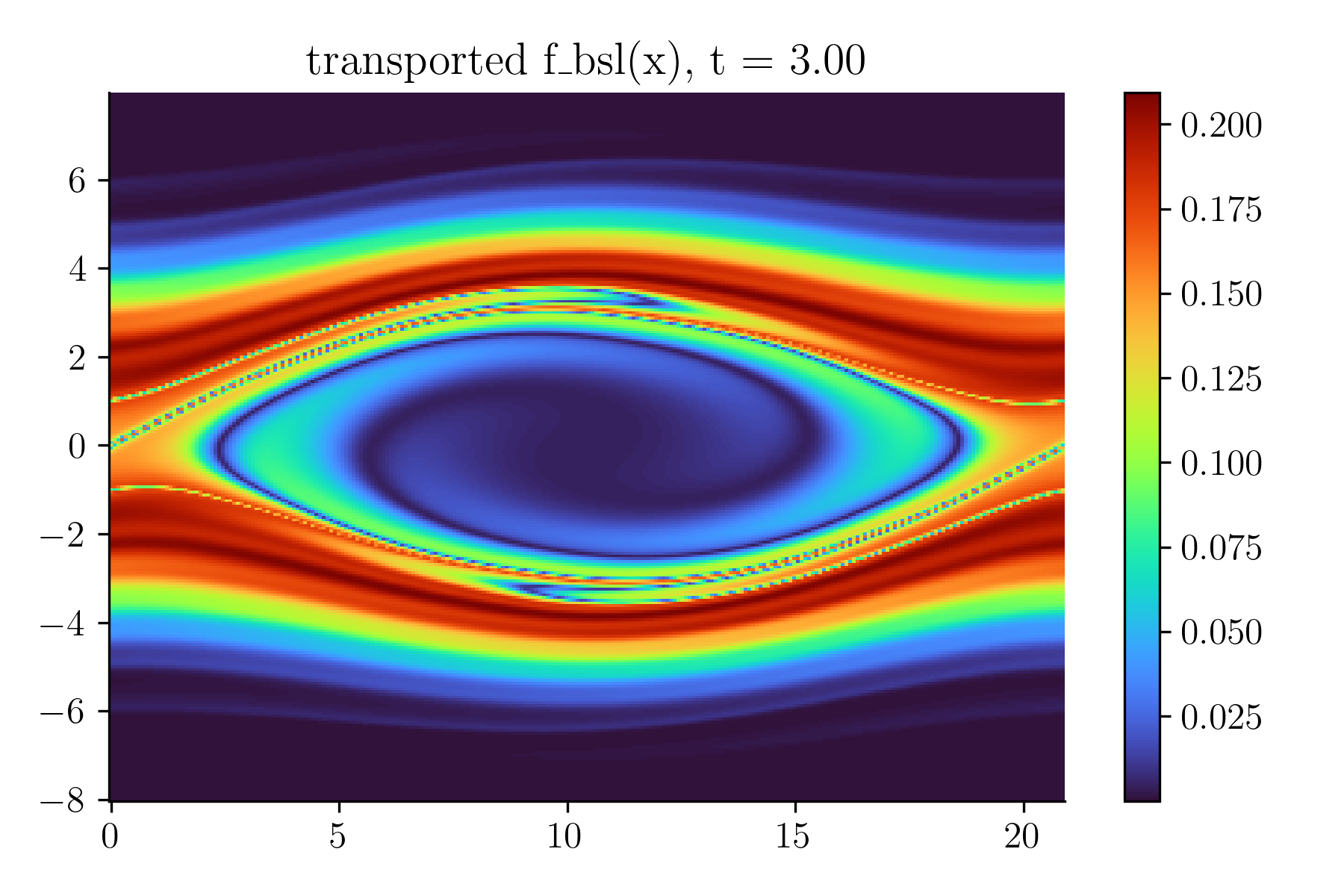}
    \end{subfigure}
    \hfill
    \begin{subfigure}[b]{0.32\textwidth}
        \centering
        \includegraphics[width=\textwidth]{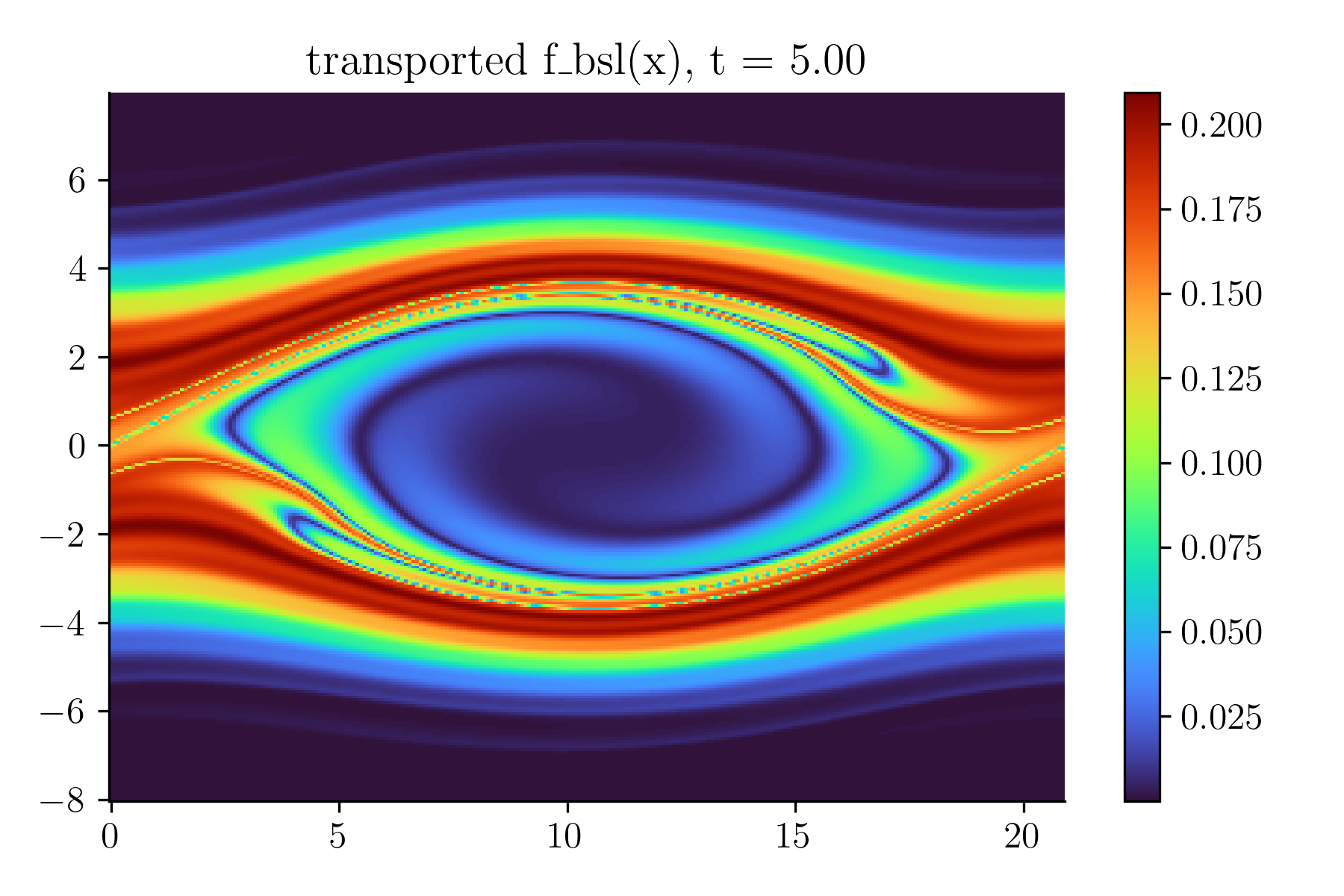}
    \end{subfigure}
    \caption{Numerical study from \cref{sec:numerical_study_flow_learning}: Characteristic flows (top row) and densities (bottom row) corresponding to the 1D1V two-stream instability solved by the BSL scheme between times
        $T_0 = 30$ and $T_1 = 35$. The left plots correspond to time $T_0$, the middle plots to time $T_0 + 3$ and the right plots to time $T_1$.}
    \label{fig:bsl_flow_f}
\end{figure}

In \cref{fig:bsl_flow_f} we illustrate the characteristic flows $\Psi_{[T_0,T_0 + t]}$ for different times $t$, together with the densities $f(t)$ obtained by advancing the density $f(T_0)$ using a fine grid BSL scheme. Here the flow is illustrated by plotting the isolines of two passive distributions transported by the BSL scheme, starting from the affine distributions $f_x = x$ and $f_v = v$ at time $T_0$:
the resulting isolines show the grid being transported forward by the flow and the evolving complexity as time increases, with the development of fine structures in phase space.
Note that close to the maximal and minimal velocities the BSL scheme uses a (non-physical) periodic boundary condition which results in strong gradients in the transported grid isolines -- this is of little importance for the transport of density, as the latter vanishes close to these extremal velocities.

\begin{figure}[!ht]
    \centering
    \begin{subfigure}[b]{0.32\textwidth}
        \centering
        \includegraphics[width=\textwidth]{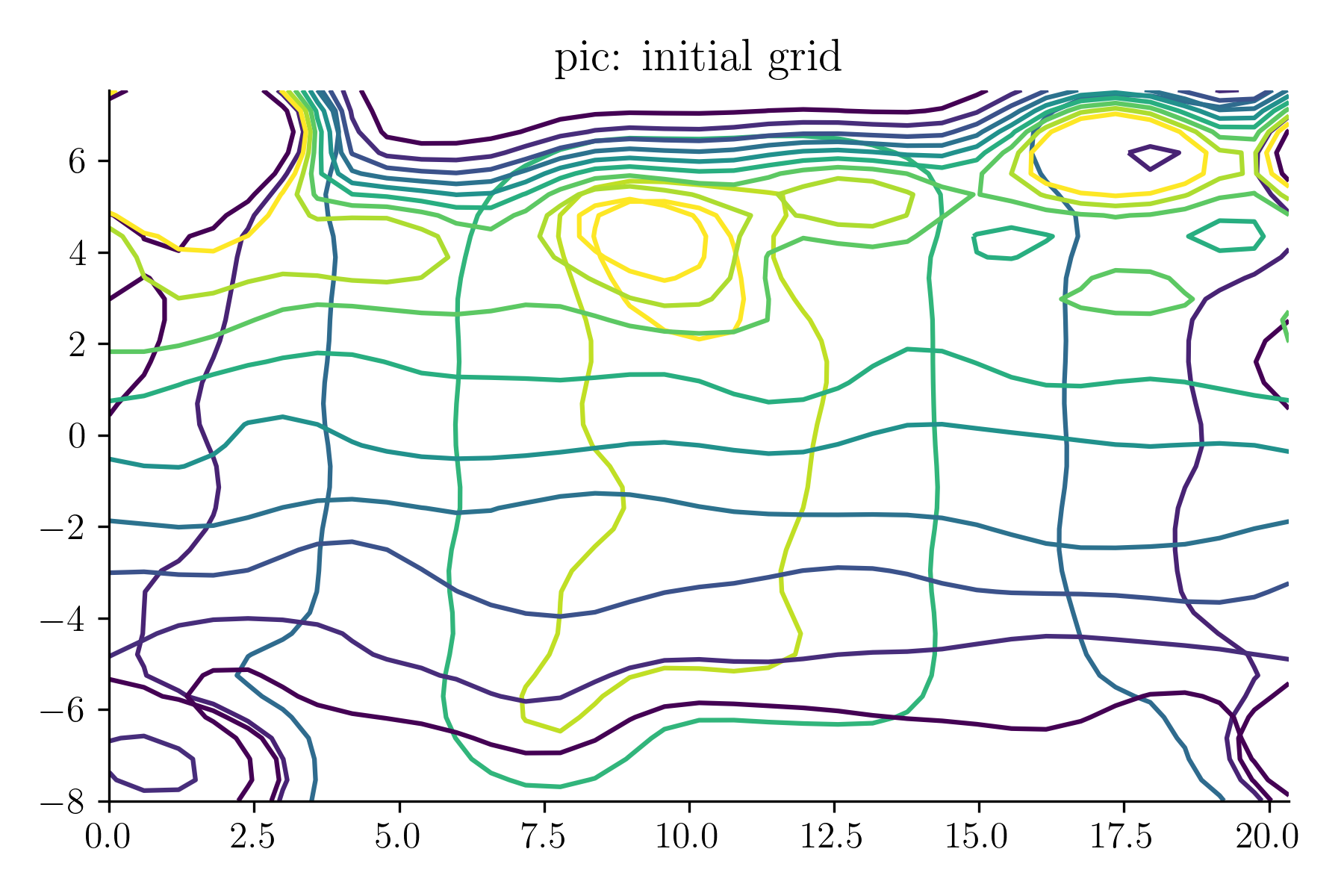}
    \end{subfigure}
    \hfill
    \begin{subfigure}[b]{0.32\textwidth}
        \centering
        \includegraphics[width=\textwidth]{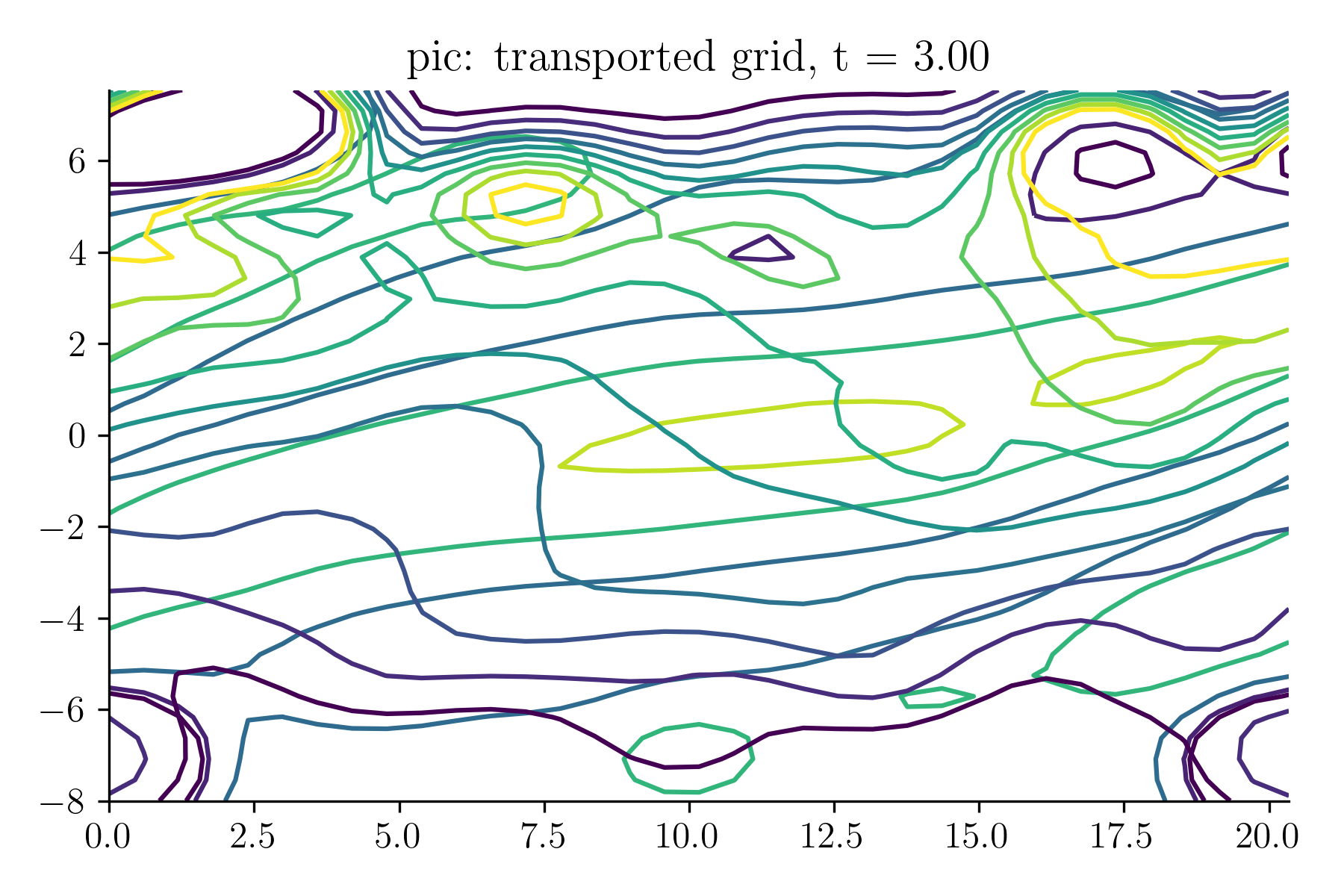}
    \end{subfigure}
    \hfill
    \begin{subfigure}[b]{0.32\textwidth}
        \centering
        \includegraphics[width=\textwidth]{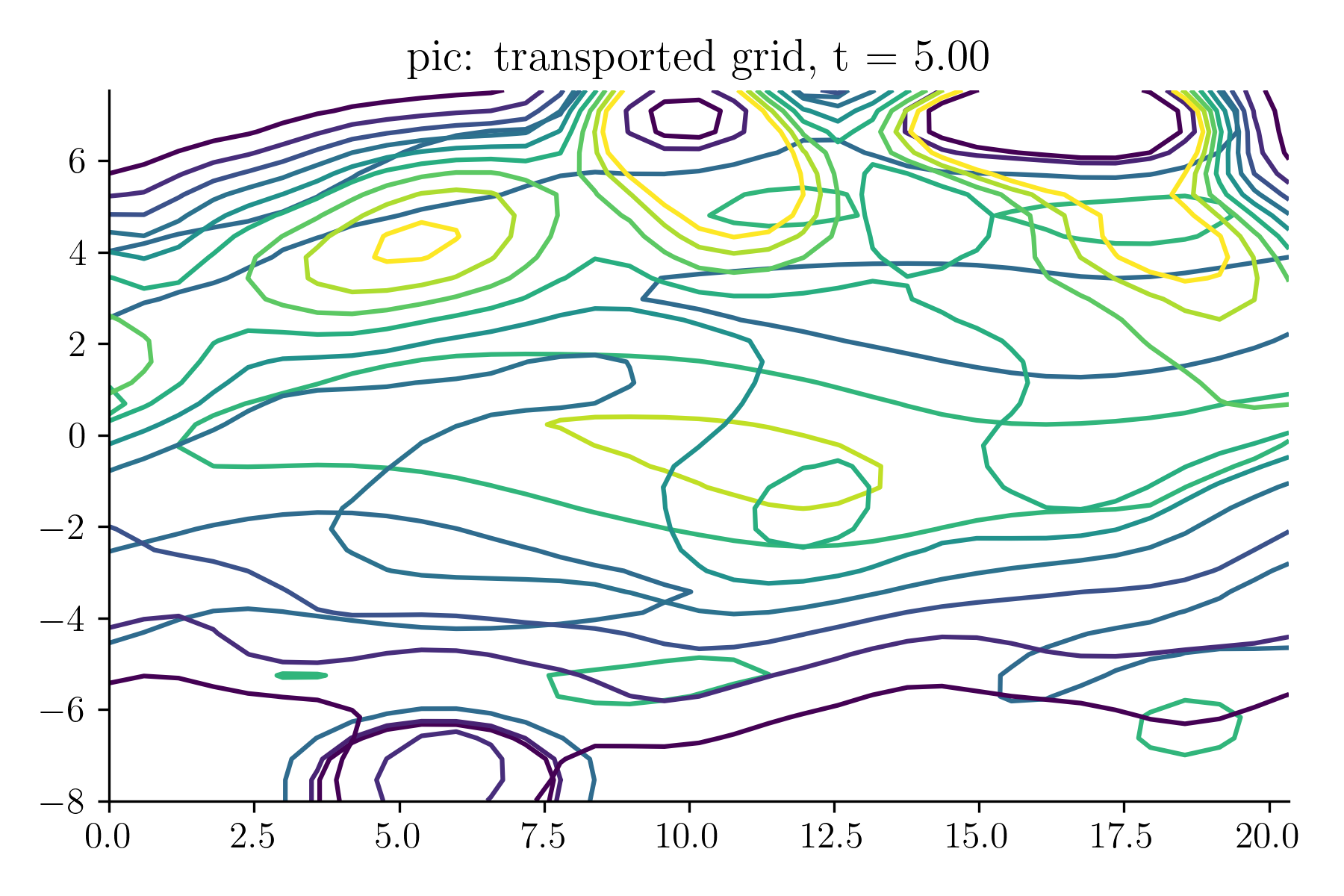}
    \end{subfigure}
    \\
    \begin{subfigure}[b]{0.32\textwidth}
        \centering
        \includegraphics[width=\textwidth]{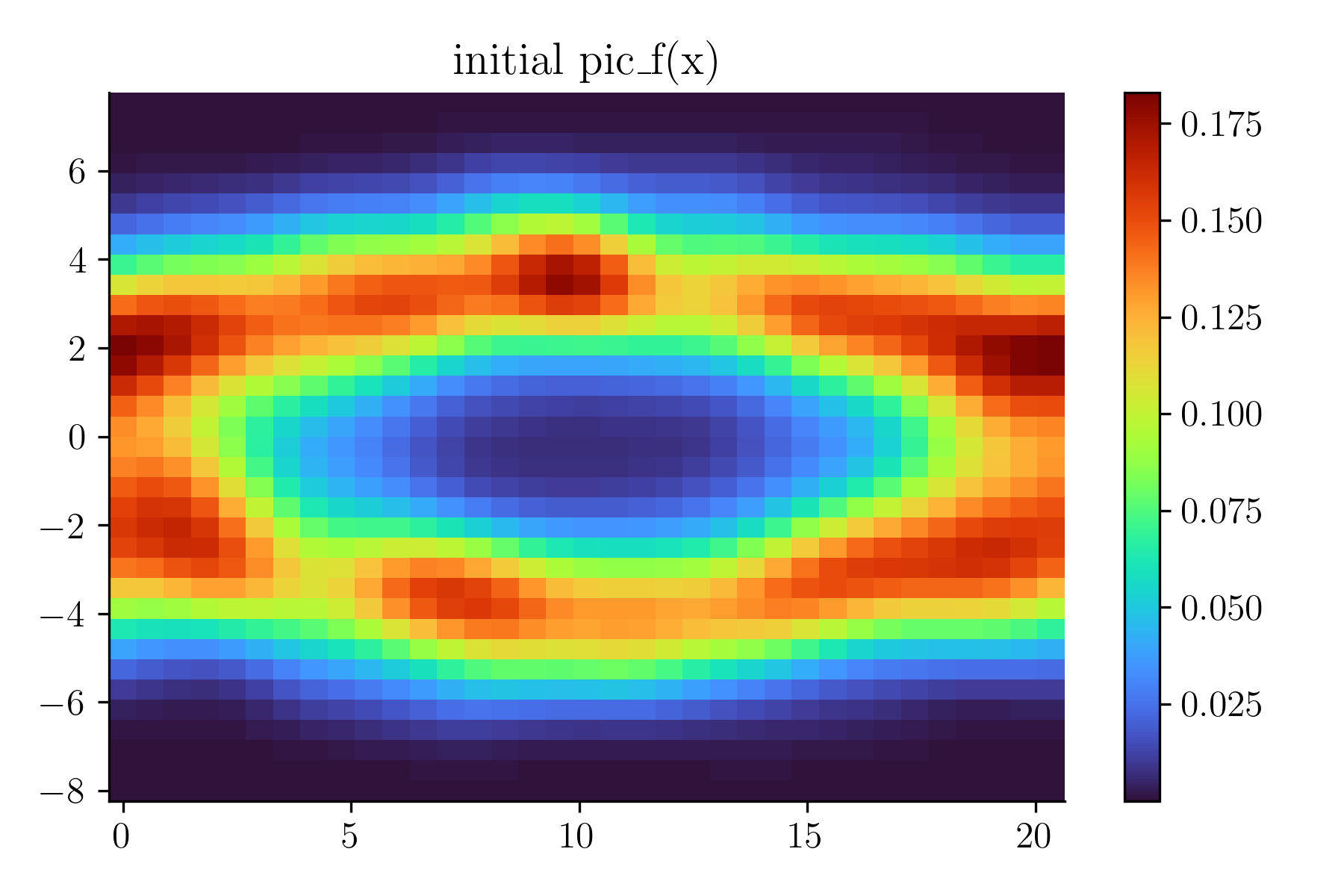}
    \end{subfigure}
    \hfill
    \begin{subfigure}[b]{0.32\textwidth}
        \centering
        \includegraphics[width=\textwidth]{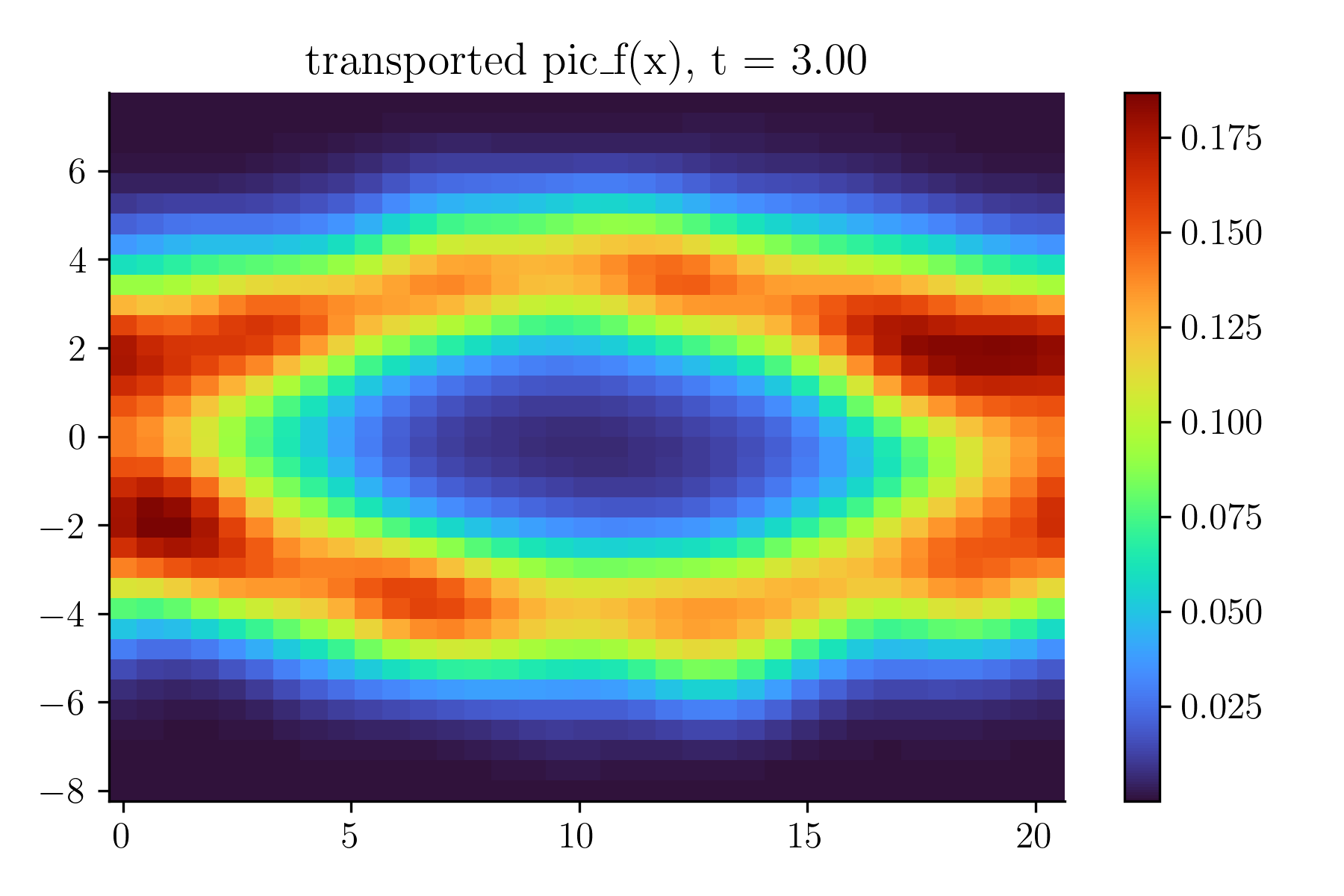}
    \end{subfigure}
    \hfill
    \begin{subfigure}[b]{0.32\textwidth}
        \centering
        \includegraphics[width=\textwidth]{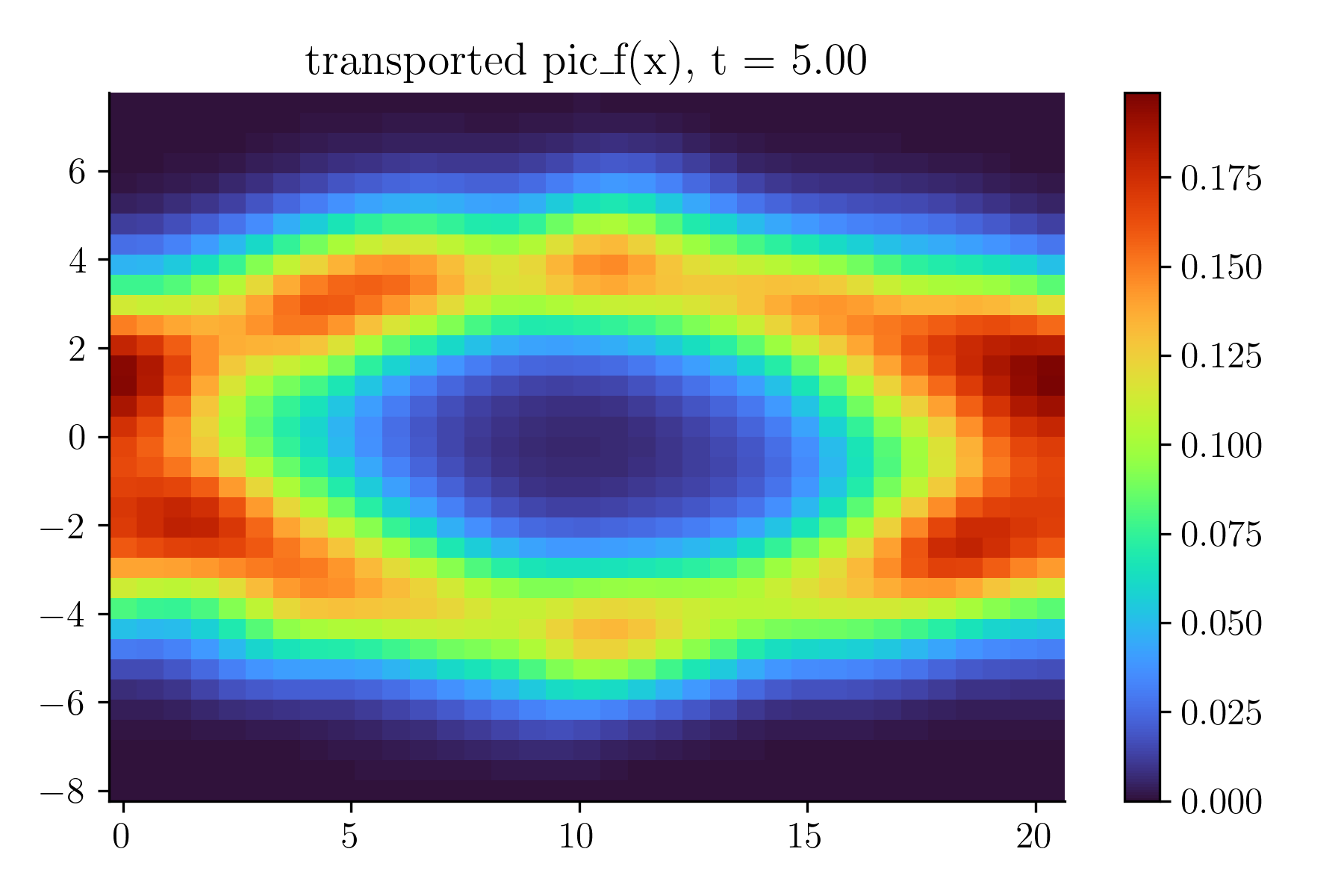}
    \end{subfigure}

    \caption{Numerical study from \cref{sec:numerical_study_flow_learning}: Visualisation of the characteristic flows (top) and densities (bottom) associated with a PIC approximation of the two-stream instability between times $T_0 = 30$ and $T_1 = 35$ with $N = 5000$ particles. Here the flows and the densities are visualized by evaluating
        smoothed particle distributions with appropriate weights, as described in the text.
    }
    \label{fig:pic_flow}
\end{figure}

In \cref{fig:pic_flow} we show the densities transported by the PIC scheme, obtained by smoothing each particle with cubic B-splines
in both the $x$ and $v$ directions, 
as in \eqref{eq:shape_function}. 
We also plot the isolines of the passive distributions $f_x = x$ and $f_v = v$ as transported by the PIC scheme (using the
transported particles with new weights associated with these distributions).

\begin{figure}
    \centering
    \begin{subfigure}[b]{0.47\textwidth}
        \centering
        \includegraphics[width=.49\textwidth]{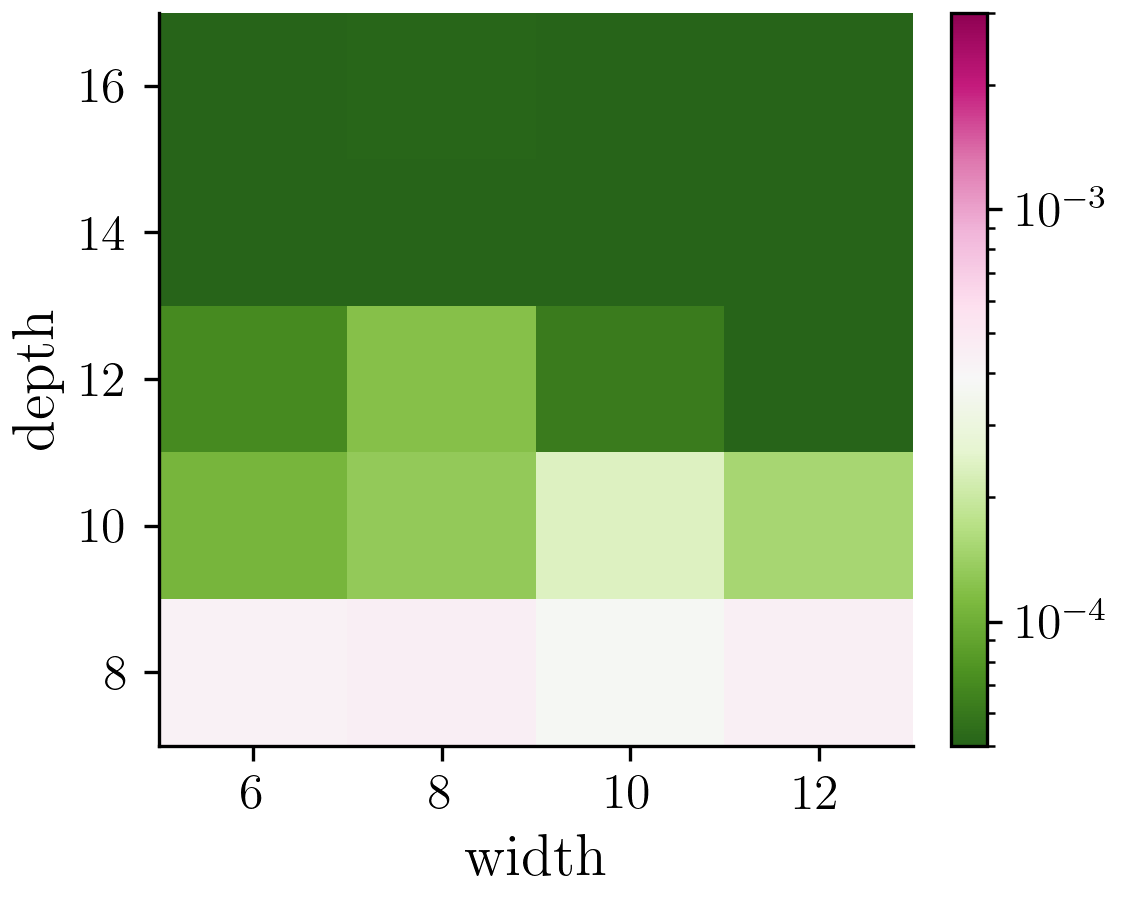}
        \includegraphics[width=.49\textwidth]{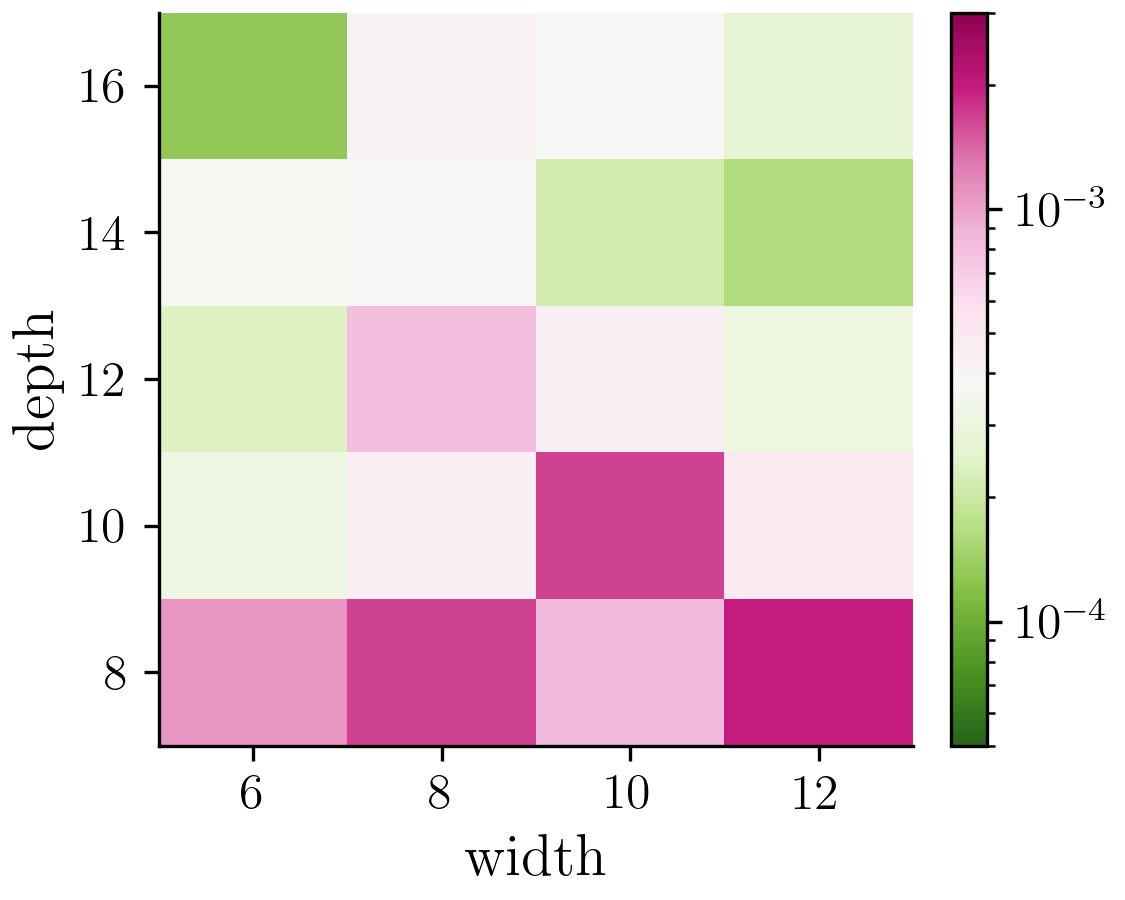}
        \caption{Direct training, 600 Adam and 600 natural gradient steps}
        \label{fig:direct_training_600_600}
    \end{subfigure}
    \hfill
    \begin{subfigure}[b]{0.47\textwidth}
        \centering
        \includegraphics[width=.49\textwidth]{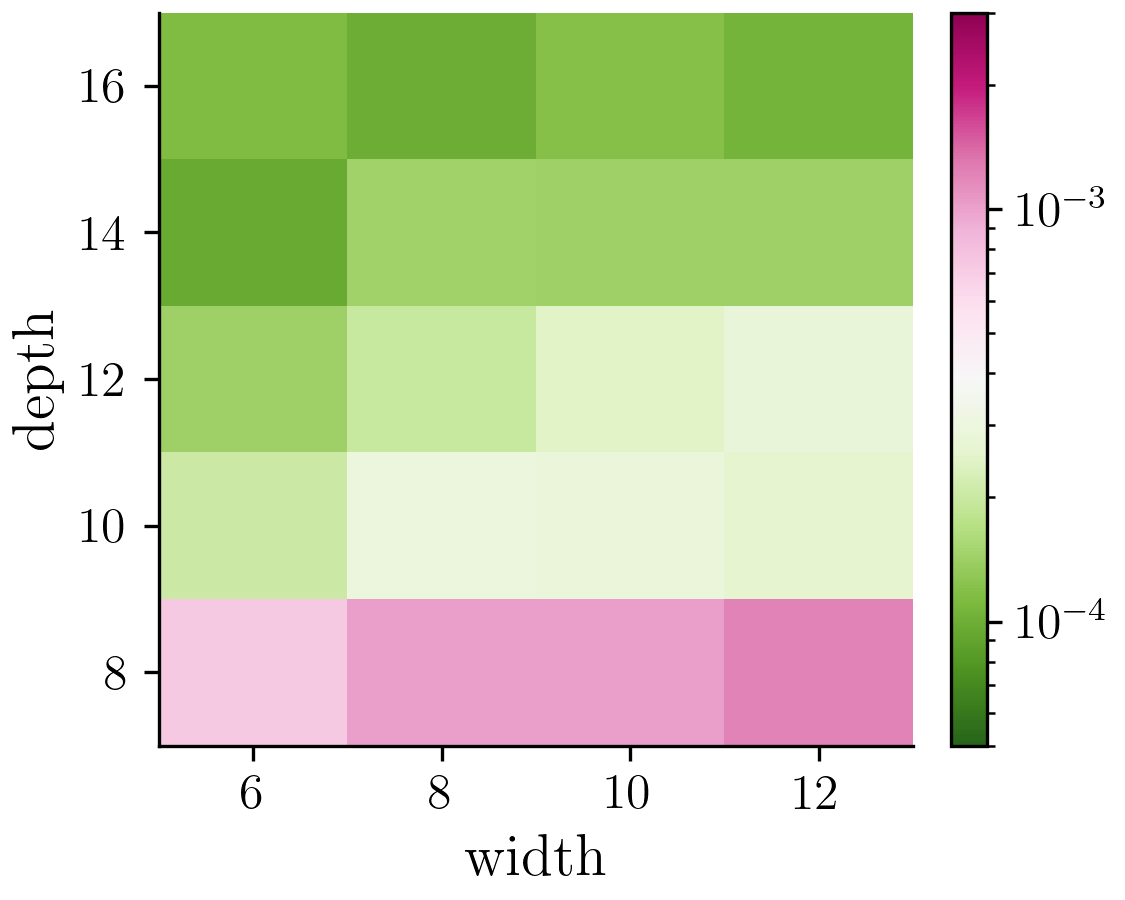}
        \includegraphics[width=.49\textwidth]{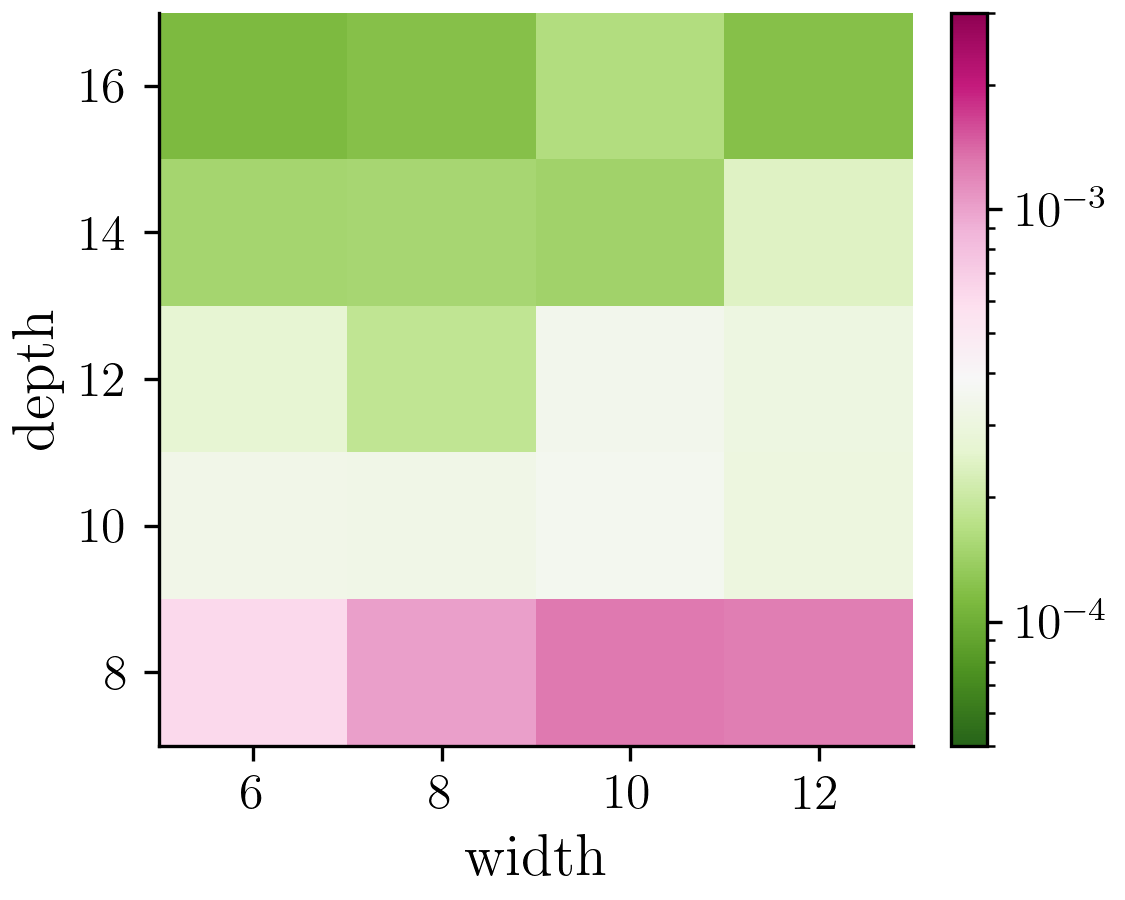}
        \caption{Incremental training, 600 Adam and 600 natural gradient steps}
        \label{fig:incremental_training_600_600}
    \end{subfigure}
    \\
    \begin{subfigure}[b]{0.47\textwidth}
        \centering
        \includegraphics[width=.49\textwidth]{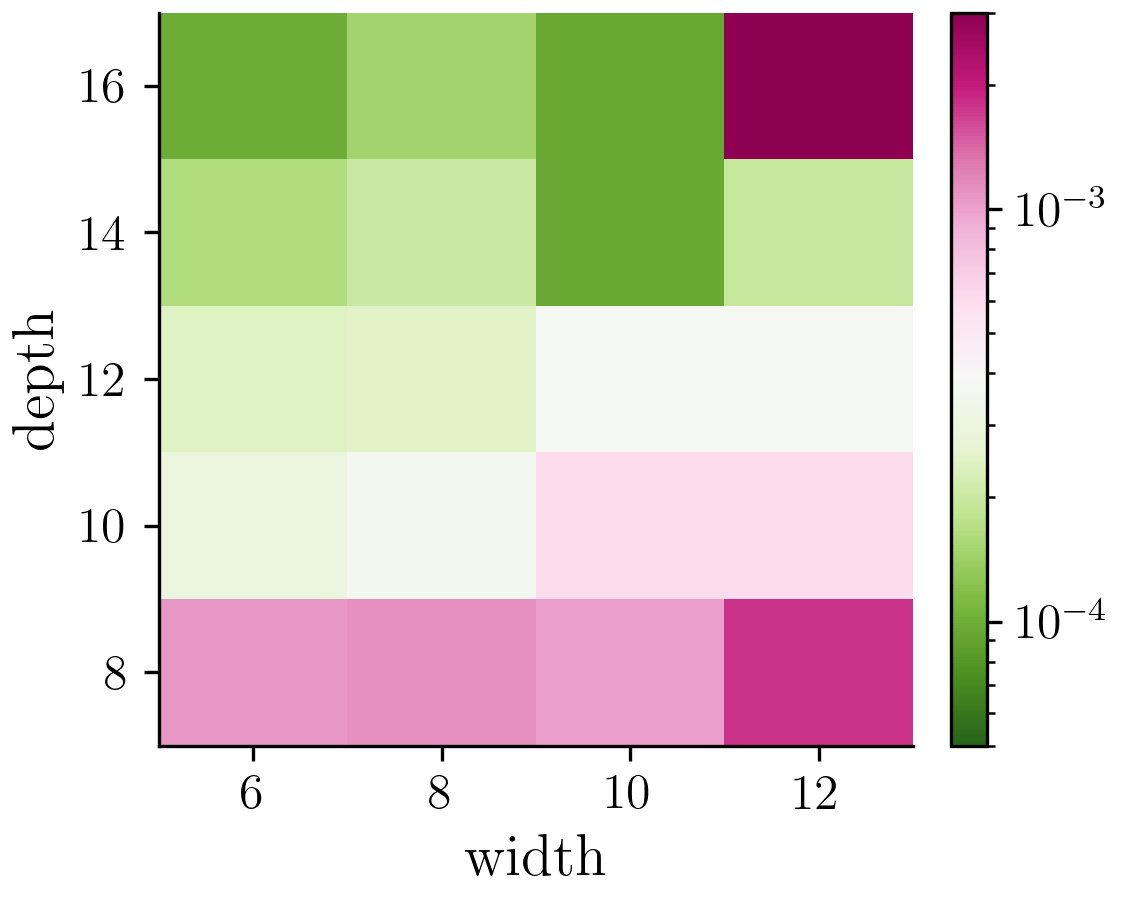}
        \includegraphics[width=.49\textwidth]{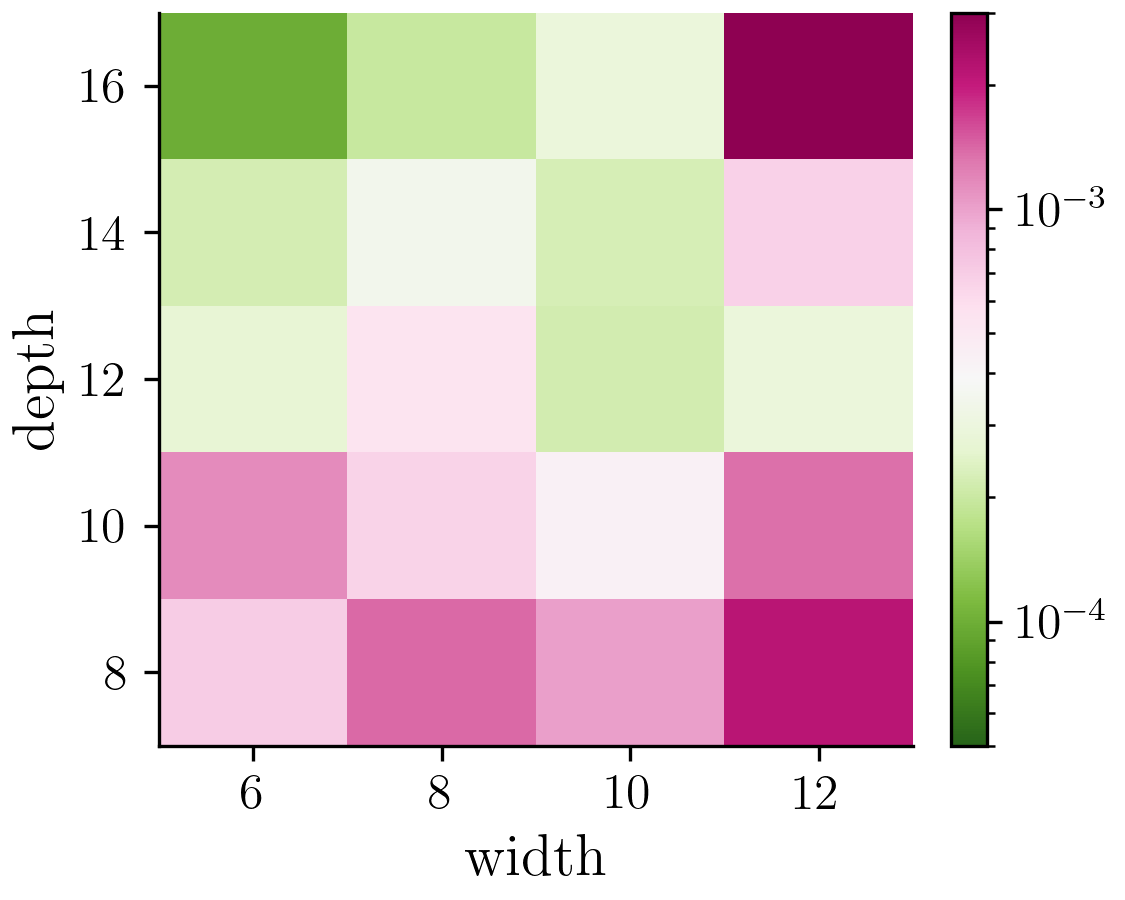}
        \caption{Direct training, 600 Adam and 900 natural gradient steps}
        \label{fig:direct_training_600_900}
    \end{subfigure}
    \hfill
    \begin{subfigure}[b]{0.47\textwidth}
        \centering
        \includegraphics[width=.49\textwidth]{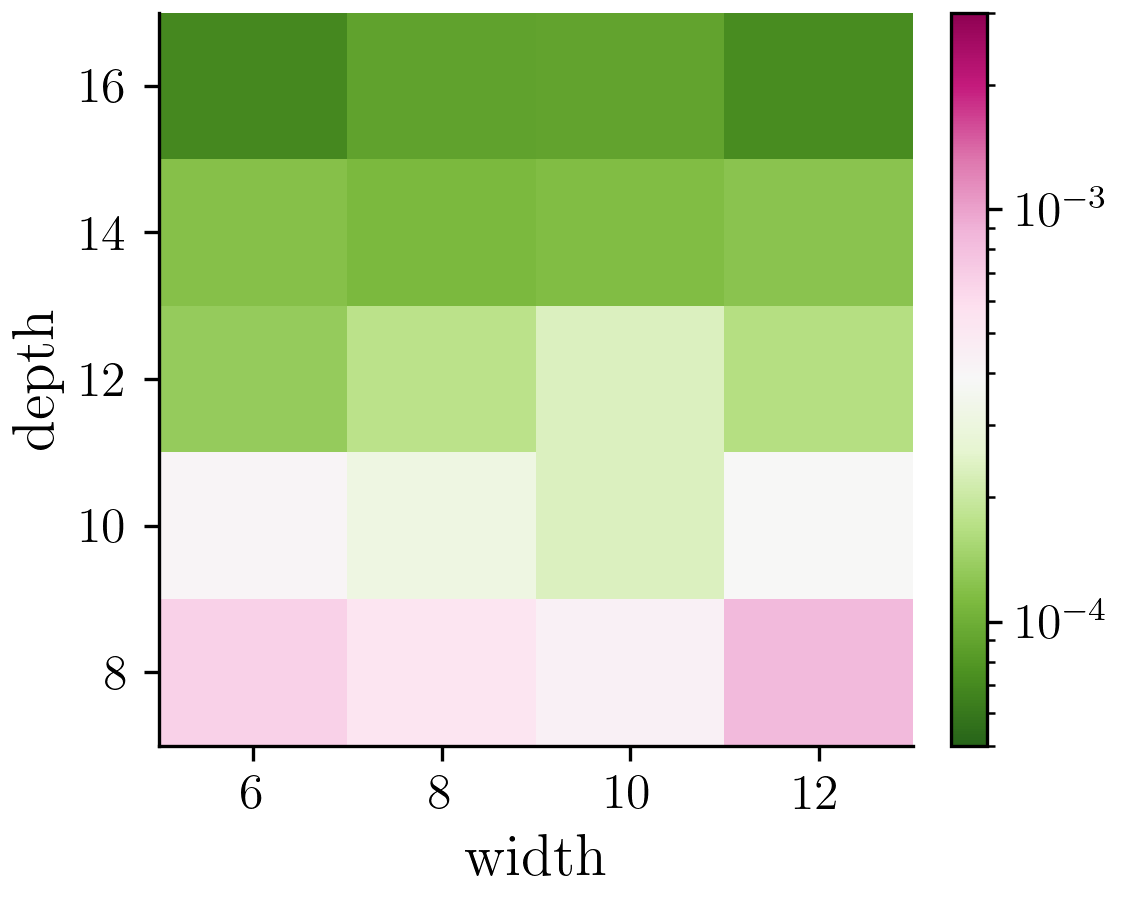}
        \includegraphics[width=.49\textwidth]{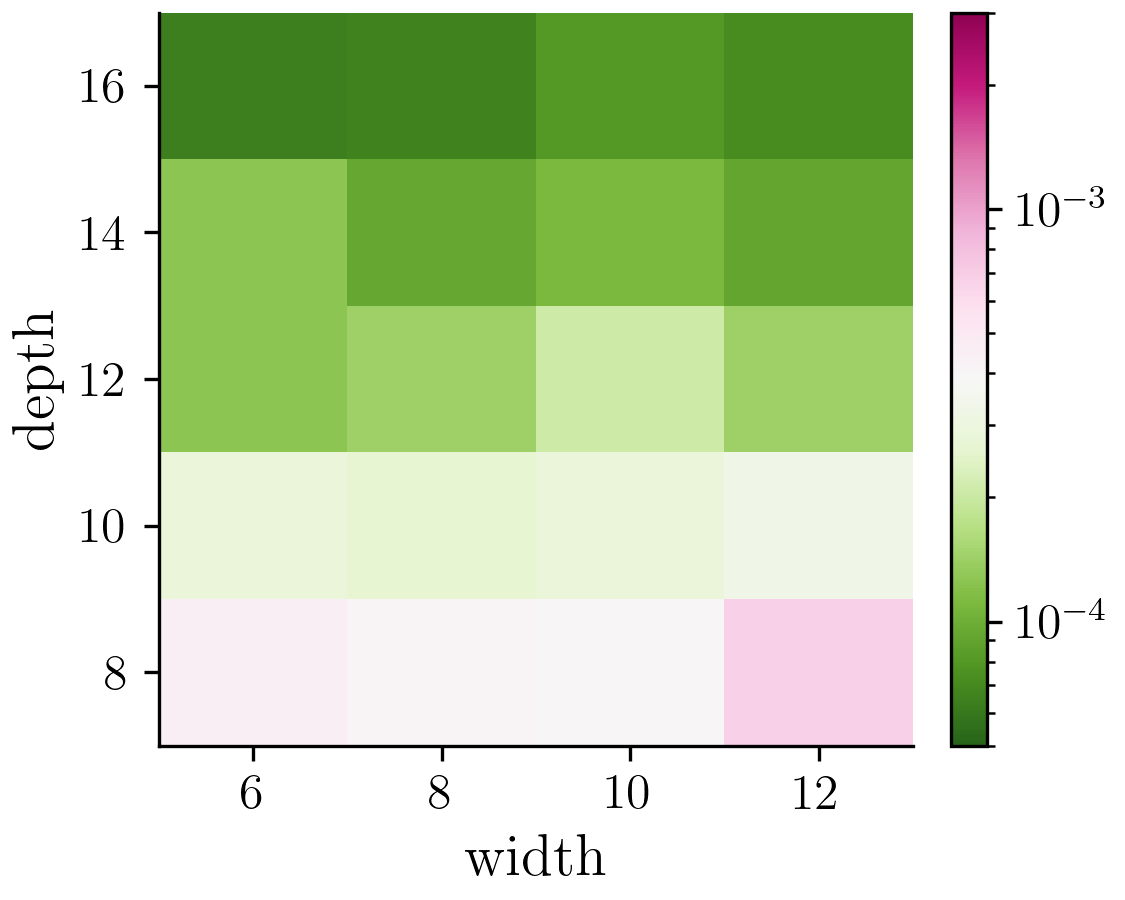}
        \caption{Incremental training, 600 Adam and 900 natural gradient steps}
        \label{fig:incremental_training_600_900}
    \end{subfigure}

    \caption{Numerical study from \cref{sec:numerical_study_flow_learning}: Quadratic errors of the flows learned by networks of different widths and depths,
        using a direct training strategy for the left panels ((a) and (c))
        and an incremental training strategy for the right panels ((b) and (d)).
        For each case, we show two plots: the left ones correspond to the errors measured on the training dataset while the right ones to a set of test particles not used in the training. 
        For each network the errors have been averaged over 10 training runs in order to take the variability of the training process into account.
    }
    \label{fig:flow_learning_diags}
\end{figure}

In \cref{fig:flow_learning_diags} we then plot the quadratic flow errors obtained by SympNets of different sizes (width and depth), using different numbers of training epochs (as indicated) and different training strategies. Here all the errors correspond to the training of the flow $\Psi_{[T_0,T_1]}$, but in the left plots we use a direct training strategy (approximating directly the flow on the time range $[T_0,T_1]$), while in the right plots we use an incremental training strategy (approximating the flow on the time range $[T_0,T_0 + 1]$, then $[T_0,T_0 + 2]$ starting from the previously trained model, and so on until $[T_0,T_1]$).
The top row corresponds to a total of 600 Adam and 600 natural gradient steps, while the bottom row corresponds to 600 Adam and 900 natural gradient steps: in the direct strategy we use all these epochs to train the flow on the whole time interval $[T_0, T_1]$, while in the incremental strategy we use a budget of 100 Adam and 100 (resp. 150) natural gradient steps when training for the flows on $[T_0, T_0 + t]$ with $t \in \{1, \dots 4\}$, and use the remaining budget of 200 Adam and 200 (resp. 300) natural gradient steps for the last training on $[T_0, T_1]$.
Finally, for each training strategy and number of epochs, we show on the left the quadratic errors measured on the training set, and on the right the quadratic errors measured on the test set. From these plots we see that the incremental training strategy yields  more stable results: a monotonic convergence as the number of layers increases (which is not the case for the direct strategy, a sign that larger networks are harder to train in a direct manner), and very good agreement between the errors measured on the training and testing datasets (which again is not the case with the direct strategy, a sign that the latter is overfitting). We also note that increasing the number of epochs does not significantly improve the results, which indicates that the networks are able to learn the flow with a relatively small number of epochs.

\begin{figure}[!ht]
    \centering
    \begin{subfigure}[b]{0.32\textwidth}
        \centering
        \includegraphics[width=\textwidth]{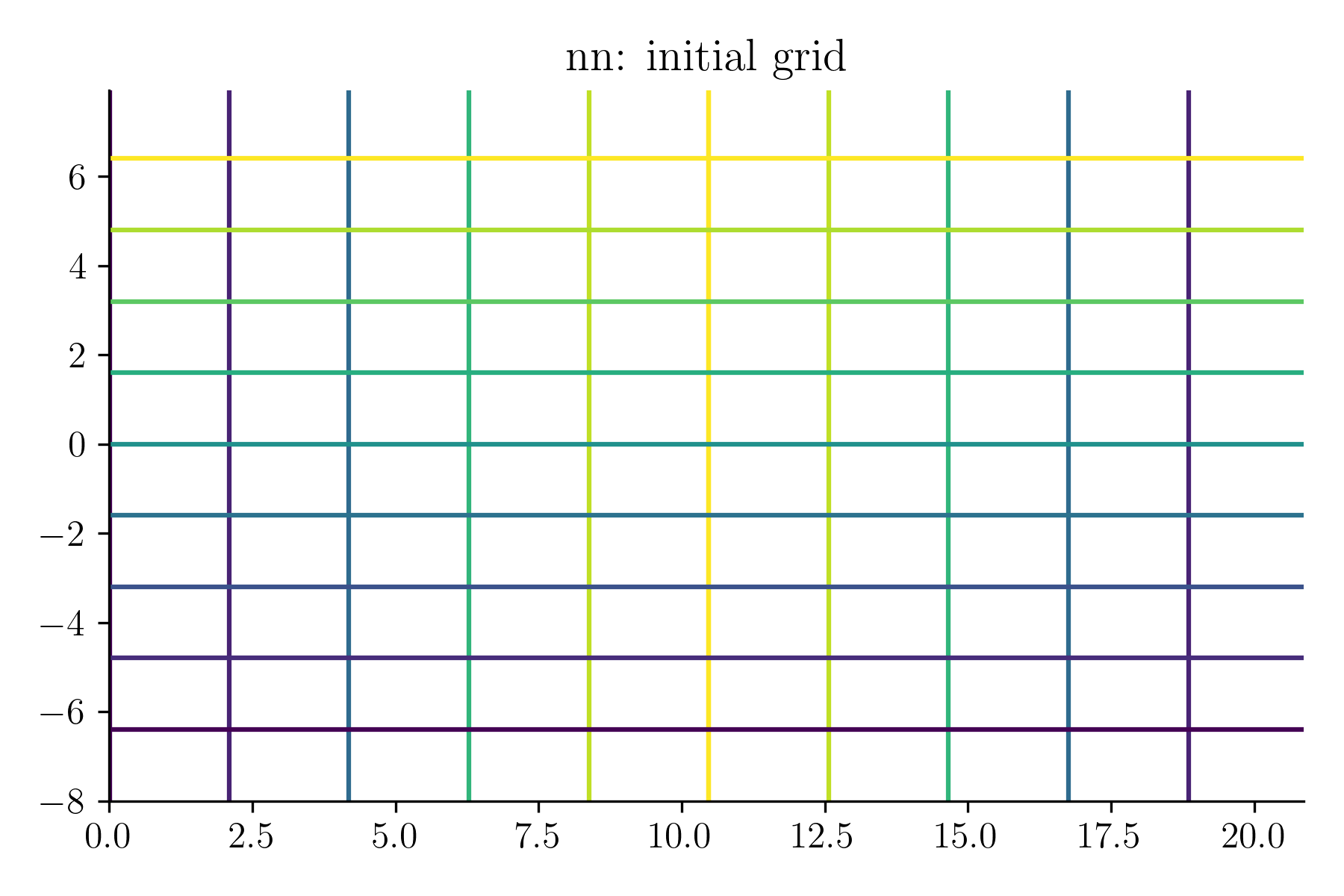}
    \end{subfigure}
    \hfill
    \begin{subfigure}[b]{0.32\textwidth}
        \centering
        \includegraphics[width=\textwidth]{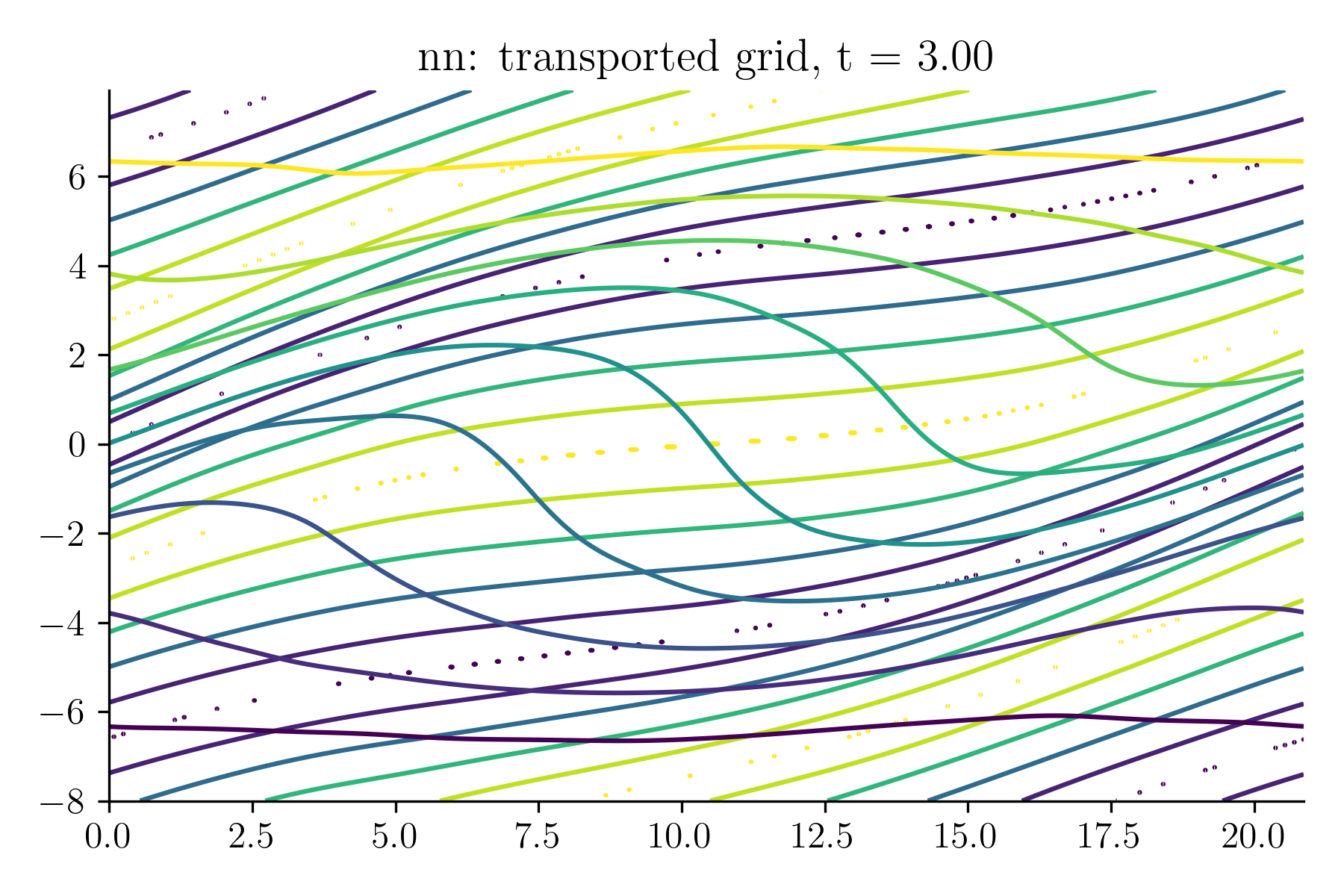}
    \end{subfigure}
    \hfill
    \begin{subfigure}[b]{0.32\textwidth}
        \centering
        \includegraphics[width=\textwidth]{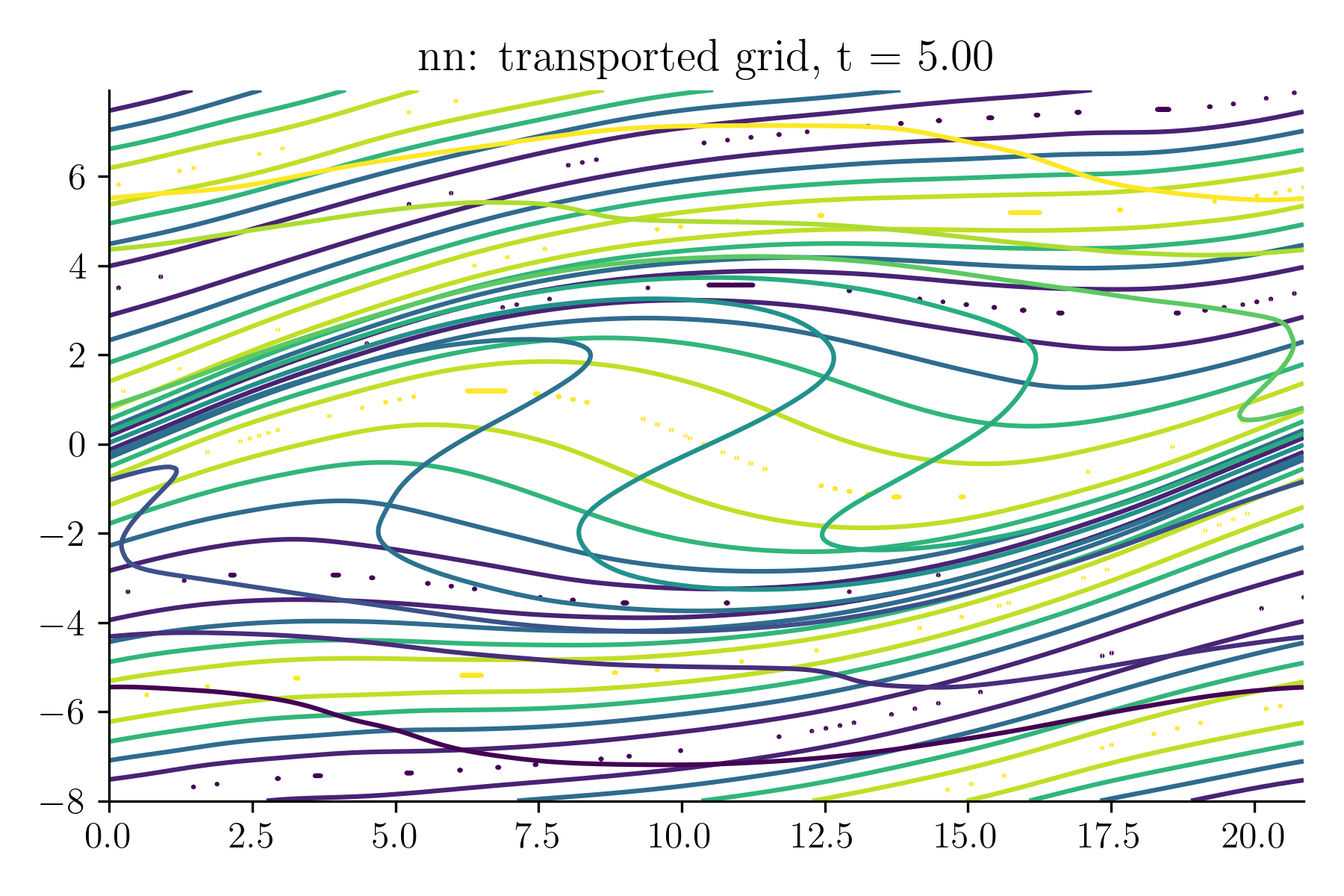}
    \end{subfigure}
    \\
    \begin{subfigure}[b]{0.32\textwidth}
        \centering
        \includegraphics[width=\textwidth]{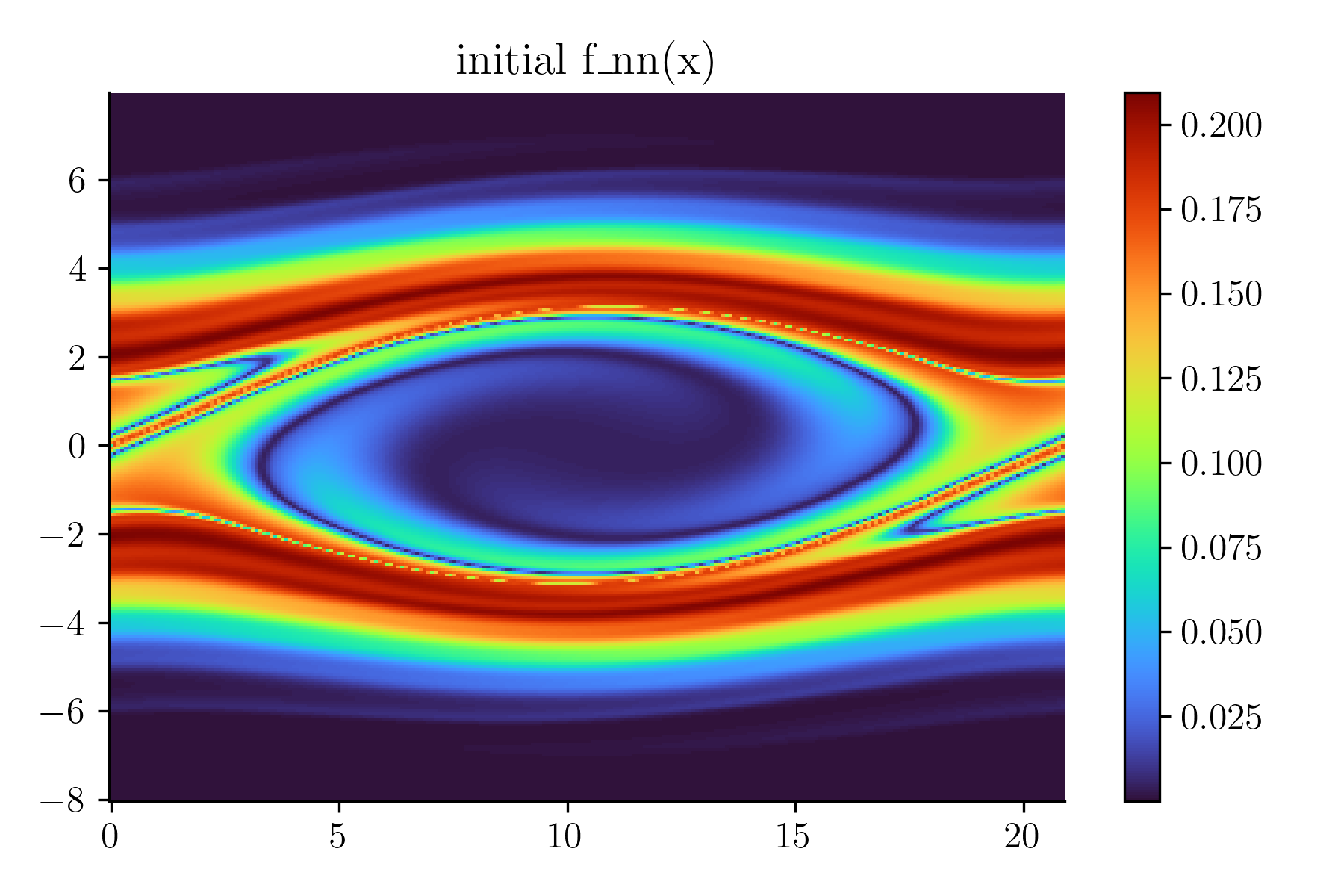}
    \end{subfigure}
    \hfill
    \begin{subfigure}[b]{0.32\textwidth}
        \centering
        \includegraphics[width=\textwidth]{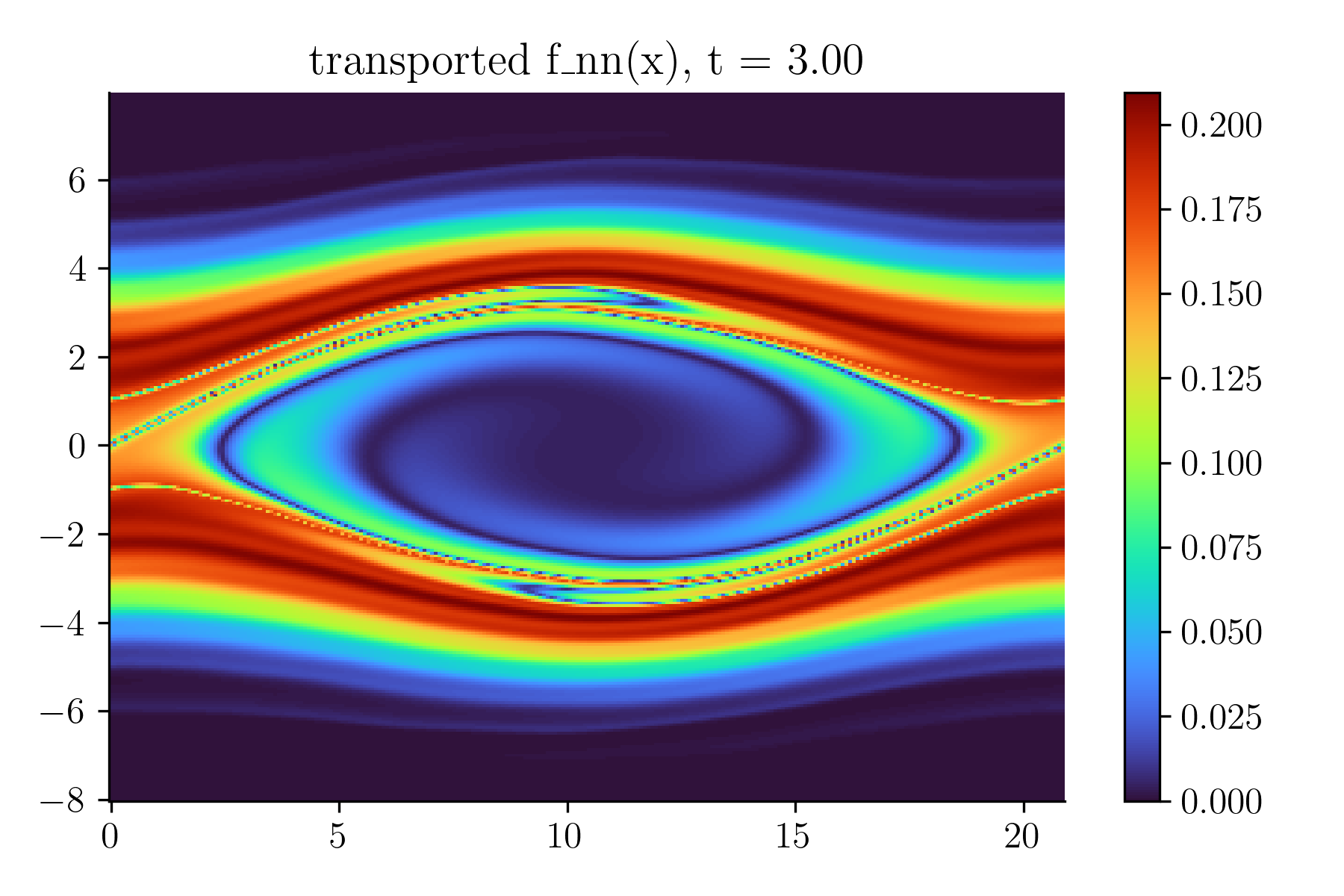}
    \end{subfigure}
    \hfill
    \begin{subfigure}[b]{0.32\textwidth}
        \centering
        \includegraphics[width=\textwidth]{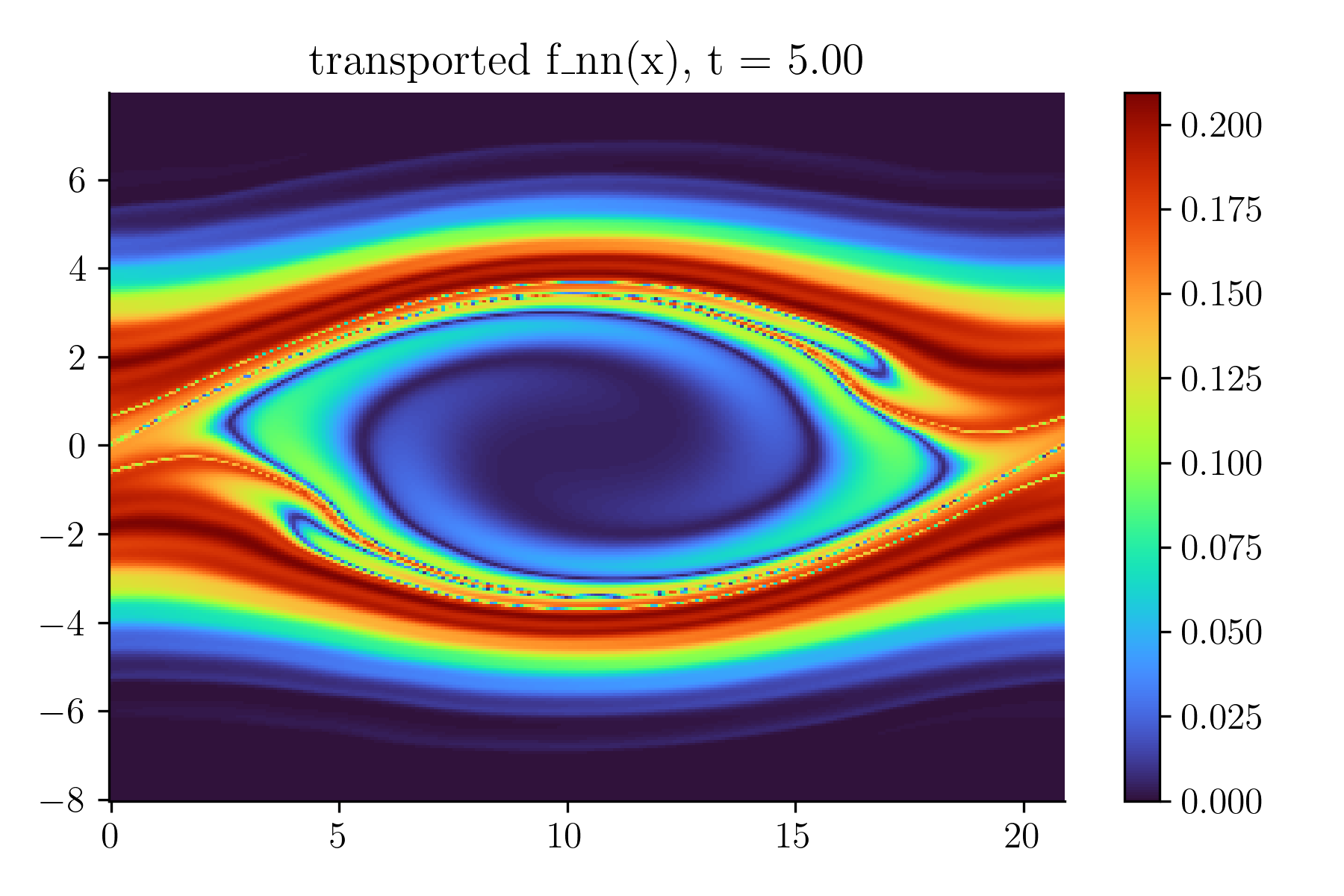}
    \end{subfigure}

    \caption{Numerical study from \cref{sec:numerical_study_flow_learning}: Top: characteristic flows learned by a SympNet of width $w=8$ and depth $\ell=14$ using
        a training dataset of 5000 particles as represented in \cref{fig:pic_flow}. Bottom: densities obtained by transporting the initial density with the learned flow.}
    \label{fig:nn_flow}
\end{figure}

In \cref{fig:nn_flow} we then show the trained flows and associated densities obtained by a SympNet of width $w=8$ and depth $\ell=14$ trained with the incremental strategy. Here the plots are obtained in a similar way as for \cref{fig:bsl_flow_f}, by transporting the isolines of the passive distributions $f_x = x$ and $f_v = v$ (to visualize the flow) and the density $f(T_0)$ (to visualize the density $f(T_0+t)$) with the trained flow. These results may be compared to the ones of the reference BSL scheme in \cref{fig:bsl_flow_f}, keeping in mind that they are obtained by learning the flow using a set of particles of the same resolution as the one used in \cref{fig:pic_flow}.

\subsection{1D1V test cases}

\subsubsection{1D1V two-stream instability}
\label{sec:1D1V_two-stream_instability}

\label{sec:tsi}
Our first test is a typical two-stream instability in 1D1V. The initial condition reads
\begin{equation}
    \finit(x,v) =   ( 1+ \varepsilon \cos(kx)) \frac{1}{2 \sqrt{2\pi}} \bigl(e^{-(v - v_0)^2/2} + e^{-(v + v_0)^2/2}\bigr),
    \label{eq:tsi_init}
\end{equation}
with $\varepsilon=0.05$, $k=0.3$ and $v_0=3$. In this situation, two beams with opposite velocities interact. This eventually leads to an instability, creating thin phase-space
filaments. The computational domain is $[0,2\pi /k ]\times[-9,9]. $ We use $\Np = 10^4$ particles, the time
step $\dt = 0.05$, and the bulk-update
period $N_\Psi = 20$. (which means that the bulk is updated every 1 time unit).

\begin{figure}[!ht]
    \centering
    \begin{subfigure}[b]{\textwidth}
        \centering
        \includegraphics[width=\textwidth]{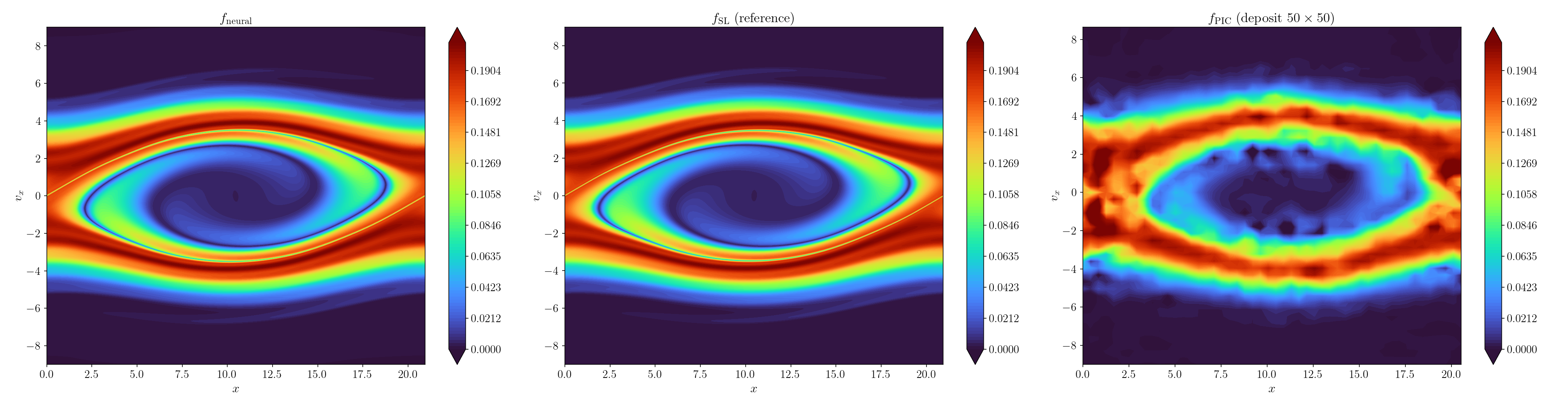}
        \caption{$t = 25$}
    \end{subfigure}

    \vspace{0.3cm}
    \begin{subfigure}[b]{\textwidth}
        \centering
        \includegraphics[width=\textwidth]{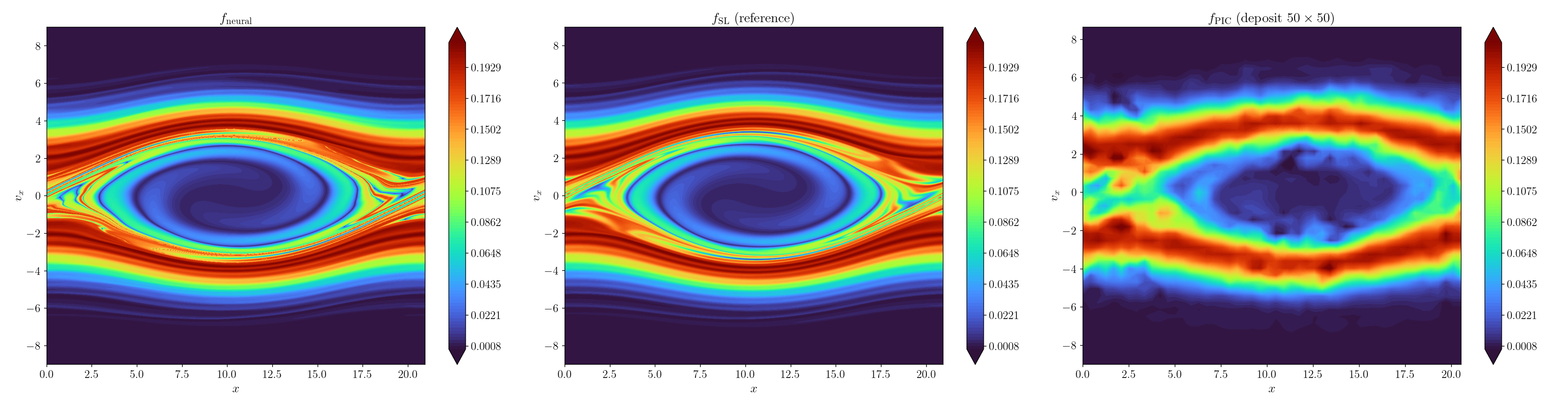}
        \caption{$t = 50$}
    \end{subfigure}

    \vspace{0.3cm}
    \begin{subfigure}[b]{\textwidth}
        \centering
        \includegraphics[width=\textwidth]{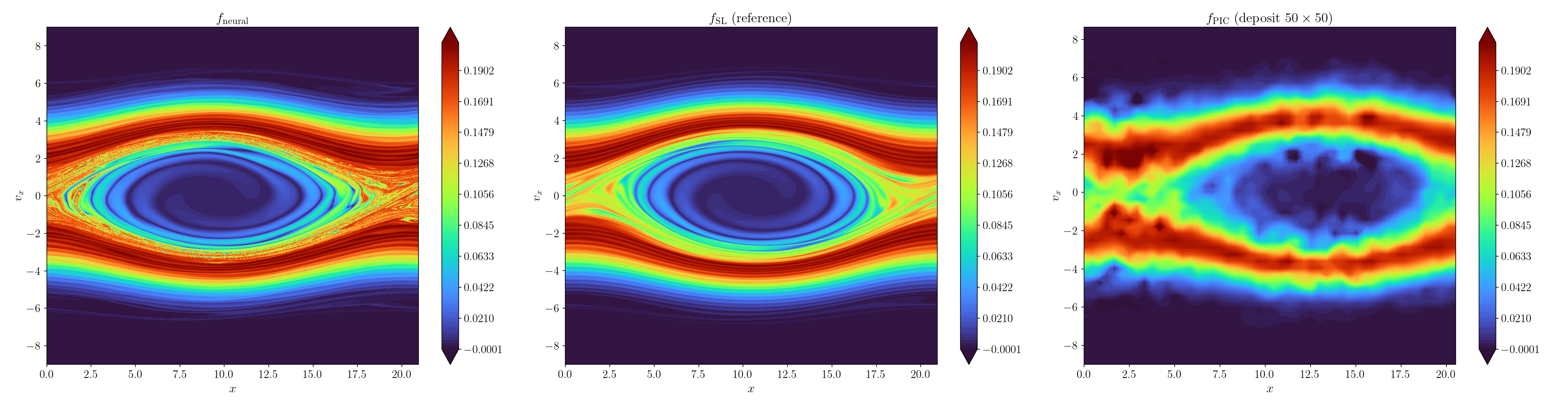}
        \caption{$t = 99$}
    \end{subfigure}
    \caption{1D1V two-stream instability from \cref{sec:1D1V_two-stream_instability}: Comparison between the density given by the Neural $\df$ scheme (left), the BSL scheme (middle) and the standard $\df$ scheme (right), at $t = 25$, $t = 50$ and $t = 99$. The Neural $\df$ and the standard $\df$ schemes use $N_p = 10^4$ particles, while the BSL scheme uses a grid of size $1024 \times 1024$.}
    \label{fig:tsi_comparaison}
\end{figure}

\begin{figure}[!ht]
    \centering
    \centering
    \includegraphics[width=\textwidth]{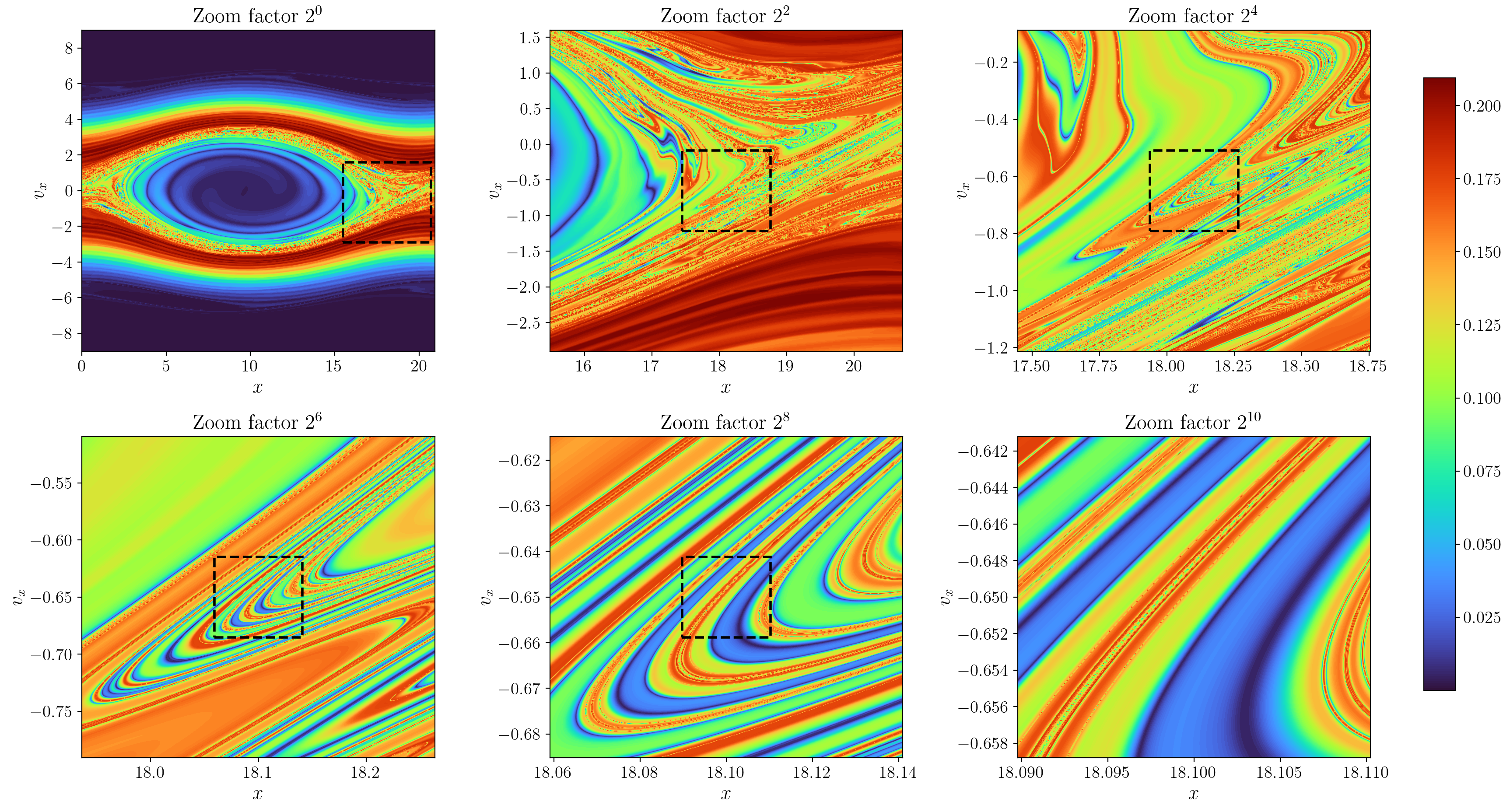}

    \caption{1D1V two-stream instability from \cref{sec:1D1V_two-stream_instability}: Zoom on the phase space density $f(t,x,v)$ at $t=99$ for the 1D1V two-stream instability given by the Neural $\df$ scheme with $N_p = 40 000$ particles. On each zoom, the density is evaluated on a $1024 \times 1024$ grid.}
    \label{fig:tsi_zoom}
\end{figure}

\Cref{fig:tsi_comparaison} compares the Neural $\df$ scheme with the BSL reference and the standard $\df$ scheme at $t = 25$, $50$ and $99$, in the nonlinear saturation regime. The characteristic vortex structure and the thin phase-space filaments generated by the filamentation of the distribution function are clearly visible. The Neural $\df$-PIC method reproduces the vortex structure and the filaments seen in the BSL reference with $\Np = 10^4$ particles, while drastically reducing the noise of the phase-space density compared to the standard $\df$ scheme.
Unlike grid-based or particle methods, the neural bulk is a function that can be evaluated at any point of phase space. Figure~\ref{fig:tsi_zoom} illustrates this feature: successive zooms, reveal filamentary structures at scales far below the reach of a grid-based scheme, and without any numerical diffusion. This figure is meant to be illustrative rather than quantitative: we make no claim that these fine filaments are accurately located, they are in fact likely mispositioned, but they show the ability of the representation to resolve arbitrarily fine, non-dissipative structures.


\begin{figure}[!ht]
    \centering
    \begin{subfigure}[b]{\textwidth}
        \centering
        \includegraphics[width=\textwidth]{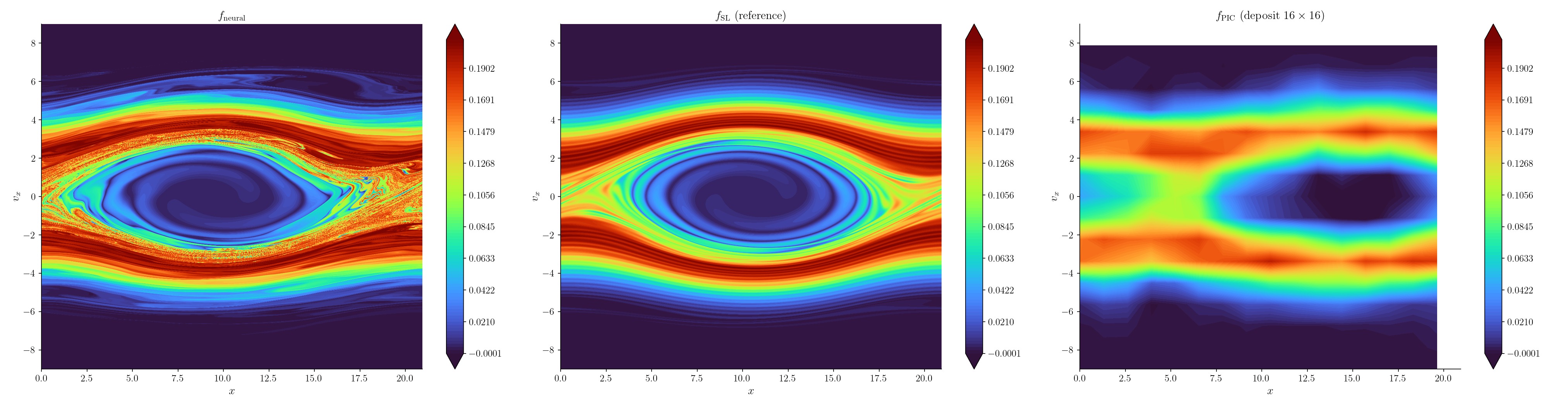}
        \caption{$N_p = 10^3$}
    \end{subfigure}

    \vspace{0.3cm}
    \begin{subfigure}[b]{\textwidth}
        \centering
        \includegraphics[width=\textwidth]{plot_TSI10k/compare_neural_sl_pic_step_1980.png}
        \caption{$N_p = 10^4$}
    \end{subfigure}

    \vspace{0.3cm}
    \begin{subfigure}[b]{\textwidth}
        \centering
        \includegraphics[width=\textwidth]{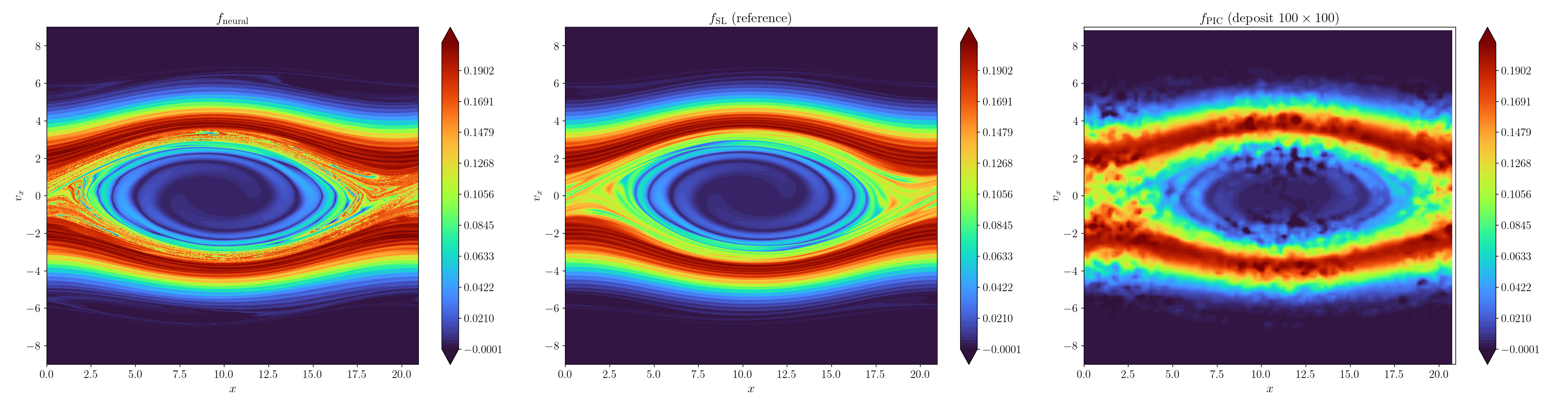}
        \caption{$N_p = 4\times10^4$}
    \end{subfigure}
    \caption{1D1V two-stream instability from \cref{sec:1D1V_two-stream_instability}: Comparison between the density given by the Neural $\df$ scheme (left), the BSL scheme (middle) and the standard $\df$ scheme (right) at $t = 99$, for an increasing number of particles $N_p = 10^3$, $10^4$ and $4\times10^4$. The BSL scheme uses a grid of size $1024 \times 1024$.}
    \label{fig:tsi_comparaison_np}
\end{figure}

\Cref{fig:tsi_comparaison_np} compares the three schemes at the final time $t = 99$ for an increasing number of particles, $N_p \in \{10^3, 10^4, 4\times10^4\}$. While the standard $\df$ scheme is strongly polluted by noise at low particle counts, the Neural $\df$ scheme remains close to the BSL reference even with $N_p = 10^3$ particles.

\begin{figure}[!ht]
    \centering
    \begin{subfigure}[b]{0.32\textwidth}
        \centering
        \includegraphics[width=\textwidth]{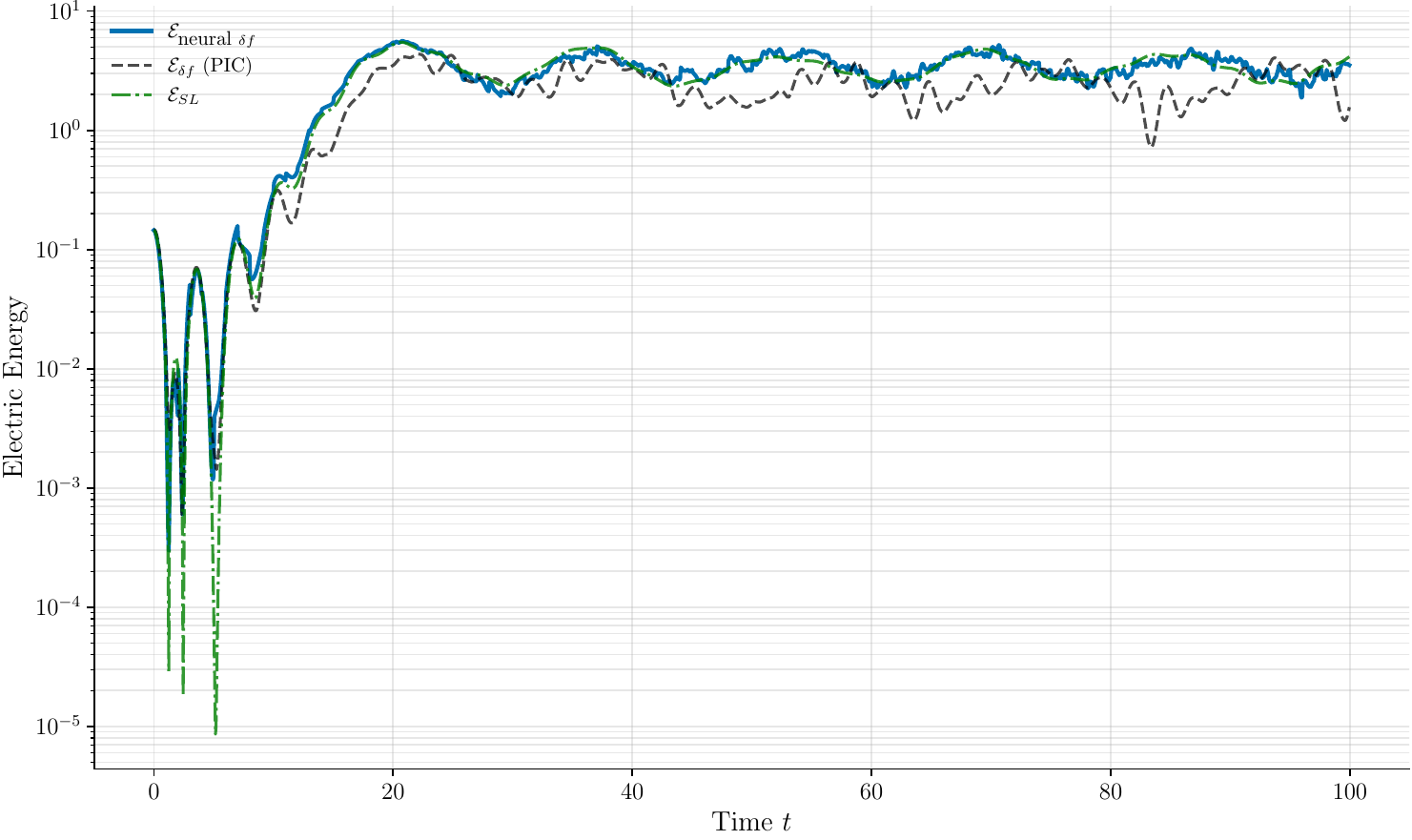}
        \caption{Electric energy, $N_p = 10^3$.}
    \end{subfigure}
    \hfill
    \begin{subfigure}[b]{0.32\textwidth}
        \centering
        \includegraphics[width=\textwidth]{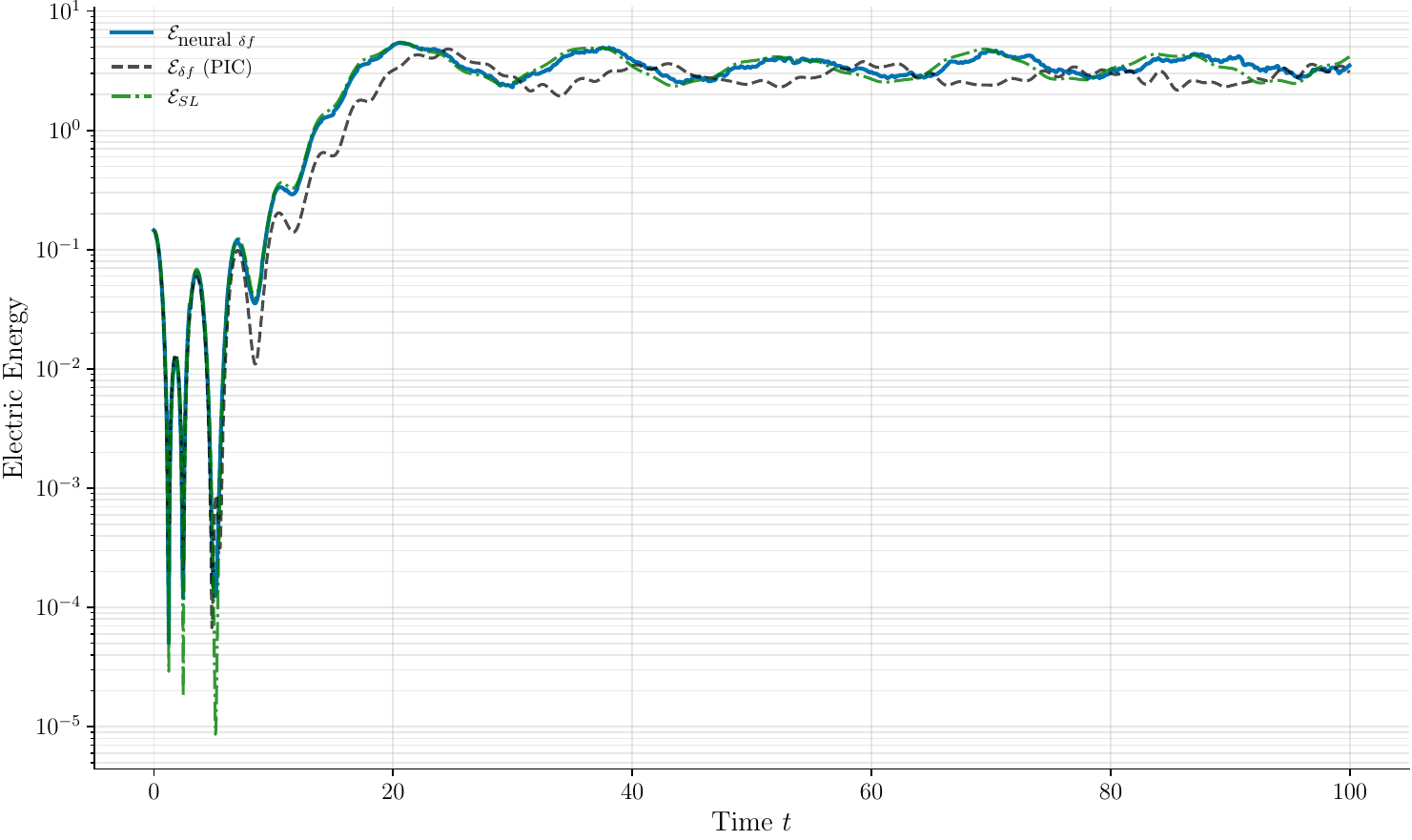}
        \caption{Electric energy, $N_p = 10^4$.}
    \end{subfigure}
    \hfill
    \begin{subfigure}[b]{0.32\textwidth}
        \centering
        \includegraphics[width=\textwidth]{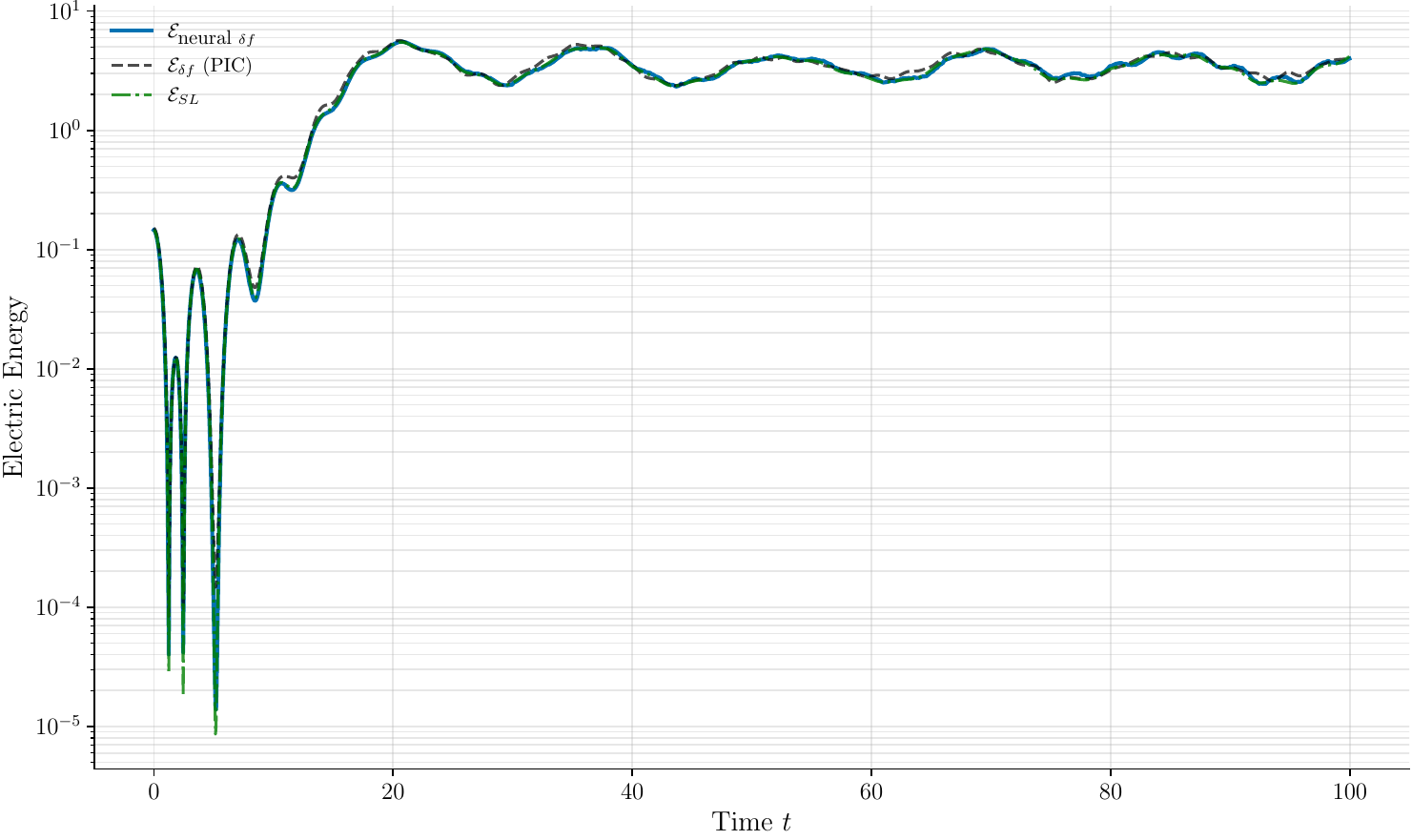}
        \caption{Electric energy, $N_p = 4\times10^4$.}
    \end{subfigure}

    \vspace{0.4cm}
    \begin{subfigure}[b]{0.32\textwidth}
        \centering
        \includegraphics[width=\textwidth]{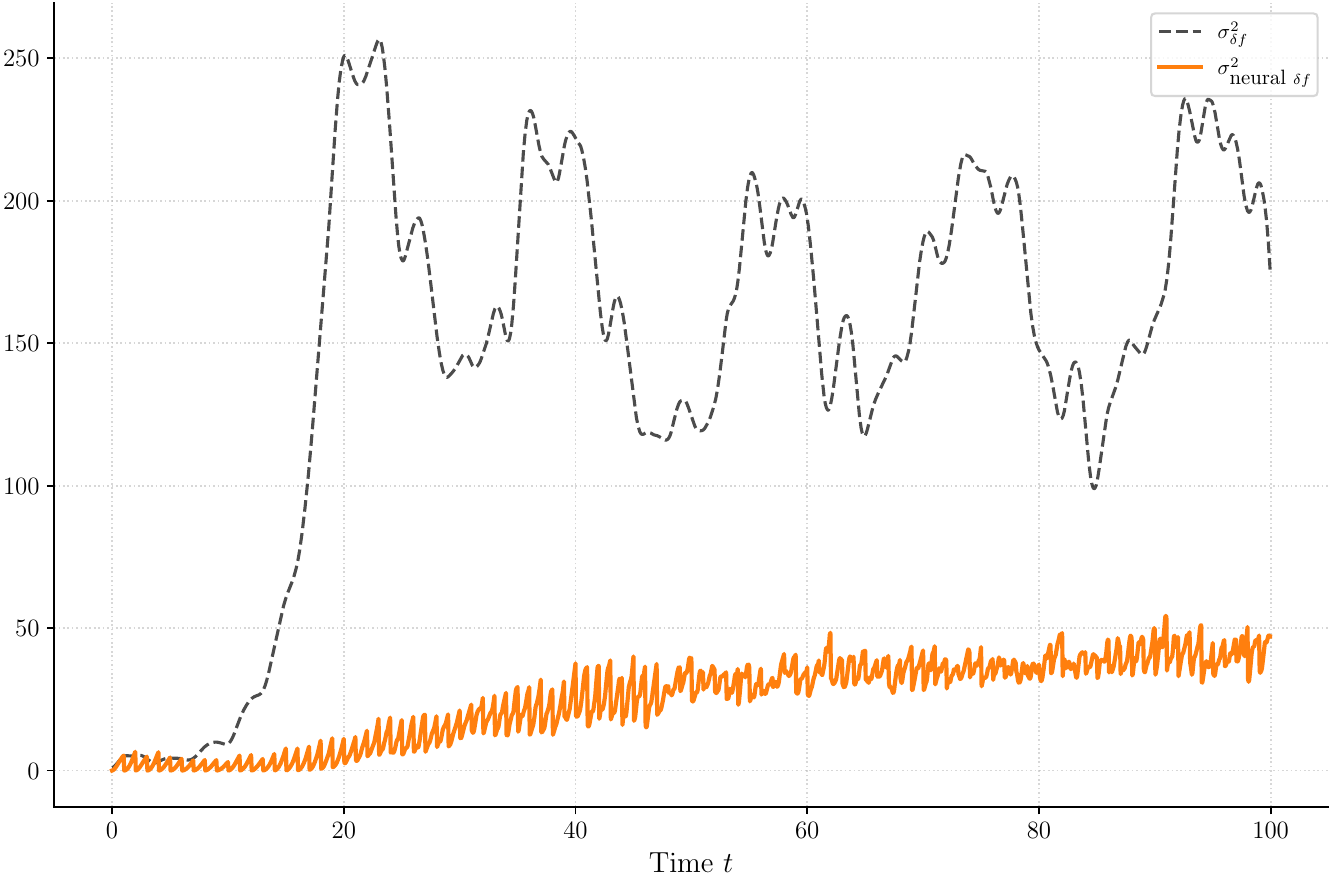}
        \caption{Weight variance, $N_p = 10^3$.}
    \end{subfigure}
    \hfill
    \begin{subfigure}[b]{0.32\textwidth}
        \centering
        \includegraphics[width=\textwidth]{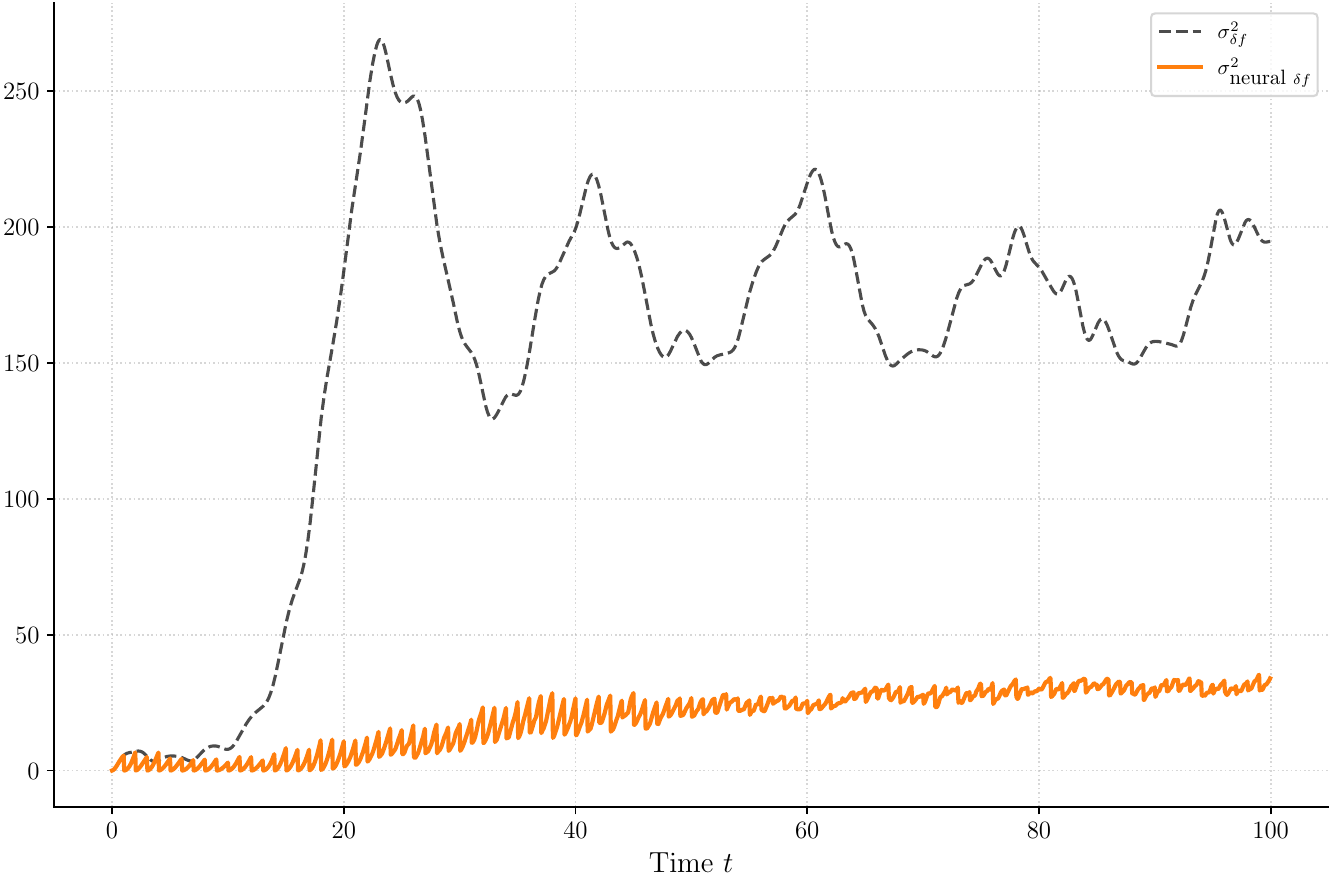}
        \caption{Weight variance, $N_p = 10^4$.}
    \end{subfigure}
    \hfill
    \begin{subfigure}[b]{0.32\textwidth}
        \centering
        \includegraphics[width=\textwidth]{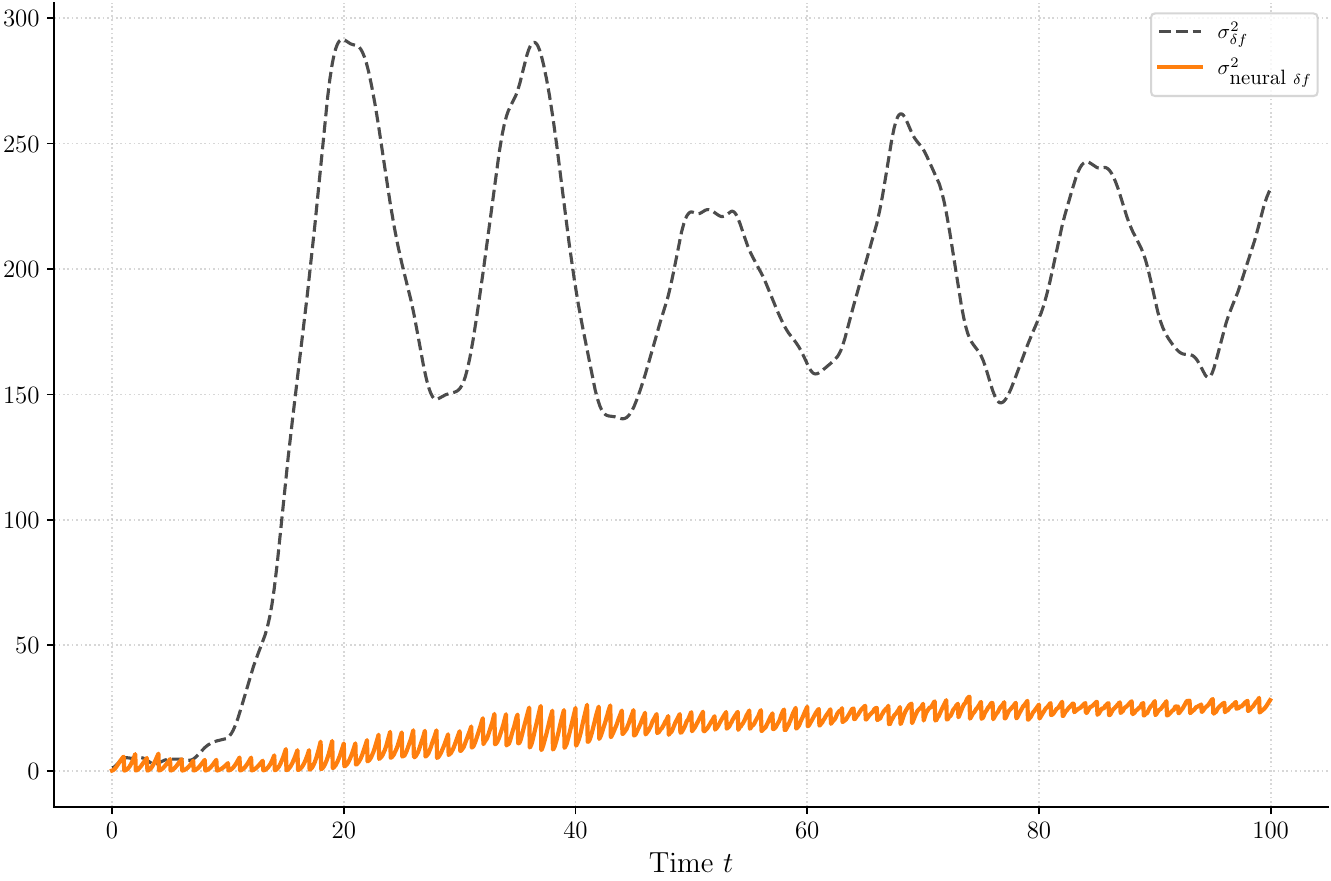}
        \caption{Weight variance, $N_p = 4\times10^4$.}
    \end{subfigure}
    \caption{1D1V two-stream instability from \cref{sec:1D1V_two-stream_instability}: time evolution of the electric energy (top row) and of the empirical weight variance $\sigma^2_{\df}$ (bottom row) for an increasing number of particles $N_p = 10^3$, $10^4$ and $4\times10^4$.}
    \label{fig:tsi_energy_variance_np}
\end{figure}

The top row of \cref{fig:tsi_energy_variance_np} shows the time evolution of the electric energy
$\mathcal{E}(t) =  \int_{L\T^d} |E(t,x) | ^2\,\mathrm{d}x$ for the Neural
$\df$-PIC method, the standard $\df$-PIC method and the BSL method, denoted by $\mathcal{E}_{\mathrm{neural } \ \df}$,  $\mathcal{E}_{\df }$ and  $\mathcal{E}_{\mathrm{SL}}$ respectively.
The Neural $\df$ method correctly captures the evolution of the electric energy and stays very close to the energy given by the BSL method, for all three particle counts.

The bottom row of \cref{fig:tsi_energy_variance_np} shows the time evolution of the empirical weight variance (which correspond to the case $\alpha(z) = 1$ from \Cref{sec:the_df_method}) for the Neural $\df$ and the standard $\df$ method. It is computed on the unnormalised weights $w_k := \Np\,\dw_k = \bigl(\finit(z_k^0)-{\fbulk^m}(z_k^n)\bigr)/g(z_k^0)$, by
\begin{equation}
    | \sigma_{\df}^n| ^2 = \frac{1}{\Np}\sum_{k=1}^{\Np}|w_k^n|^2 - \overline{w}^2,
    \text{\quad with \quad} \overline{w} = \frac{1}{\Np}\sum_{k=1}^{\Np} w_k.
    \label{eq:weight_variance}
\end{equation}

Defined this way, $|\sigma_{\df}^n|^2$ is an empirical estimate of the population variance, which is independent of $\Np$ and governs the Monte-Carlo error on the $\df$ part of the moments through $\sigma_{\df}/\Np^{\nicefrac 1 2}$.
In the standard $\df$-PIC method, the variance grows rapidly as the distribution evolves away from its initial state, reaching values between $150$ and $300$ once the instability saturates around $t=20$, and then oscillates within this range; consistently with its interpretation as a population variance, these values are essentially independent of $\Np$.
The Neural $\df$-PIC method keeps the variance much lower, from about $50$ at $\Np = 10^3$ down to about $25$ at $\Np = 4\times10^4$, since a larger number of markers yields a better-trained flow. The reduction factor thus grows from roughly $5$ at $\Np = 10^3$ to about $10$ at $\Np = 4\times10^4$. The periodic oscillations correspond to the bulk updates.
Nevertheless, one can observe a growth of the value of the weight even after the computation of the new bulk. One possible way to understand this phenomenon is the following. Denoting by $\|w\|^2_{\Np} := \frac{1}{\Np}\sum_{k=1}^{\Np} w_k^2$ a scaled $\ell^2$ norm on the weights,
one has
\begin{equation*}
    \sigma_{\df}^n
    \le \left(\frac{1}{\Np}\sum_{k=1}^{\Np}|w_k^n|^2\right)^{1/2}
    = \|w^n\|_{\Np}.
\end{equation*}
Writing the weight as
\begin{equation*}
    w_k^n
    = \underbrace{\frac{\widetilde{\fbulk^m}(z_k^n)-\fbulk^m(z_k^n)}{g(z_k^0)}}_{=:\,s_k^n}
    + \underbrace{\frac{\finit(z_k^0)-\widetilde{\fbulk^m}(z_k^n)}{g(z_k^0)}}_{=:\,e_k^n},
\end{equation*}
one obtains\begin{equation*}
    \sigma_{\df}^n \le \|w^n\|_{\Np} \le \|s^n\|_{\Np} + \|e^n\|_{\Np},
\end{equation*}
where $s^n$ measures the projection error of the bulk on the coarse B-spline grid,
and $e^n$ the transport error resulting from the approximation of the backward trajectories by the neural networks.

The first term $s^n$ is determined by the smoothness of the fine representation, the size of the grid and the order of the splines used.
The second term $e^n$ is determined by the accuracy of each trained flow, and by the number of flows in the composition. By composing several flows, each small error made on each flow gets amplified by the number of composed flows, resulting in a growth of the weights. A way to reduce this growth would be to reset the flows by a remapping of the bulk density, at the cost of an error on the approximation of the density.
With the incremental training strategy of \cref{sec:training}, the bulk at the final time $t=99$ is a composition of $35$, $40$ and $39$ networks for $\Np = 10^3$, $10^4$ and $4\times10^4$ respectively (against $14$, $17$ and $15$ at $t=50$), to be compared with the $99$ bulk updates performed over the run.

For the lowest particle count $N_p = 10^3$, the density of particles per grid cell for the Poisson solver is much lower, a little less than 1 particle per cell versus more than 3 per cell when $N_p = 10^4$ (we recall that we use 32 grid cells per direction in all test cases). As seen in \Cref{fig:tsi_energy_variance_np}, this has the consequence that the standard $\df$ scheme is unable to track the correct evolution of the electric energy, while the Neural $\df$ scheme is able to roughly follow the semi-Lagrangian scheme despite also having a low number of sample points to train the networks.

\begin{figure}[!ht]
    \centering
    \includegraphics[width=0.495\textwidth]{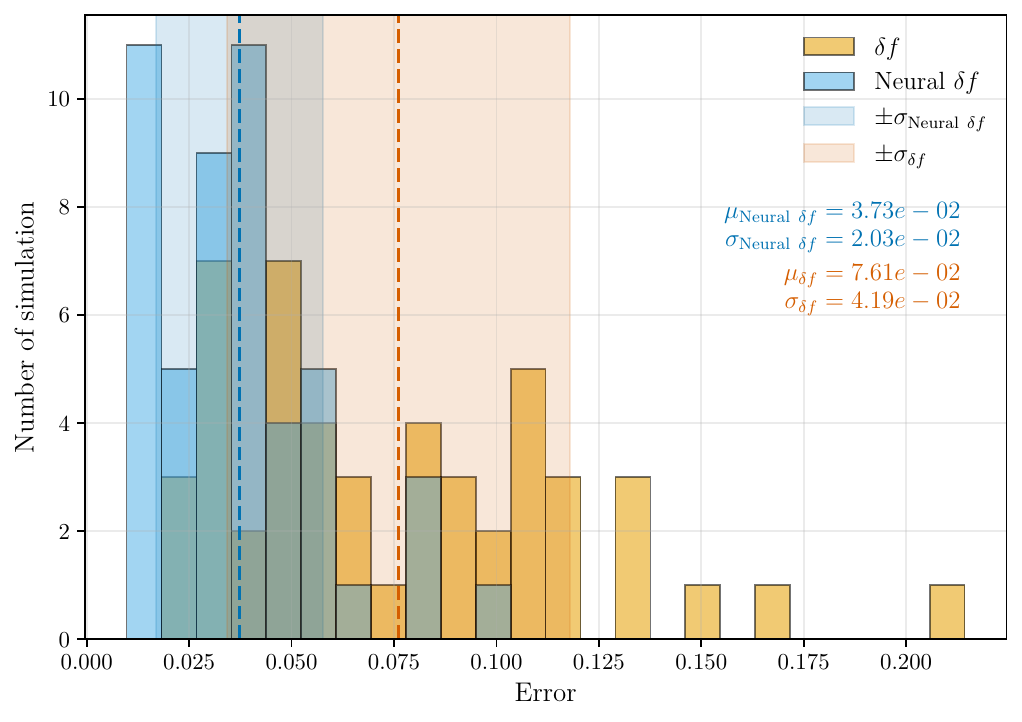}
    \includegraphics[width=0.495\textwidth]{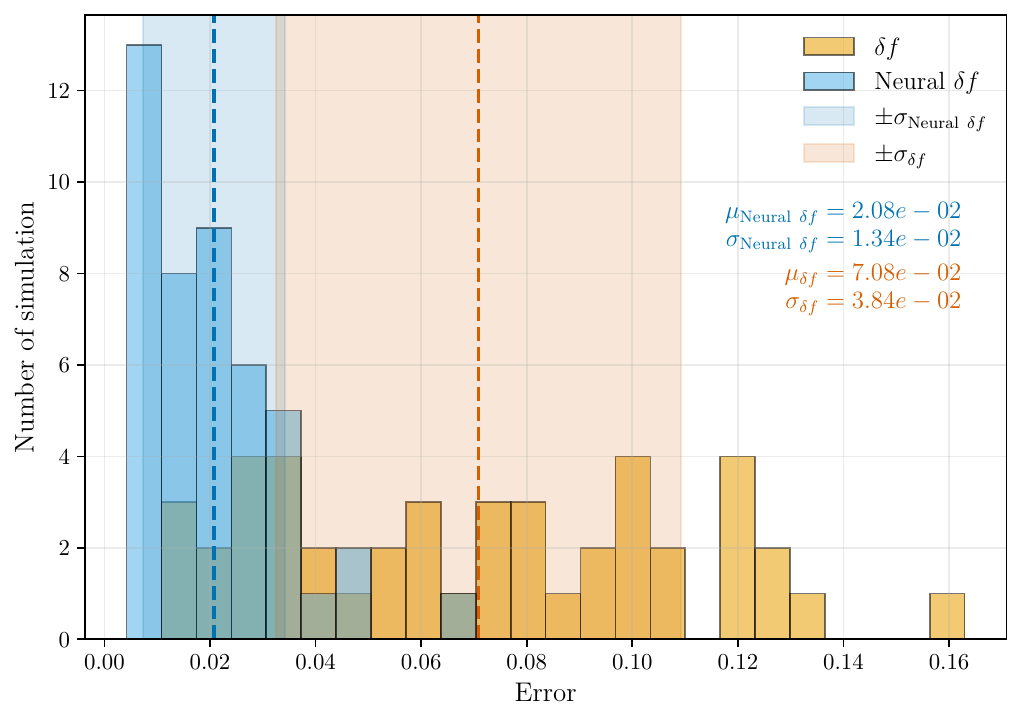}
    \caption{1D1V two-stream instability from \cref{sec:1D1V_two-stream_instability}: Error distribution of the error on the electric energy. In blue are the error for Neural $\df$ scheme, in orange are the error for the standard $\df$ scheme. On the left the simulations are done with $N_p = 20000$ particles, and on the right with $N_p = 30000$ particles.}
    \label{fig:tsi_stats}
\end{figure}

In \Cref{fig:tsi_stats}, we assessed the noise reduction in a statistically robust manner, by performing
50 independent realizations of both the standard and Neural $\df$-PIC
methods with the same number of particles.
Among these simulations, the only change is the initial marker draw.
The error metric is the
$L^2$-in-time error of the electric energy relative to the BSL reference:
\begin{equation}
    \left(\int_0^T \abs{\mathcal{E}(t) - \mathcal{E}_{\mathrm{BSL}}(t)}^2\,\mathrm{d}t\right)^{1/2}.
    \label{eq:error_metric}
\end{equation}
Overall, we notice a reduction of both the mean and of the variance of the error.
We emphasize that the present study is a \emph{proof of concept} on the Vlasov-Poisson system. In this setting, the Neural $\df$-PIC method is, in terms of computational time, far more expensive than the standard $\df$-PIC or the BSL scheme, and we do not claim it to be competitive with a standard $\df$ scheme using a larger number of particles to reach a comparable accuracy. The additional memory cost, on the other hand, remains low, since the stored networks are small. Our objective is rather to assess whether symplectic neural networks can effectively denoise the density by dynamically evolving the bulk. This is a prerequisite for the regimes we ultimately target, such as flux-driven or edge gyrokinetic simulations, where the distribution departs strongly from any static or Maxwellian equilibrium, so that evolving the bulk is genuinely necessary, and where simply increasing the number of particles is prohibitively expensive.

\subsubsection{1D1V bump-on-tail instability}
\label{sec:bot}

The bump-on-tail instability consists of an initial centered Maxwellian together with a beam of particles with a positive velocity:
\begin{equation}
    \finit(x,v) = \left(
    \frac{0.9}{\sqrt{2\pi}}\,e^{-v^2/2}
    + \frac{0.2}{\sqrt{10\pi}}\,e^{-(v-3.8)^2/10}
    \right)
    \bigl(1 + 0.03\cos(0.4x)\bigr).
    \label{eq:bot_init}
\end{equation}
The computational domain is $[0,10\pi]\times[-6,6]$ with $\Np = 10^4$, $\dt = 0.05$, and $N_\Psi = 10$. As in the two-stream instability, we also run this test with $\Np = 10^3$ and $\Np = 4\times10^4$ particles to assess the effect of the number of particles.
This test case is more challenging because the solution creates two vortices that move in time, yielding a more complex dynamics that the network has to reconstruct.

\begin{figure}[!ht]
    \centering
    \begin{subfigure}[b]{\textwidth}
        \centering
        \includegraphics[width=\textwidth]{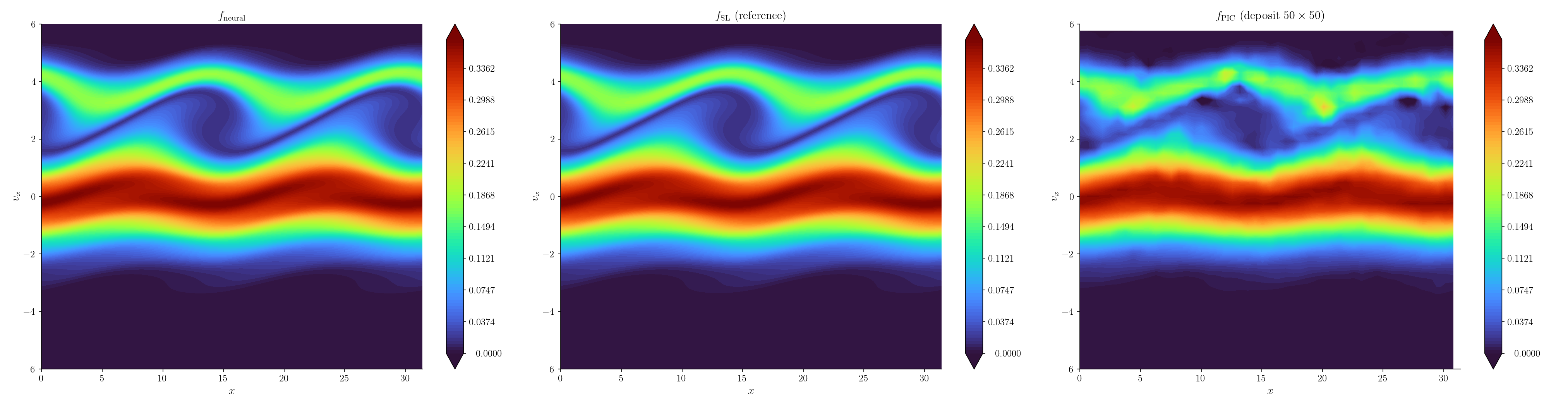}
        \caption{$t = 15$}
    \end{subfigure}

    \vspace{0.3cm}
    \begin{subfigure}[b]{\textwidth}
        \centering
        \includegraphics[width=\textwidth]{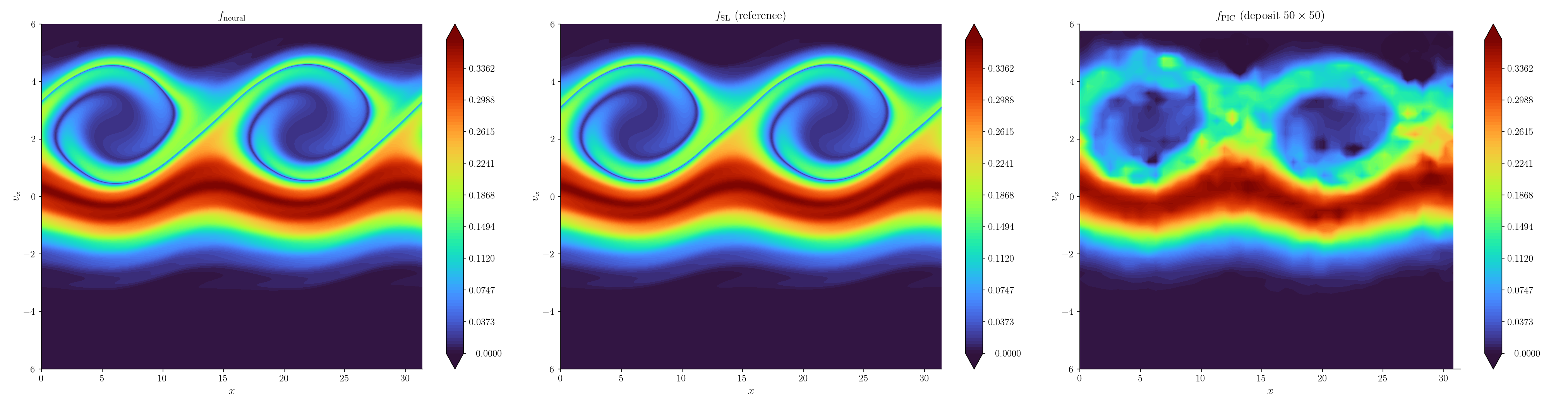}
        \caption{$t = 24$}
    \end{subfigure}

    \vspace{0.3cm}
    \begin{subfigure}[b]{\textwidth}
        \centering
        \includegraphics[width=\textwidth]{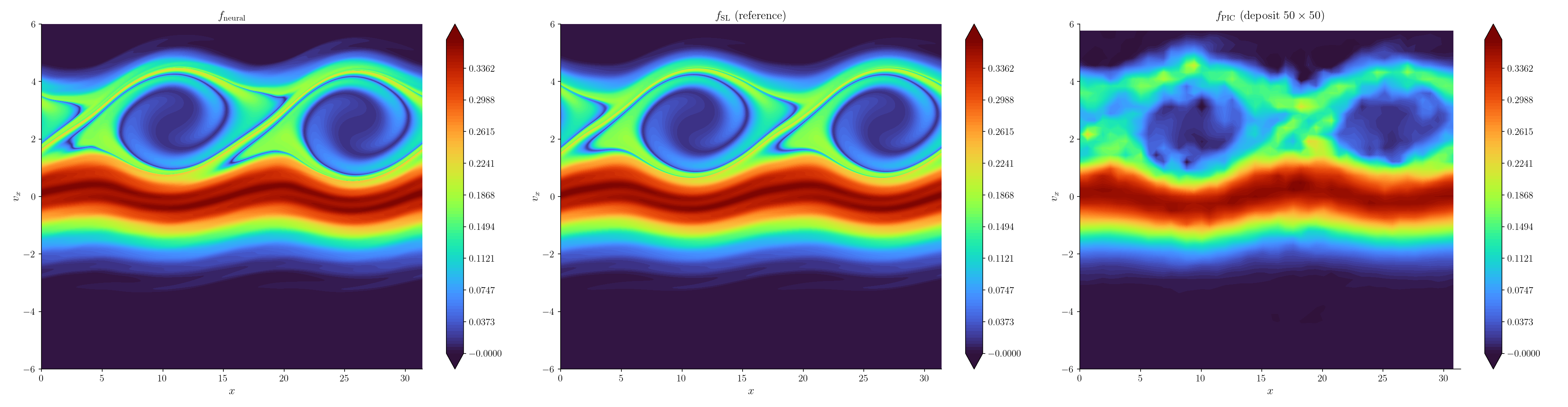}
        \caption{$t = 32$}
    \end{subfigure}
    \caption{1D1V bump-on-tail instability from \cref{sec:bot}: Comparison between the density given by the Neural $\df$ scheme (left), the BSL scheme (middle) and the standard $\df$ scheme (right), at $t = 15$, $t = 24$ and $t = 32$. The Neural $\df$ and the standard $\df$ schemes use $\Np = 10^4$ particles.}
    \label{fig:bot_comparaison}
\end{figure}

\Cref{fig:bot_comparaison} compares the Neural $\df$ scheme with the BSL reference and the standard $\df$ scheme at $t = 15$, $24$ and $32$, with $\Np = 10^4$ particles. The bump-on-tail instability generates a more complex phase-space structure than the two-stream case: because the domain is larger, two vortices appear, making the flow more difficult to approximate. The Neural $\df$ scheme nonetheless reproduces this structure accurately and strongly reduces the noise of the phase-space density compared to the standard $\df$ scheme.

\begin{figure}[!ht]
    \centering
    \begin{subfigure}[b]{\textwidth}
        \centering
        \includegraphics[width=\textwidth]{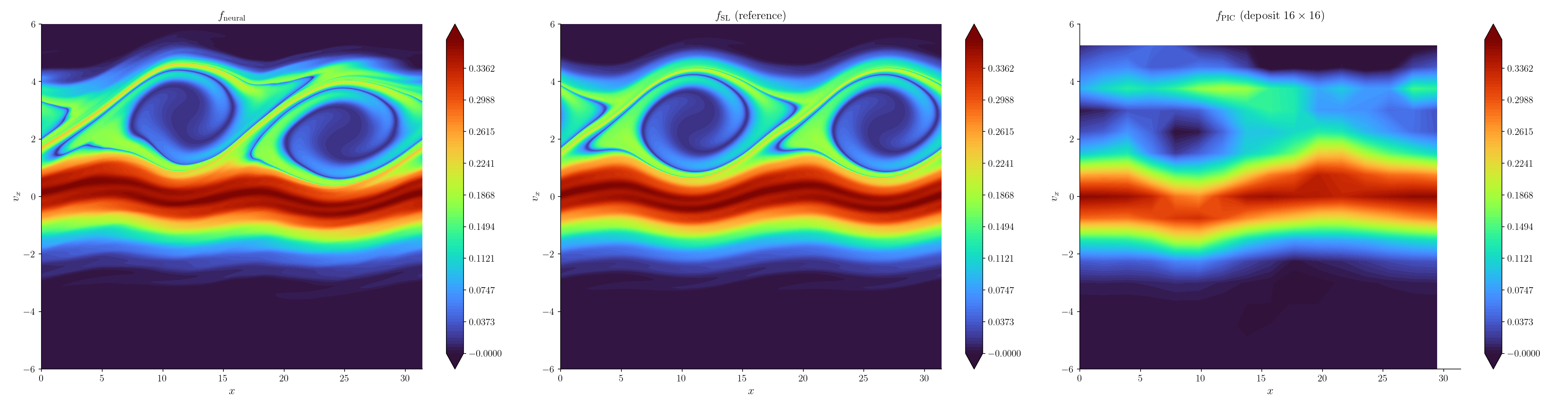}
        \caption{$\Np = 10^3$}
    \end{subfigure}

    \vspace{0.3cm}
    \begin{subfigure}[b]{\textwidth}
        \centering
        \includegraphics[width=\textwidth]{plot_BOT10k/compare_neural_sl_pic_step_640.png}
        \caption{$\Np = 10^4$}
    \end{subfigure}

    \vspace{0.3cm}
    \begin{subfigure}[b]{\textwidth}
        \centering
        \includegraphics[width=\textwidth]{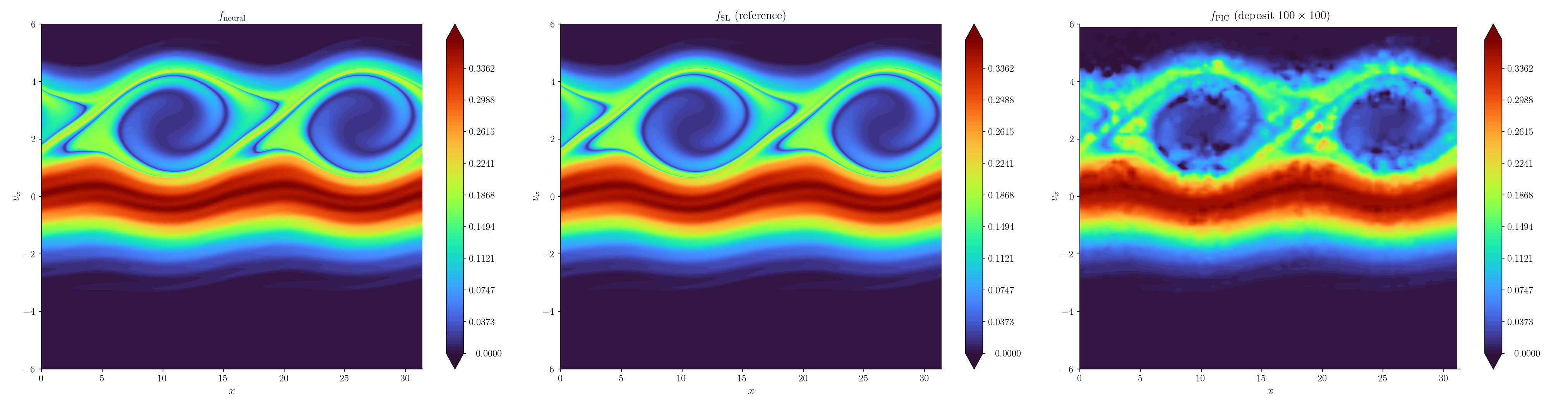}
        \caption{$\Np = 4\times10^4$}
    \end{subfigure}
    \caption{1D1V bump-on-tail instability from \cref{sec:bot}: Comparison between the density given by the Neural $\df$ scheme (left), the BSL scheme (middle) and the standard $\df$ scheme (right) at $t = 32$, for an increasing number of particles $\Np = 10^3$, $10^4$ and $4\times10^4$.}
    \label{fig:bot_comparaison_np}
\end{figure}

\Cref{fig:bot_comparaison_np} shows the same comparison at $t = 32$ for an increasing number of particles, $\Np = 10^3$, $10^4$ and $4\times10^4$. As for the two-stream case, the standard $\df$ scheme is strongly polluted by noise at low particle counts, whereas the Neural $\df$ scheme remains close to the BSL reference.

\begin{figure}[!ht]
    \centering
    \begin{subfigure}[b]{0.32\textwidth}
        \centering
        \includegraphics[width=\textwidth]{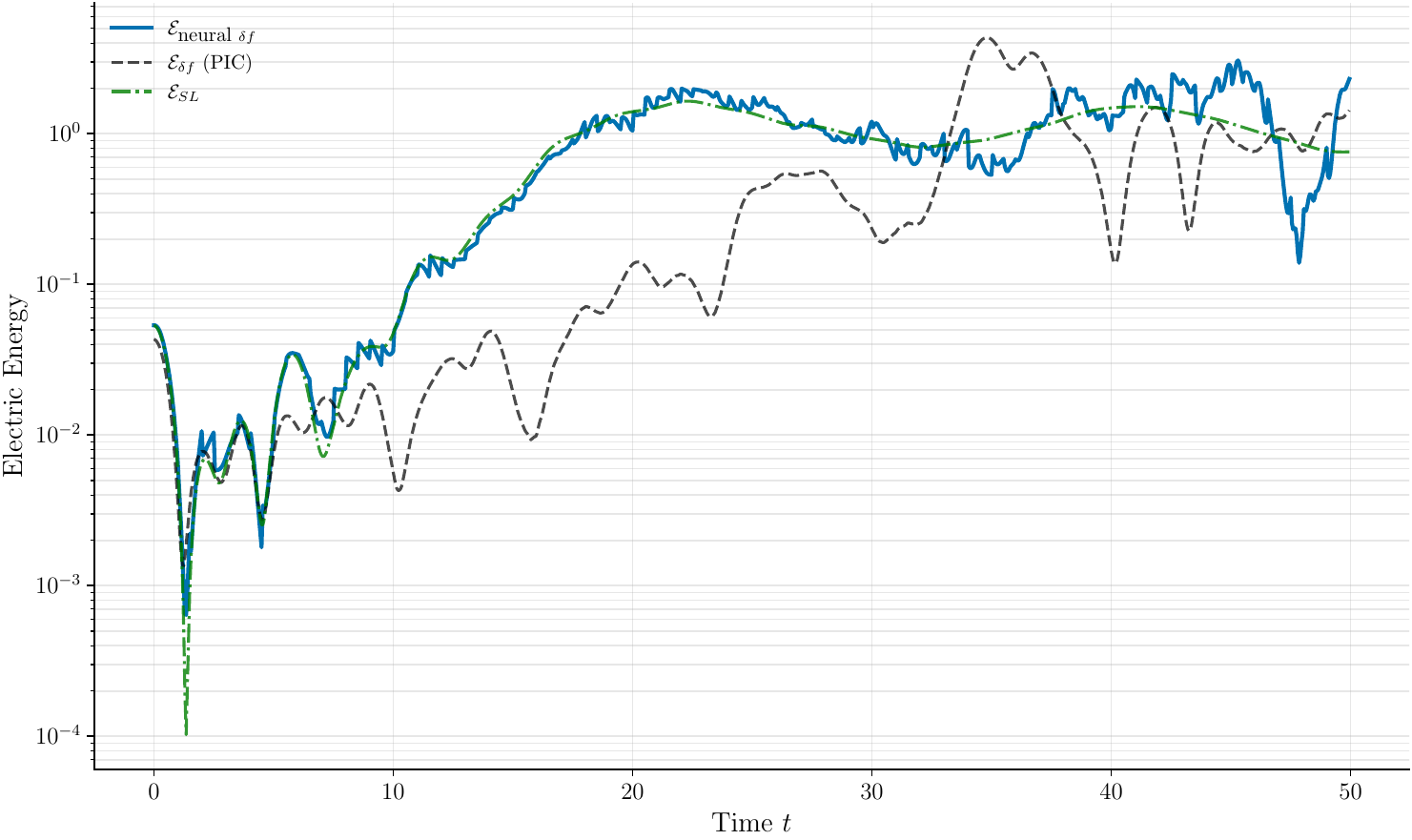}
        \caption{Electric energy, $\Np = 10^3$.}
    \end{subfigure}
    \hfill
    \begin{subfigure}[b]{0.32\textwidth}
        \centering
        \includegraphics[width=\textwidth]{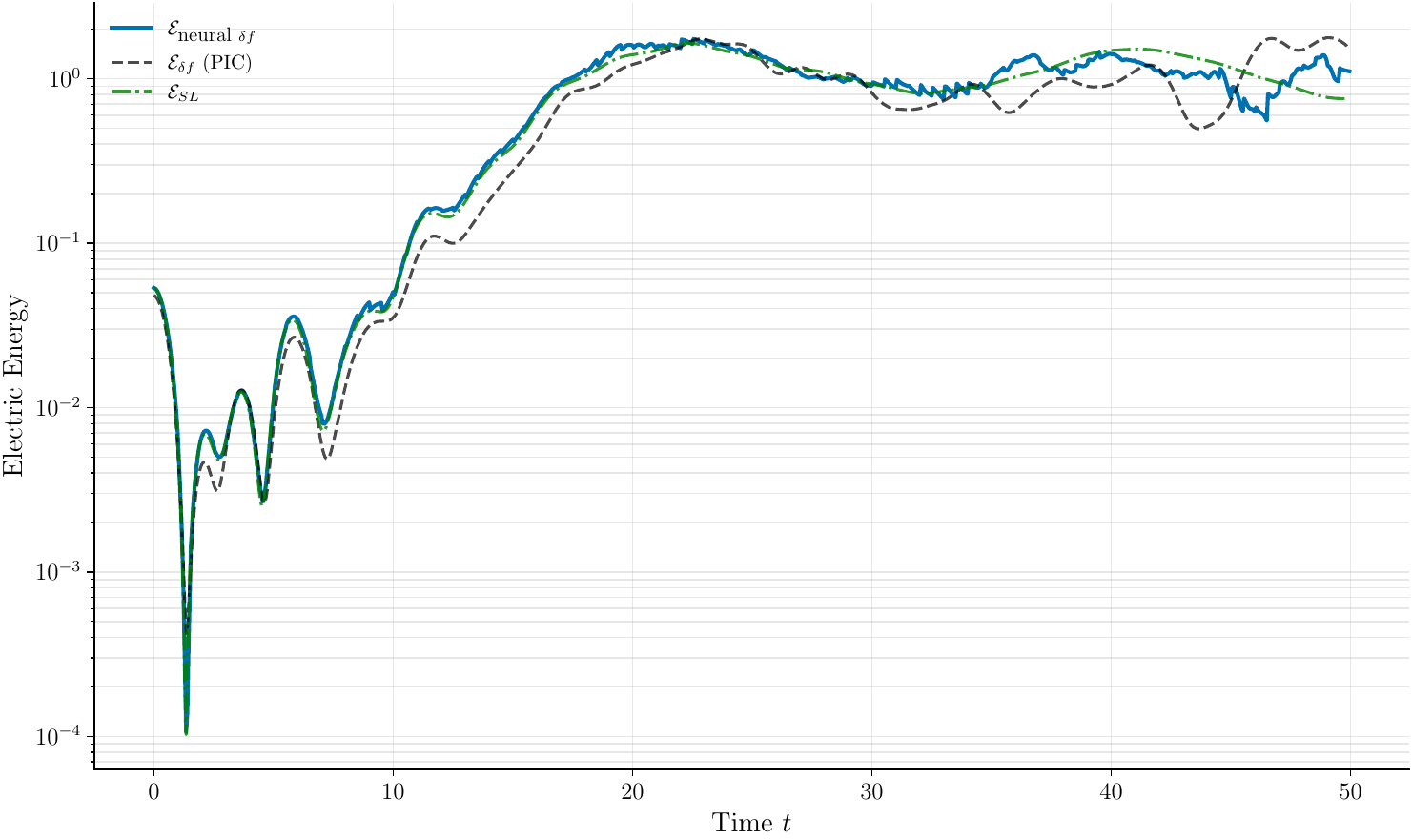}
        \caption{Electric energy, $\Np = 10^4$.}
    \end{subfigure}
    \hfill
    \begin{subfigure}[b]{0.32\textwidth}
        \centering
        \includegraphics[width=\textwidth]{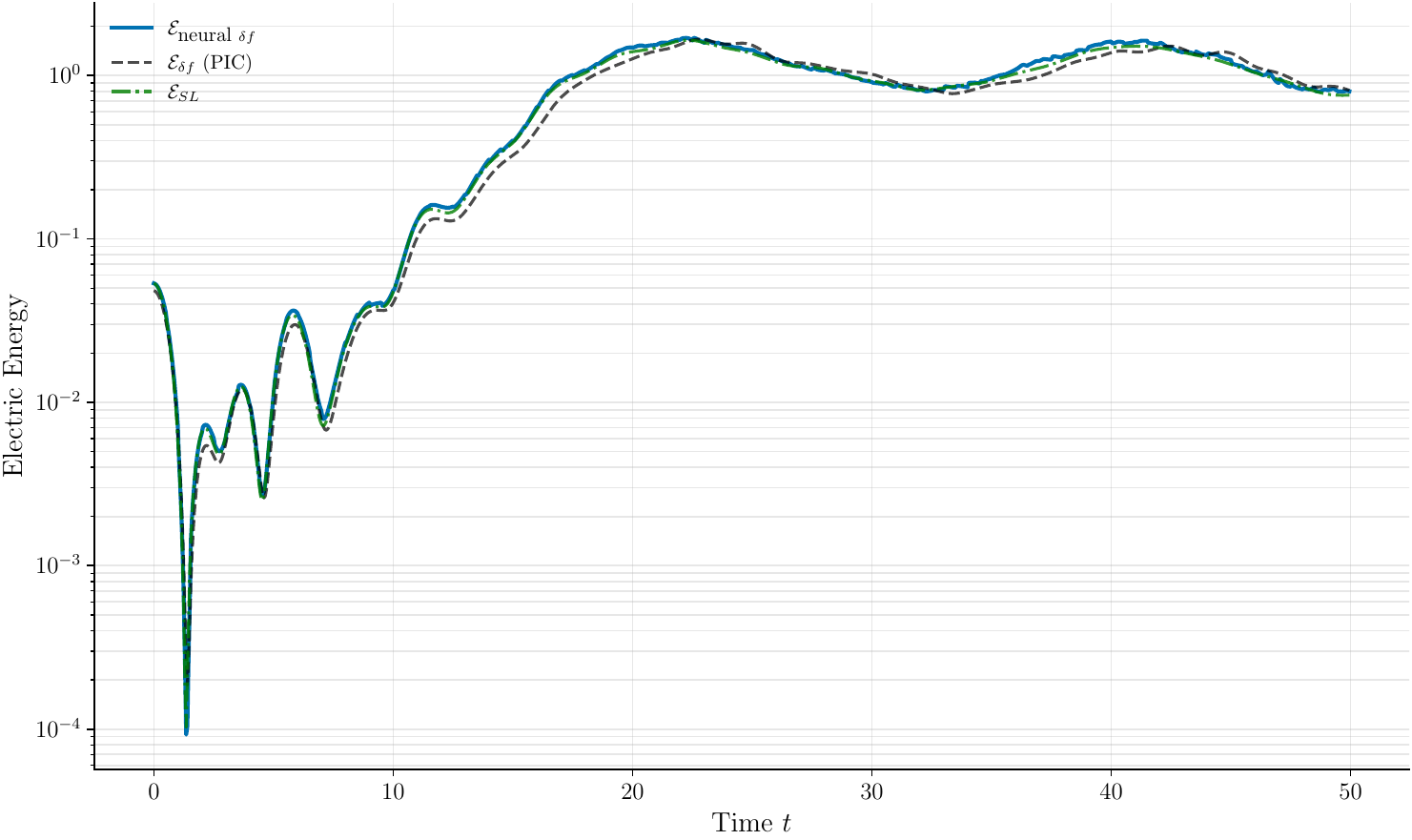}
        \caption{Electric energy, $\Np = 4\times10^4$.}
    \end{subfigure}

    \vspace{0.4cm}
    \begin{subfigure}[b]{0.32\textwidth}
        \centering
        \includegraphics[width=\textwidth]{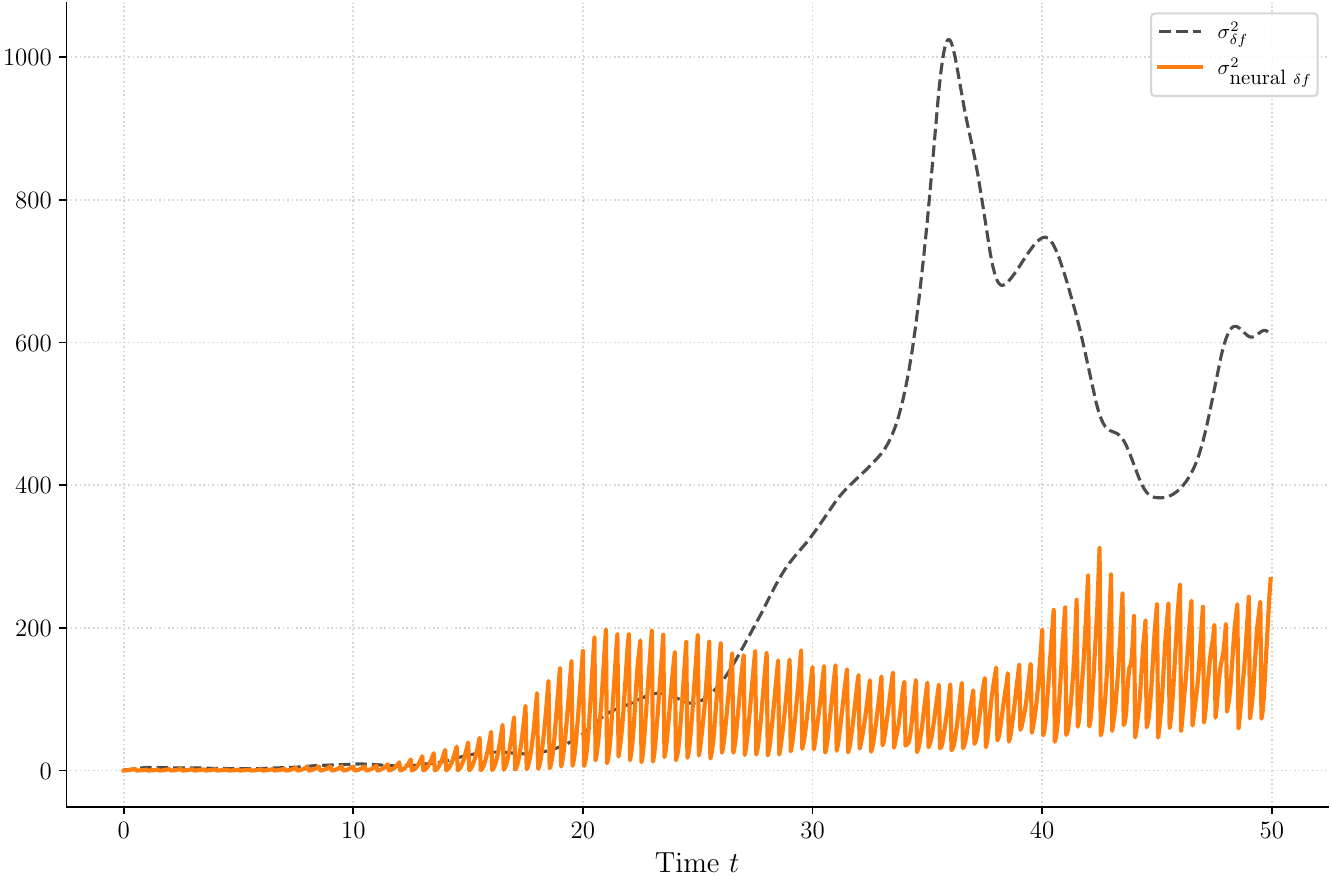}
        \caption{Weight variance, $\Np = 10^3$.}
    \end{subfigure}
    \hfill
    \begin{subfigure}[b]{0.32\textwidth}
        \centering
        \includegraphics[width=\textwidth]{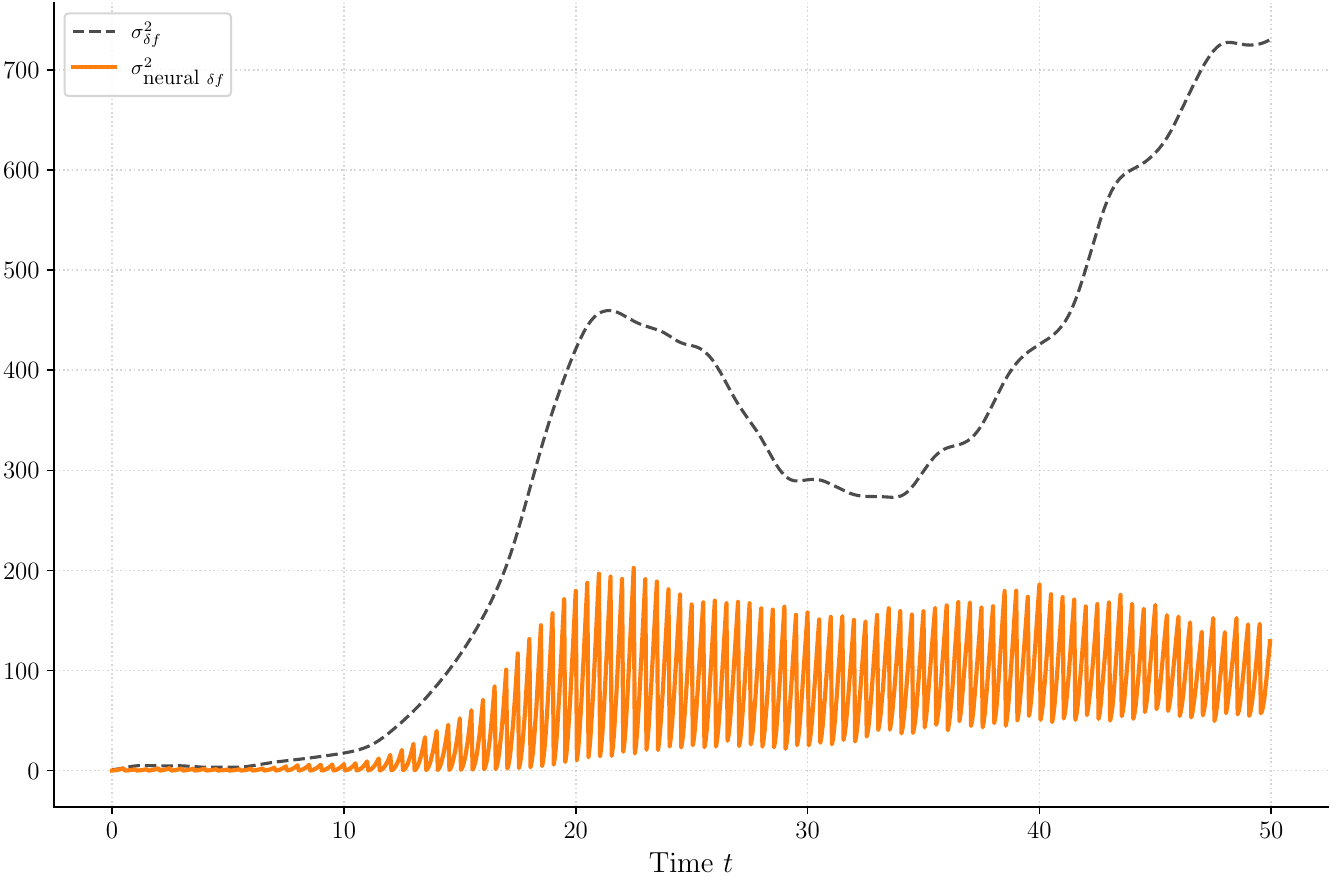}
        \caption{Weight variance, $\Np = 10^4$.}
    \end{subfigure}
    \hfill
    \begin{subfigure}[b]{0.32\textwidth}
        \centering
        \includegraphics[width=\textwidth]{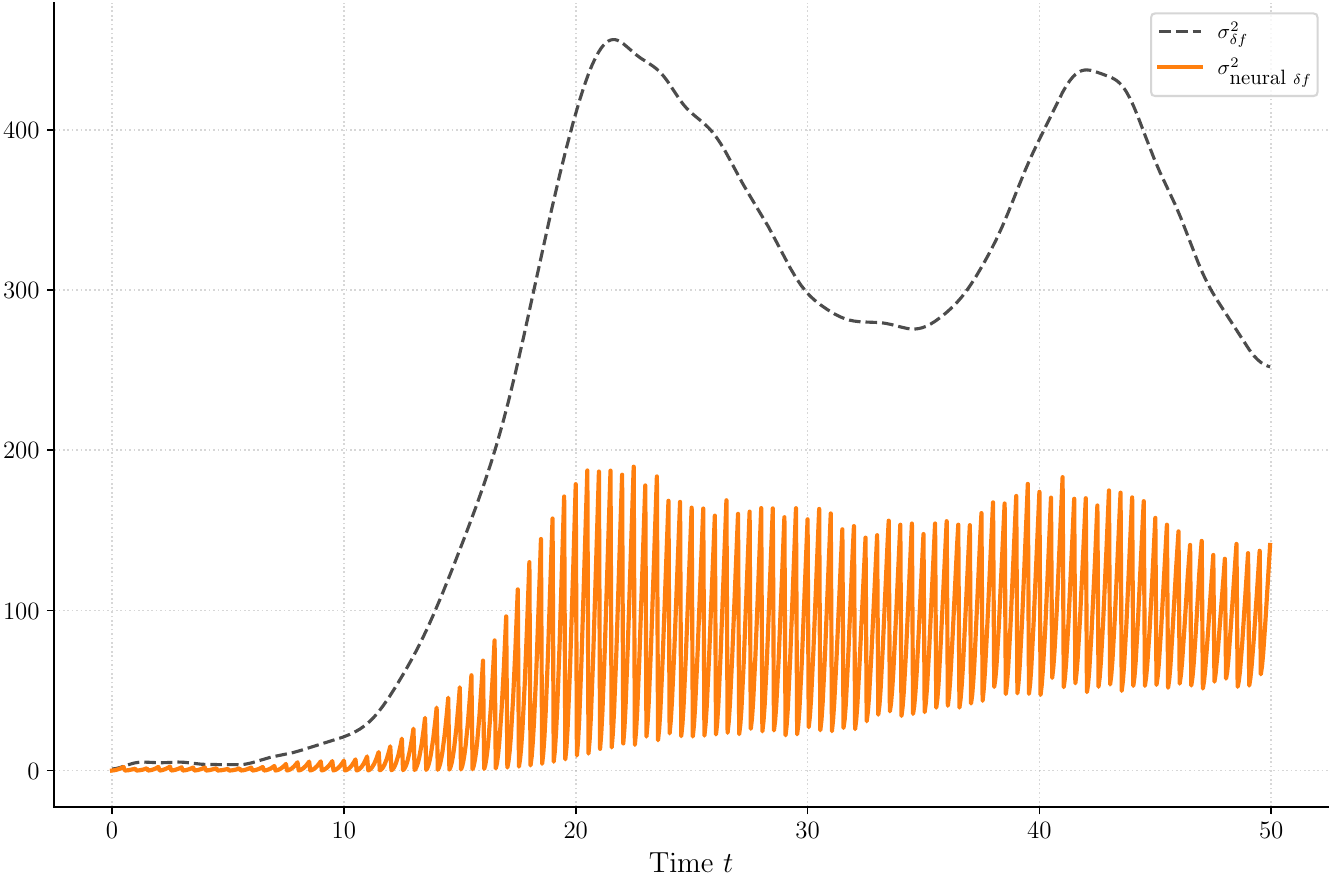}
        \caption{Weight variance, $\Np = 4\times10^4$.}
    \end{subfigure}
    \caption{1D1V bump-on-tail instability from \cref{sec:bot}: time evolution of the electric energy (top row) and of the weight variance $\sigma_{\df}^2$ (bottom row) for an increasing number of particles $\Np = 10^3$, $10^4$ and $4\times10^4$.}
    \label{fig:bot_energy_variance_np}
\end{figure}

\Cref{fig:bot_energy_variance_np} shows the time evolution of the electric energy and of the weight variance $\sigma_{\df}^2$ for the same three particle counts. The Neural $\df$-PIC method tracks the reference electric energy given by the BSL scheme at all particle counts. As in the two-stream case, the standard $\df$-PIC method with a static bulk shows a strong growth of the variance, here particularly pronounced because the solution deviates more strongly from the initial distribution, whereas the Neural $\df$-PIC method reduces it by a factor of a little less than $10$; one can again observe a growth of the weights after each bulk update. With the incremental training strategy of \cref{sec:training}, the bulk at the final time is a composition of $18$, $22$ and $21$ networks for $\Np = 10^3$, $10^4$ and $4\times10^4$ respectively, to be compared with the $100$ bulk updates performed over the run.

\begin{figure}[!ht]
    \centering
    \includegraphics[width=0.6\textwidth]{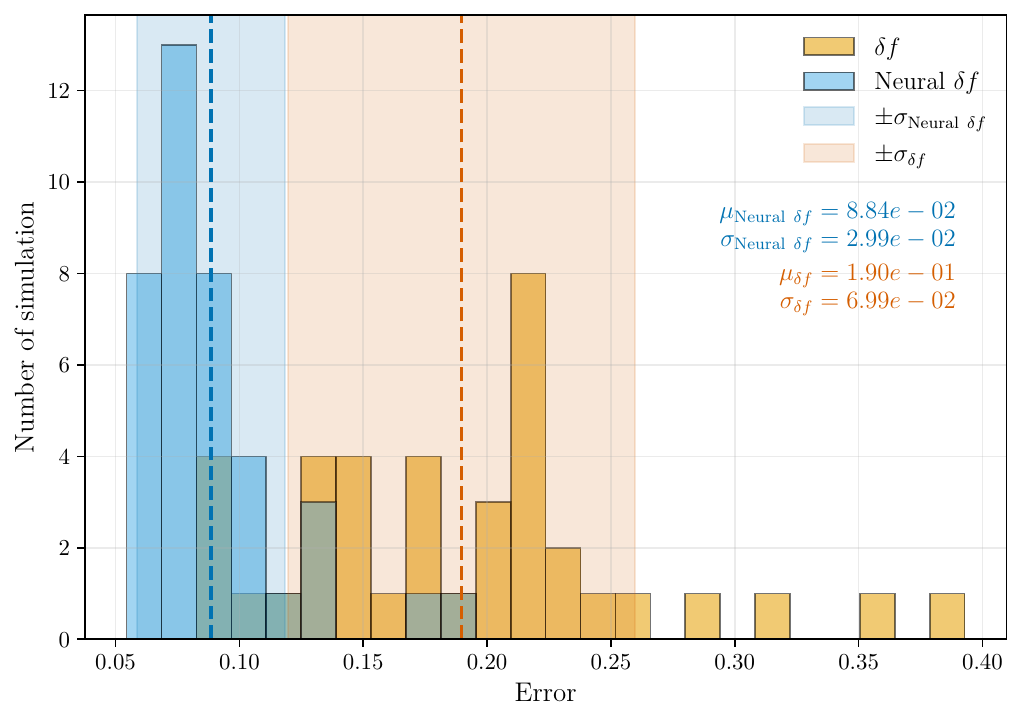}
    \caption{1D1V bump-on-tail instability from \cref{sec:bot}: Histogram of the error between the Neural $\df$ method (in blue) and the standard $\df$ method. The reference solution is given by a BSL scheme. The simulations are done with $N_p = 20000$ particles.}
    \label{fig:bot_stats}
\end{figure}

Just like in the 1D1V two-stream instability, we performed
50 independent simulations with both the standard and Neural $\df$
methods with the same number of particles. The results are illustrated in \cref{fig:bot_stats}. In this case, the reduction in the error is more important than for the two-stream instability.

\subsection{3D3V test cases}
\label{sec:3d3v}

We now present test cases for the full 3D3V Vlasov-Poisson problem.
We consider four test cases: the first three are extensions of the 1D1V two-stream instability, the last one is an extension of the bump-on-tail instability.
We use $\Np = 10^6$ particles for the two-stream and bump-on-tail cases, and $\Np = 10^8$ for the four-stream and six-stream cases.
In all 3D3V test cases, we use the same time step $\dt = 0.05$ as in the 1D1V cases, and a bulk-update period of $N_\Psi = 10$. We recall that, like in 1D1V, the SympNets have $\ell = 10$ layers of width $w=8$. However, because the total number of parameters also depends on the dimension of the inputs (the total number of parameters of one network is $3w\ell (d+1) $), the networks in 3D3V have $960$ parameters, while the networks in 1D1V have 480 parameters.
This section is meant to be exploratory: its purpose is to demonstrate that the method runs in a full six-dimensional phase space, rather than to provide a quantitative error analysis. Producing a converged reference solution in 6D is indeed very difficult: even the finest $64^3 \times 63^3$ semi-Lagrangian grid used below is likely under-resolved for the fine filamentation that develops, so the BSL6D results~\cite{schild2024performance} are used only as a qualitative cross-check and not as a converged reference. We also emphasize that all the results reported here are obtained with a non-parallelized and non optimized code running on a single GPU. This constrains the number of particles we can afford, and hence the resolution attainable in 6D; a parallel implementation, which would lift these limitations, is left for future work.

Nonetheless, we consider that the results presented here are encouraging, as they show that the Neural $\df$-PIC method is able to run 6D test cases on a single process with acceptable accuracy, and to reduce the noise of the phase-space density.

\subsubsection{3D3V Two-stream instability}
\label{sec:3D3V_two_stream}

The initial condition is
\begin{equation}
    \finit(x,y,z,v_x,v_y,v_z) = ( 1+ \varepsilon \cos(kx)) \frac{1}{2 \sqrt{2\pi}} \bigl(e^{-(v_x - v_0)^2/2} + e^{-(v_x + v_0)^2/2} \bigr) \frac{1}{2 \pi} e^{-(v_y^2+v_z^2)/2},
    \label{eq:tsi_3d_init}
\end{equation}
with $k=0.3$, $v_0 = 2.4$ and $\varepsilon= 0.05$. The computational domain is $[0,2\pi/k] \times [0,1]^2 \times [-7,7]^3.$

\begin{figure}[!ht]
    \centering

    \begin{subfigure}[b]{\textwidth}
        \centering
        \includegraphics[width=\textwidth]{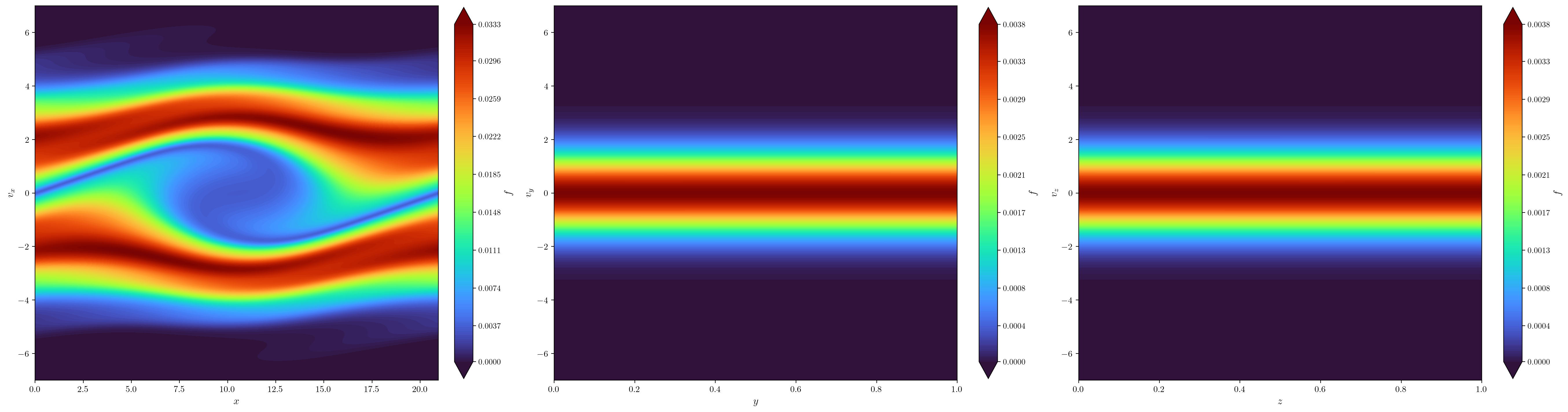}
        \caption{Cross sections of the density at $t=15$}
        \label{fig:TSI3D_1}
    \end{subfigure}

    \vspace{0.5cm}
    \begin{subfigure}[b]{\textwidth}
        \centering
        \includegraphics[width=\textwidth]{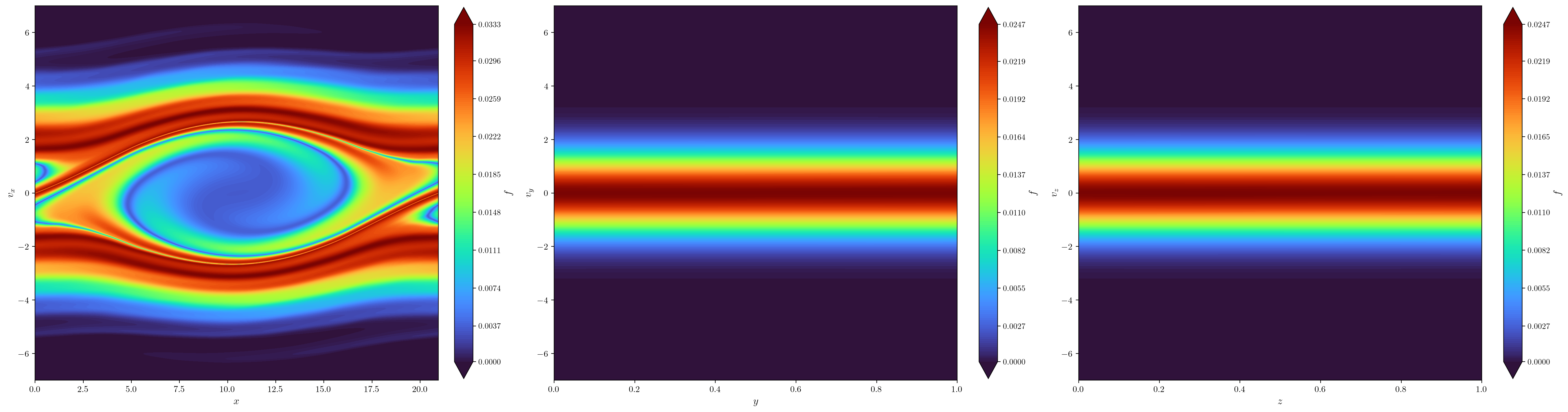}
        \caption{Cross sections of the density at $t=30$}
        \label{fig:TSI3D_2}
    \end{subfigure}

    \vspace{0.5cm}

    \begin{subfigure}[b]{\textwidth}
        \centering
        \includegraphics[width=\textwidth]{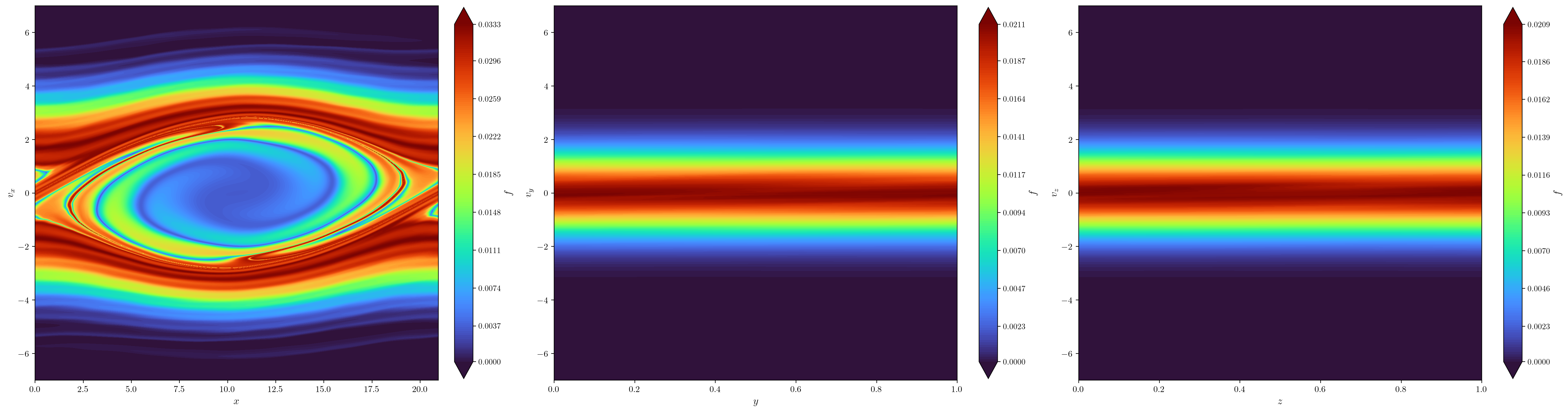}
        \caption{Cross sections of the density at $t=45$}
        \label{fig:TSI3D_3}
    \end{subfigure}

    \caption{3D3V two-stream instability from \cref{sec:3D3V_two_stream}: Cross sections of the phase-space density $f(x,y=0,z=0,v_x,v_y = 0,v_z=0)$, $f(x=0,y,z=0,v_x=0,v_y,v_z=0)$ and $f(x=0,y=0,z,v_x=0,v_y=0,v_z)$ at $t = 15$, $t=30$ and $t=45$.}
    \label{fig:TSI3D_density}
\end{figure}

\begin{figure}[!ht]
    \centering

    \begin{subfigure}[b]{0.48\textwidth}
        \centering
        \includegraphics[width=\textwidth]{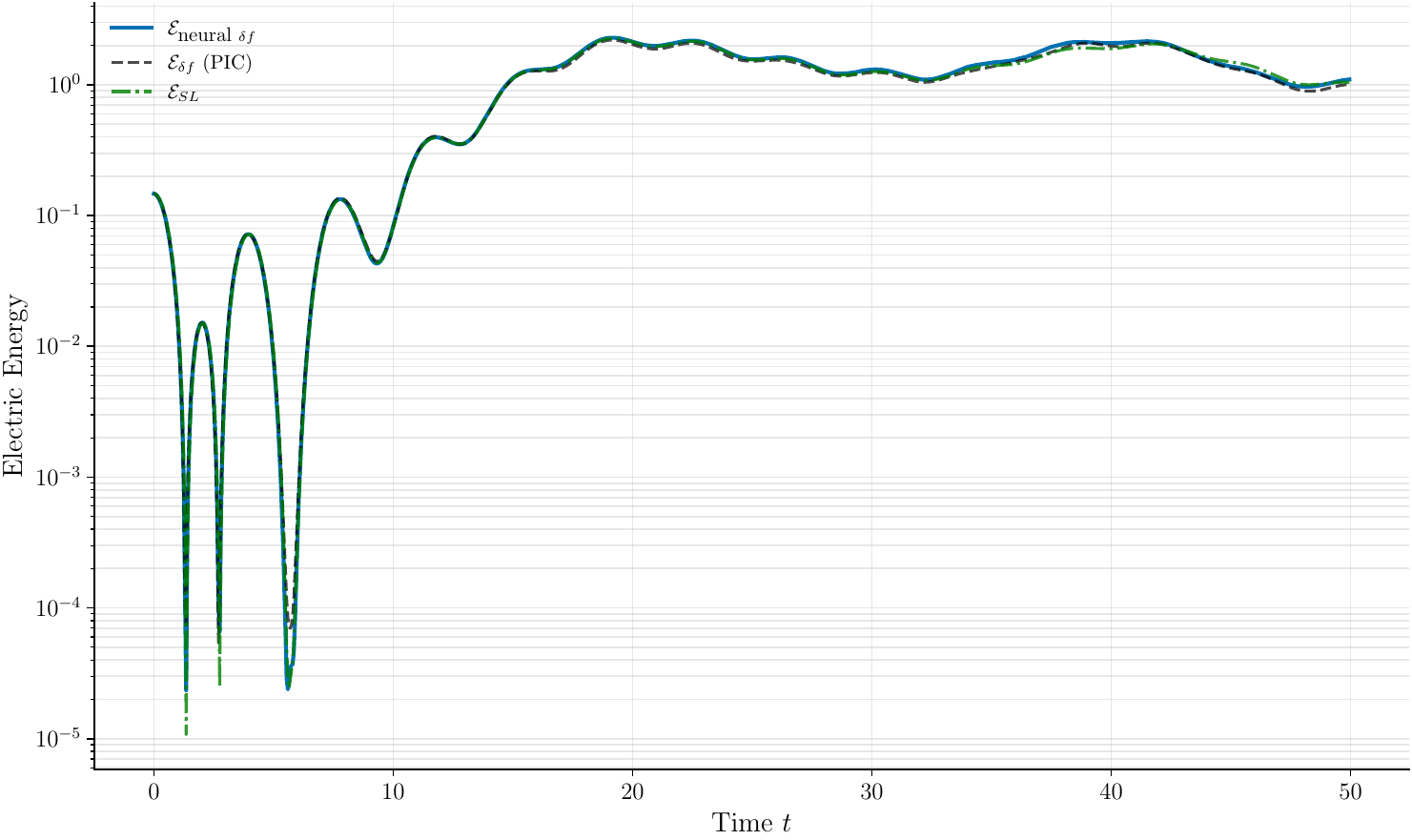}
        \caption{Evolution of the electric energy $\mathcal{E}$.}
        \label{fig:TSI3D_energy}
    \end{subfigure}
    \begin{subfigure}[b]{0.48\textwidth}
        \centering
        \includegraphics[width=\textwidth]{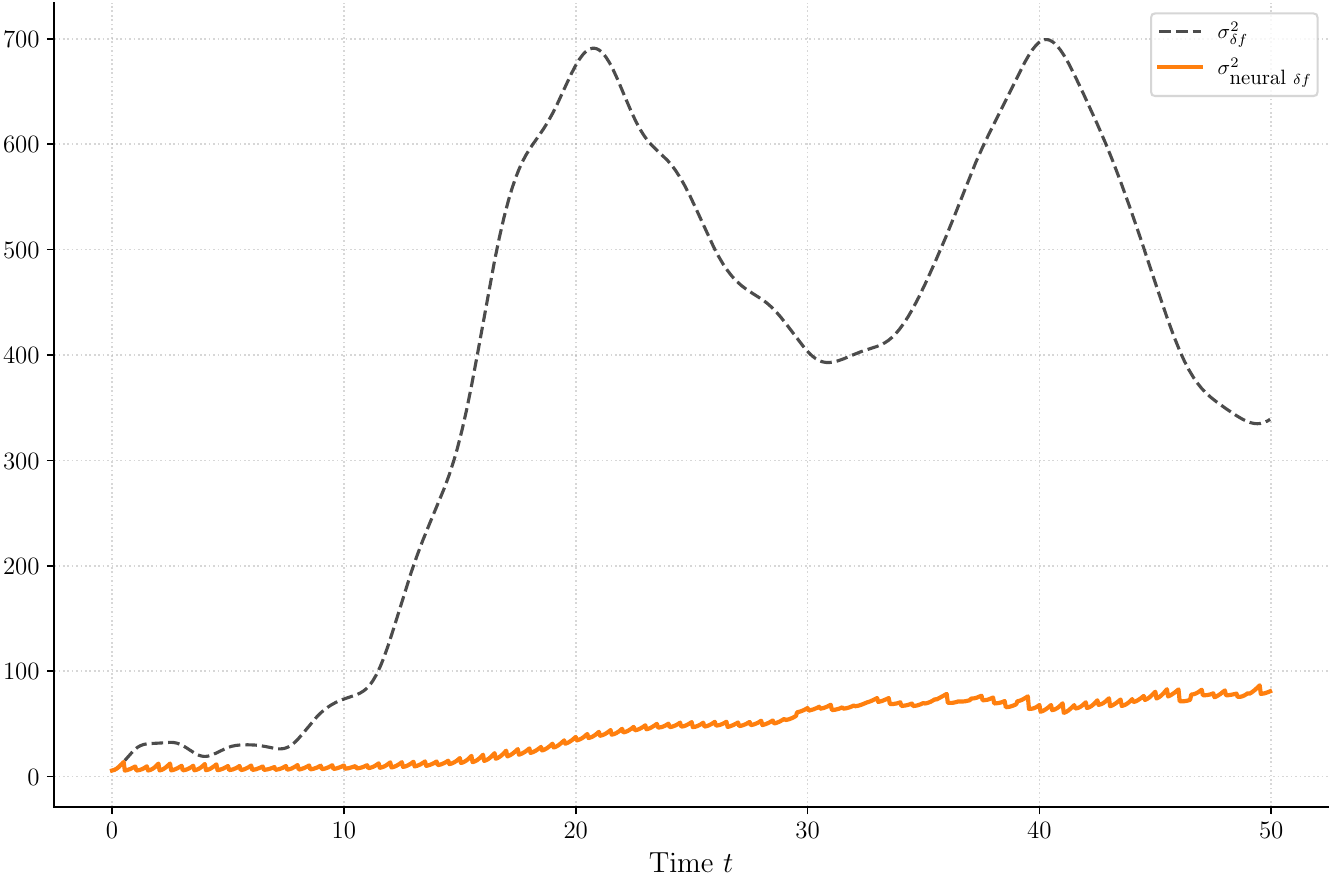}
        \caption{Evolution of the weight variance $\sigma_{\df}^2$.}
        \label{fig:TSI3D_variance}
    \end{subfigure}

    \caption{3D3V two-stream instability from \cref{sec:3D3V_two_stream}: Evolution of the electric energy and of the weight variance.}
    \label{fig:TSI_3D}
\end{figure}

\Cref{fig:TSI3D_density} shows cross sections of the phase space density at different times. The density is well reconstructed and stays symmetric despite the relatively low number of particles for a 6D domain.
\Cref{fig:TSI_3D} shows the evolution of the electric energy and of the empirical weight variance $\sigma_{\df}^2$,
both for the Neural $\df$-PIC method and for standard $\df$-PIC method. The two methods are
in qualitative agreement on the electric energy. The weight variance \eqref{eq:weight_variance}
for the Neural $\df$-PIC method is reduced by a factor of approximately 10.
The semi-Lagrangian scheme used a grid of size $64\times 32 \times 32 \times 63 \times 31 \times 31$. Over the run ($99$ bulk updates), the incremental training strategy of \cref{sec:training} produced $13$ composed networks.

\subsubsection{3D3V four-stream instability}
\label{sec:3D3V_four_stream}

The second extension of the 1D1V two-stream instability is given by the following initial condition

\begin{align}
    \finit(x,y,z,v_x,v_y,v_z) = & ( 1+ \varepsilon \cos(kx) + \varepsilon \cos(ky)) \\ & \times\frac{1}{2 \sqrt{2\pi}} \bigl(e^{-(v_x - v_0)^2/2} + e^{-(v_x + v_0)^2/2} \bigr)  \\ & \times\frac{1}{2 \sqrt{2\pi}} \bigl(e^{-(v_y - v_0)^2/2} + e^{-(v_y + v_0)^2/2} \bigr) \\ & \times\frac{1}{\sqrt{2 \pi}} e^{-v_z^2/2}
    \label{eq:fsi_3d_init}
\end{align}
with $k=0.3$, $v_0 = 2.4$ and $\varepsilon= 0.05$. The computational domain is $[0,2\pi/k] \times [0,1]^2 \times [-7,7]^3.$

\begin{figure}[!ht]
    \centering

    \begin{subfigure}[b]{\textwidth}
        \centering
        \includegraphics[width=\textwidth]{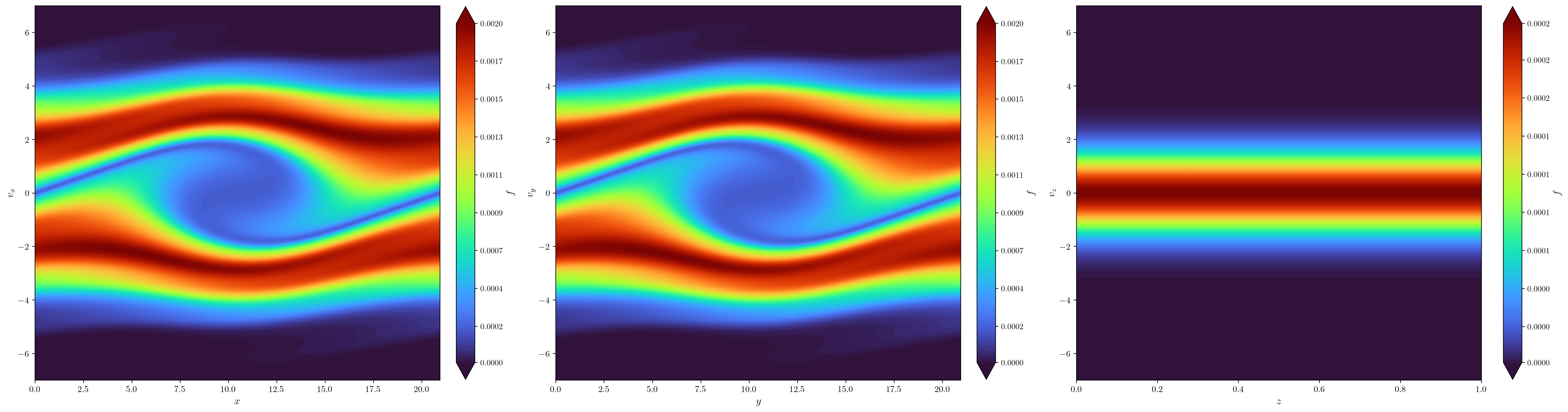}
        \caption{Cross sections of the density at $t=15$.}
        \label{fig:FSI3D_1}
    \end{subfigure}

    \vspace{0.5cm}

    \begin{subfigure}[b]{\textwidth}
        \centering
        \includegraphics[width=\textwidth]{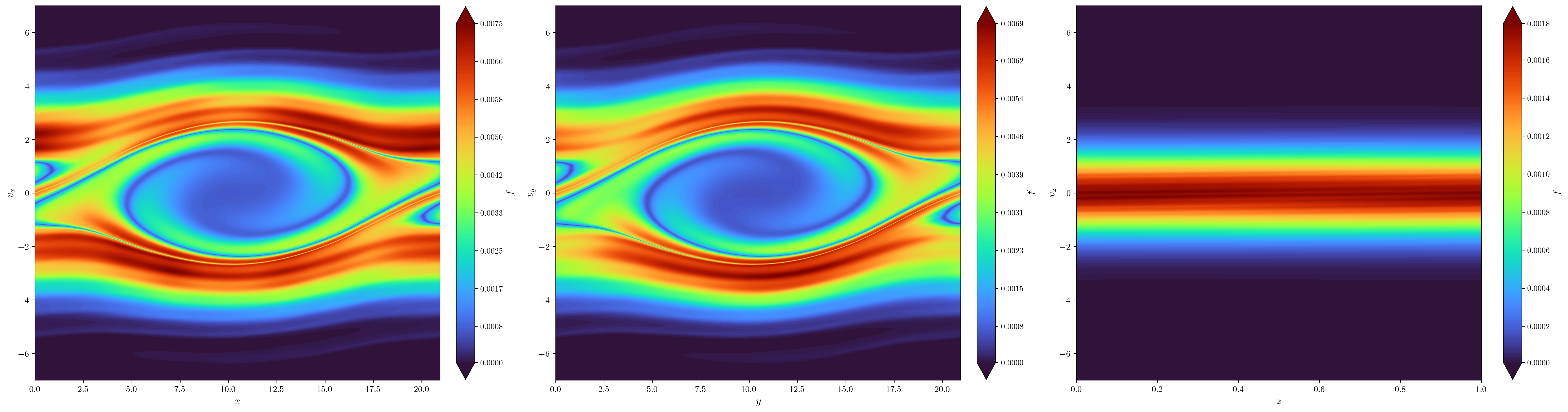}
        \caption{Cross sections of the density at $t=30$.}
        \label{fig:FSI3D_2}
    \end{subfigure}

    \vspace{0.5cm}

    \begin{subfigure}[b]{\textwidth}
        \centering
        \includegraphics[width=\textwidth]{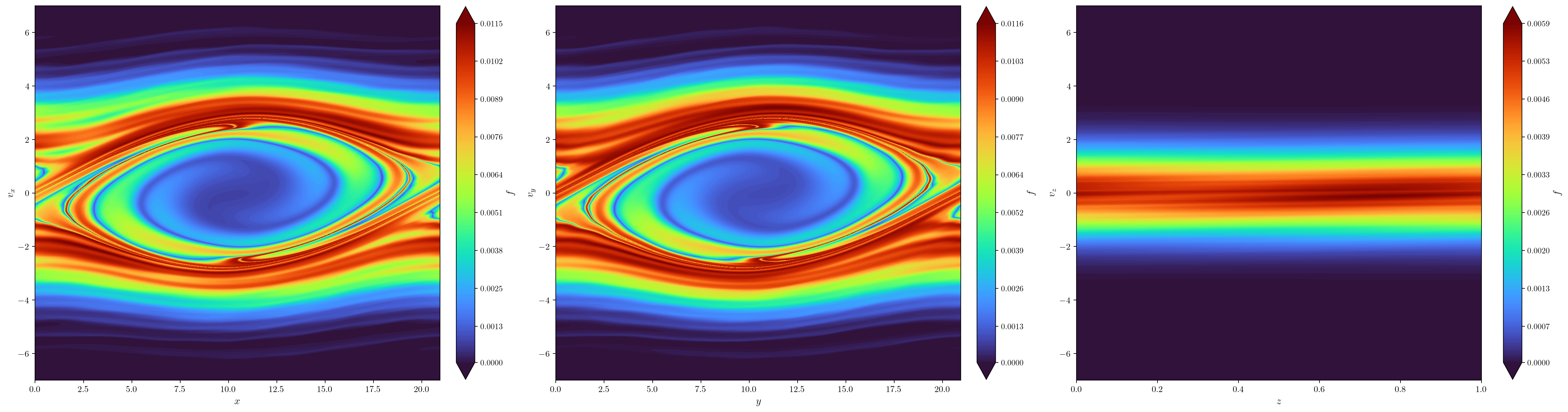}
        \caption{Cross sections of the density at $t=45$.}
        \label{fig:FSI3D_3}
    \end{subfigure}

    \caption{3D3V four-stream instability from \cref{sec:3D3V_four_stream}: Cross sections of the phase-space density $f(x,y=0,z=0,v_x,v_y = 0,v_z=0)$, $f(x=0,y,z=0,v_x=0,v_y,v_z=0)$ and $f(x=0,y=0,z,v_x=0,v_y=0,v_z)$ at $t = 15$, $t=30$ and $t=45$.}
    \label{fig:FSI3D_density}
\end{figure}

\Cref{fig:FSI3D_density} shows cross sections of the phase space density at different times. The density seems to be well reconstructed.

\begin{figure}[!ht]
    \centering

    \begin{subfigure}[b]{0.48\textwidth}
        \centering
        \includegraphics[width=\textwidth]{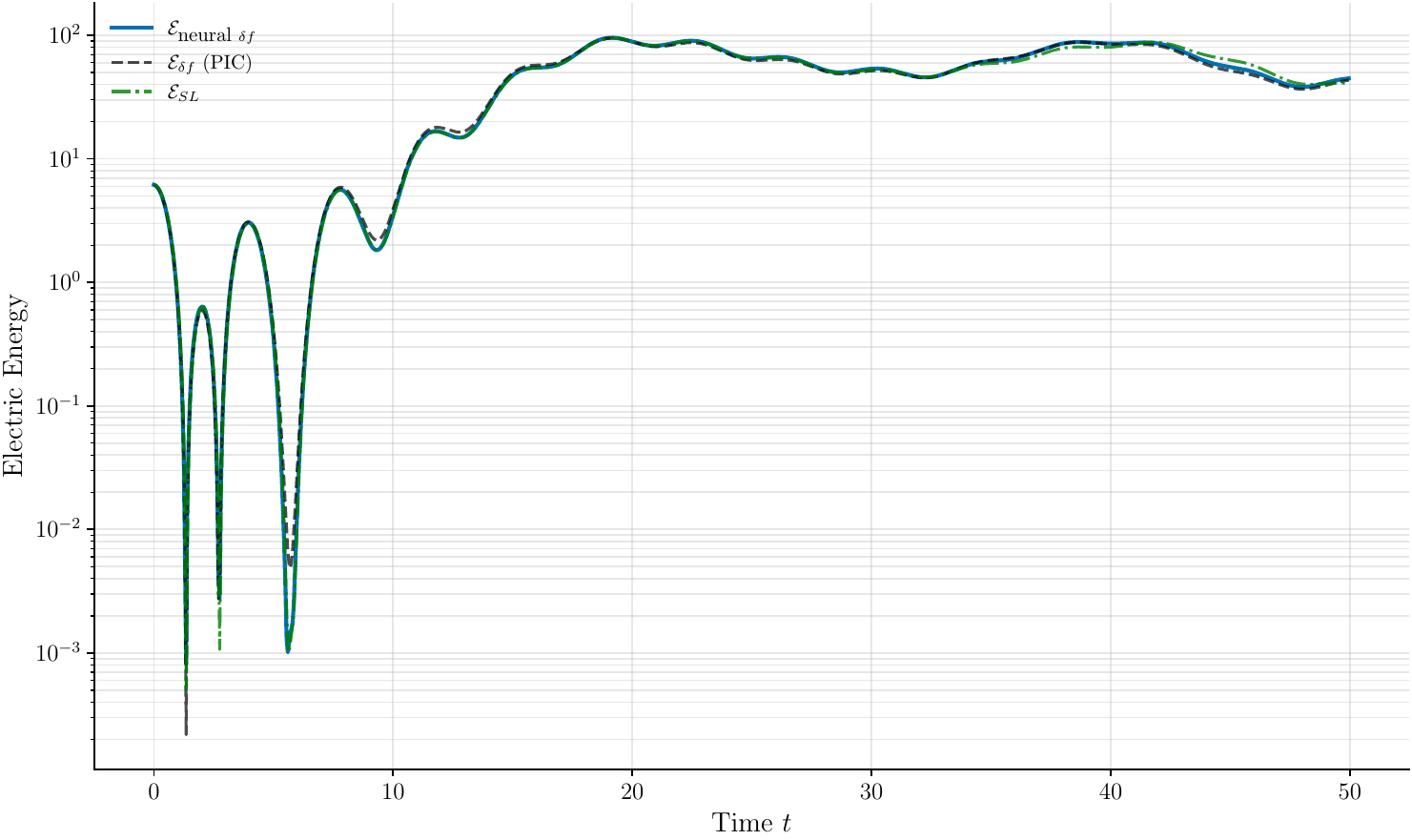}
        \caption{Evolution of the electric energy $\mathcal{E}$.}
        \label{fig:FSI3D_energy}
    \end{subfigure}
    \begin{subfigure}[b]{0.48\textwidth}
        \centering
        \includegraphics[width=\textwidth]{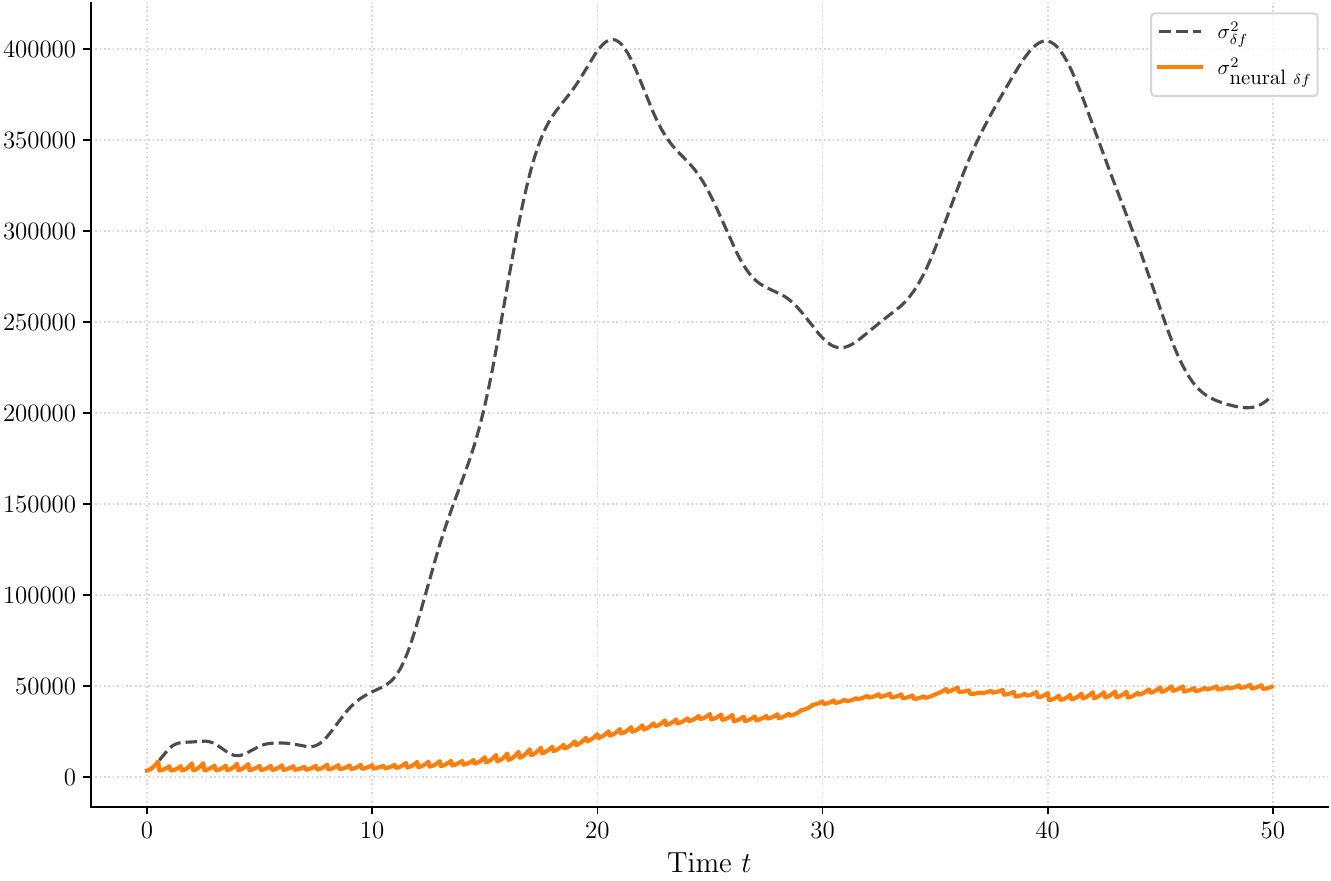}
        \caption{Evolution of the weight variance $\sigma_{\df}^2$.}
        \label{fig:FSI3D_variance}
    \end{subfigure}

    \caption{3D3V four-stream instability from \cref{sec:3D3V_four_stream}: Evolution of the electric energy and weight variance.}
    \label{fig:FSI_3D}
\end{figure}

\Cref{fig:FSI_3D} shows the evolution of the electric energy and of the empirical weight variance $\sigma_{\df}^2$,
both for the Neural $\df$-PIC method and for standard $\df$-PIC method. The two methods are
in qualitative agreement on the electric energy. The weight variance
for the Neural $\df$-PIC method is reduced by a factor of approximately 10. The semi-Lagrangian scheme used a grid of size $64\times 64 \times 32 \times 63 \times 63 \times 31$. Here, $12$ networks were trained over the run.

\subsubsection{3D3V six-stream instability}
\label{sec:3D3V_six_stream}

The third extension of the 1D1V two-stream instability is given by the following initial condition

\begin{align}
    \finit(x,y,z,v_x,v_y,v_z) = & ( 1+ \varepsilon \cos(kx) + \varepsilon \cos(ky) + \varepsilon \cos(kz) ) \\ & \times \frac{1}{2 \sqrt{2\pi}} \bigl(e^{-(v_x - v_0)^2/2} + e^{-(v_x + v_0)^2/2} \bigr) \\ &\times \frac{1}{2 \sqrt{2\pi}} \bigl(e^{-(v_y - v_0)^2/2} + e^{-(v_y + v_0)^2/2} \bigr) \\ & \times \frac{1}{2 \sqrt{2\pi}} \bigl(e^{-(v_z - v_0)^2/2} + e^{-(v_z + v_0)^2/2} \bigr)
    \label{eq:ssi_3d_init}
\end{align}
with $k=0.3$, $v_0 = 2.4$ and $\varepsilon= 0.05$. The computational domain is $[0,2\pi/k]^3 \times [-7,7]^3.$

\begin{figure}[!ht]
    \centering

    \begin{subfigure}[b]{\textwidth}
        \centering
        \includegraphics[width=\textwidth]{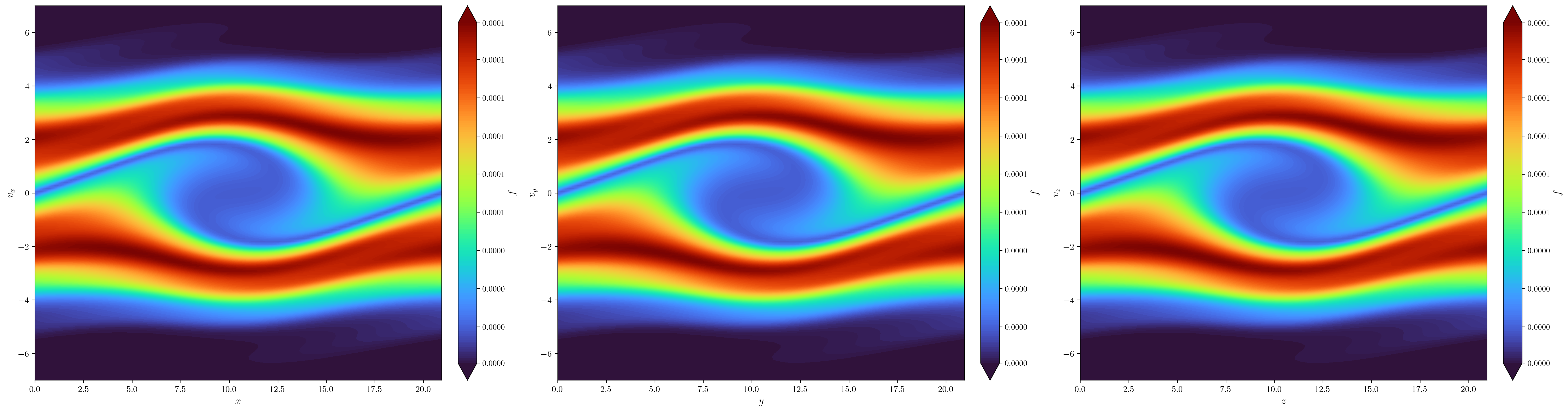}
        \caption{Cross sections of the density at $t=15$.}
        \label{fig:SSI3D_1}
    \end{subfigure}

    \vspace{0.5cm}
    \begin{subfigure}[b]{\textwidth}
        \centering
        \includegraphics[width=\textwidth]{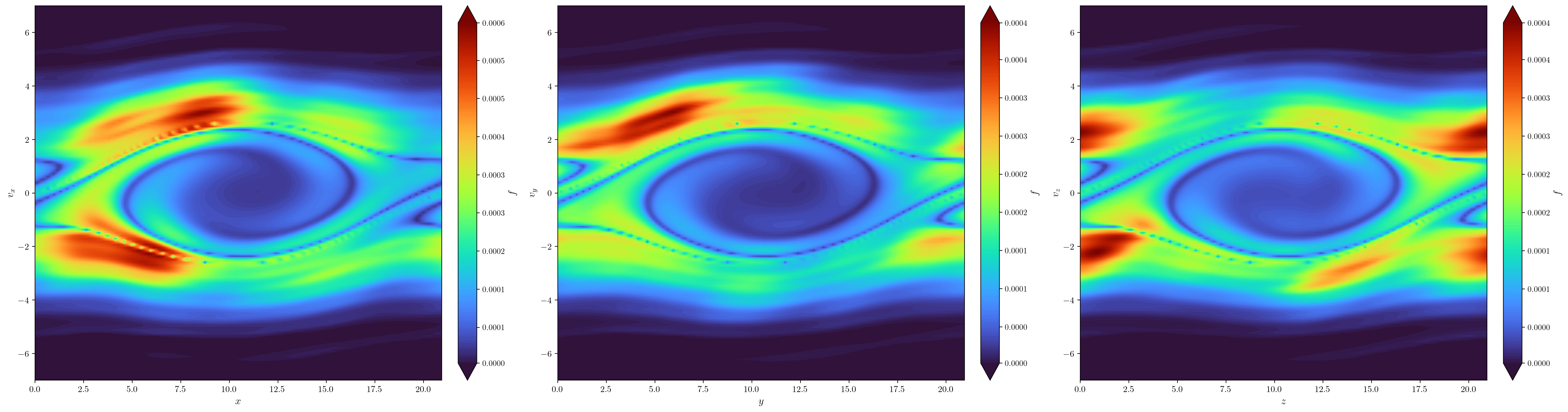}
        \caption{Cross sections of the density at $t=30$.}
        \label{fig:SSI3D_2}
    \end{subfigure}

    \vspace{0.5cm}

    \begin{subfigure}[b]{\textwidth}
        \centering
        \includegraphics[width=\textwidth]{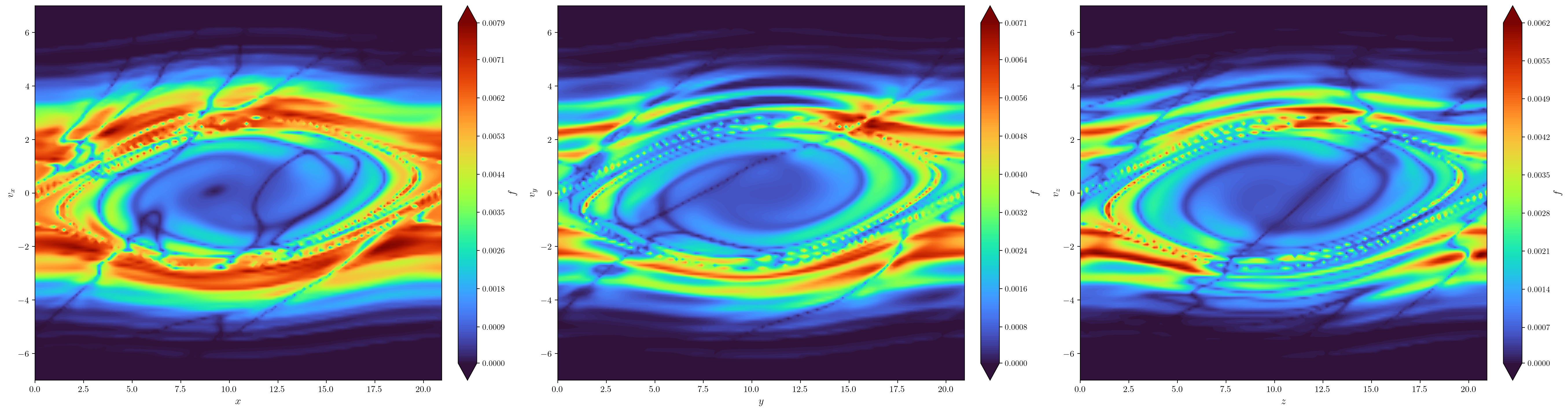}
        \caption{Cross sections of the density at $t=45$.}
        \label{fig:SSI3D_3}
    \end{subfigure}

    \caption{3D3V six-stream instability from \cref{sec:3D3V_six_stream}: Cross sections of the phase-space density $f(x,y=0,z=0,v_x,v_y = 0,v_z=0)$, $f(x=0,y,z=0,v_x=0,v_y,v_z=0)$ and $f(x=0,y=0,z,v_x=0,v_y=0,v_z)$ at $t = 15$, $t=30$ and $t=45$.}
    \label{fig:SSI3D_density}
\end{figure}

\Cref{fig:SSI3D_density} shows cross sections of the phase space density at different times. The density is well-reconstructed and symmetric at the beginning of the simulation, but the solution eventually develops spurious filamentation and loses its symmetry, even though the main structure remains well-captured. This case is the most demanding of all: compared with the four-stream instability, the dynamics evolve in all three spatial directions and the computational domain is larger (the extent in $z$ is $1$ for the four-stream case versus $2\pi/0.3$ here). As a result, the number of particles is too low for this 6D domain, and some regions of phase space contain no particles even where the distribution should be non-zero, which we believe is the cause of the spurious filamentation. Properly resolving this case would require substantially more particles, hence a parallel implementation; it is therefore at the limit of what the present single-GPU code can handle, and we report it as an illustration of the method's reach in full 6D rather than as a quantitatively converged result.

\begin{figure}[!ht]
    \centering

    \begin{subfigure}[b]{0.48\textwidth}
        \centering
        \includegraphics[width=\textwidth]{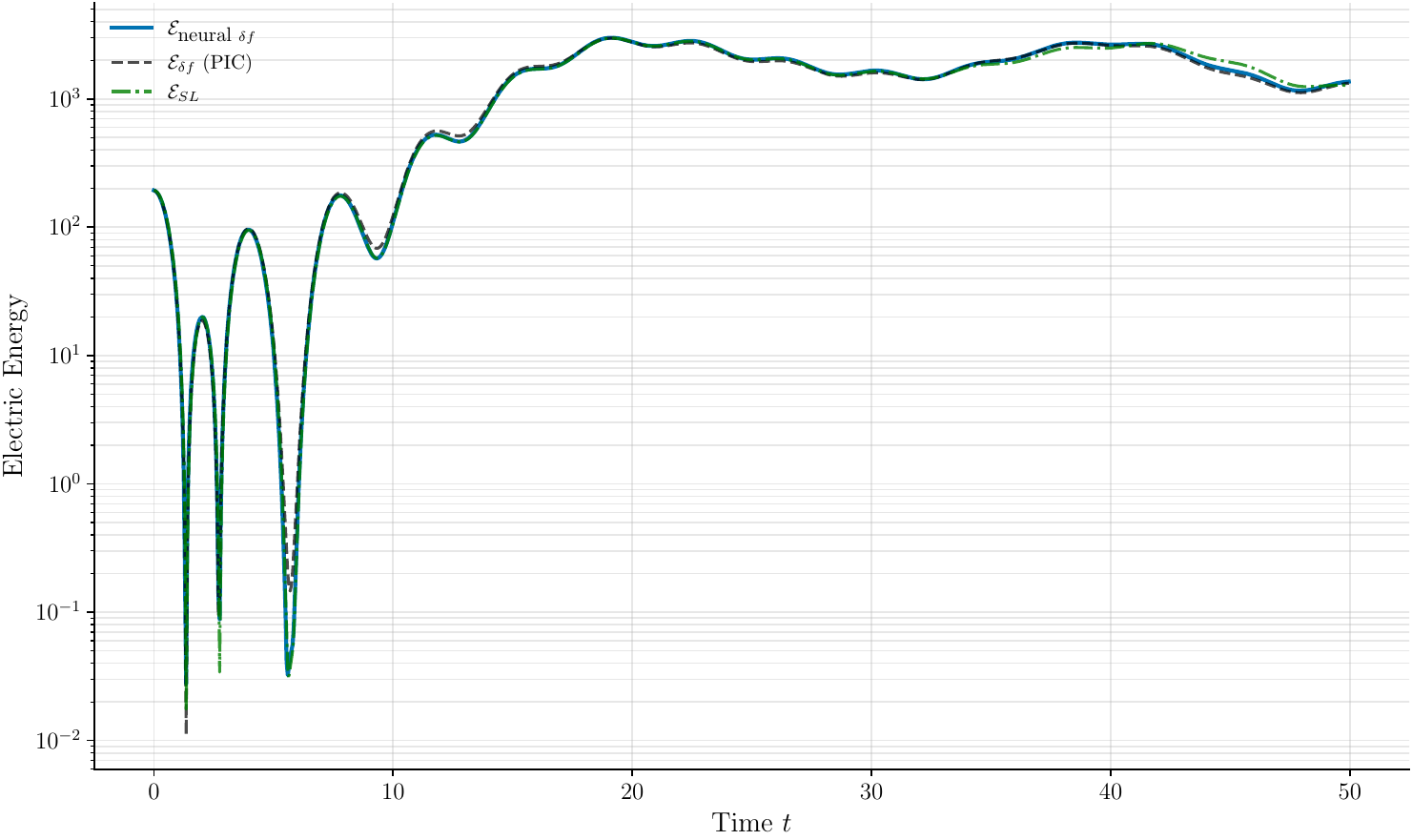}
        \caption{Evolution of the electric energy $\mathcal{E}$.}
        \label{fig:SSI3D_energy}
    \end{subfigure}
    \begin{subfigure}[b]{0.48\textwidth}
        \centering
        \includegraphics[width=\textwidth]{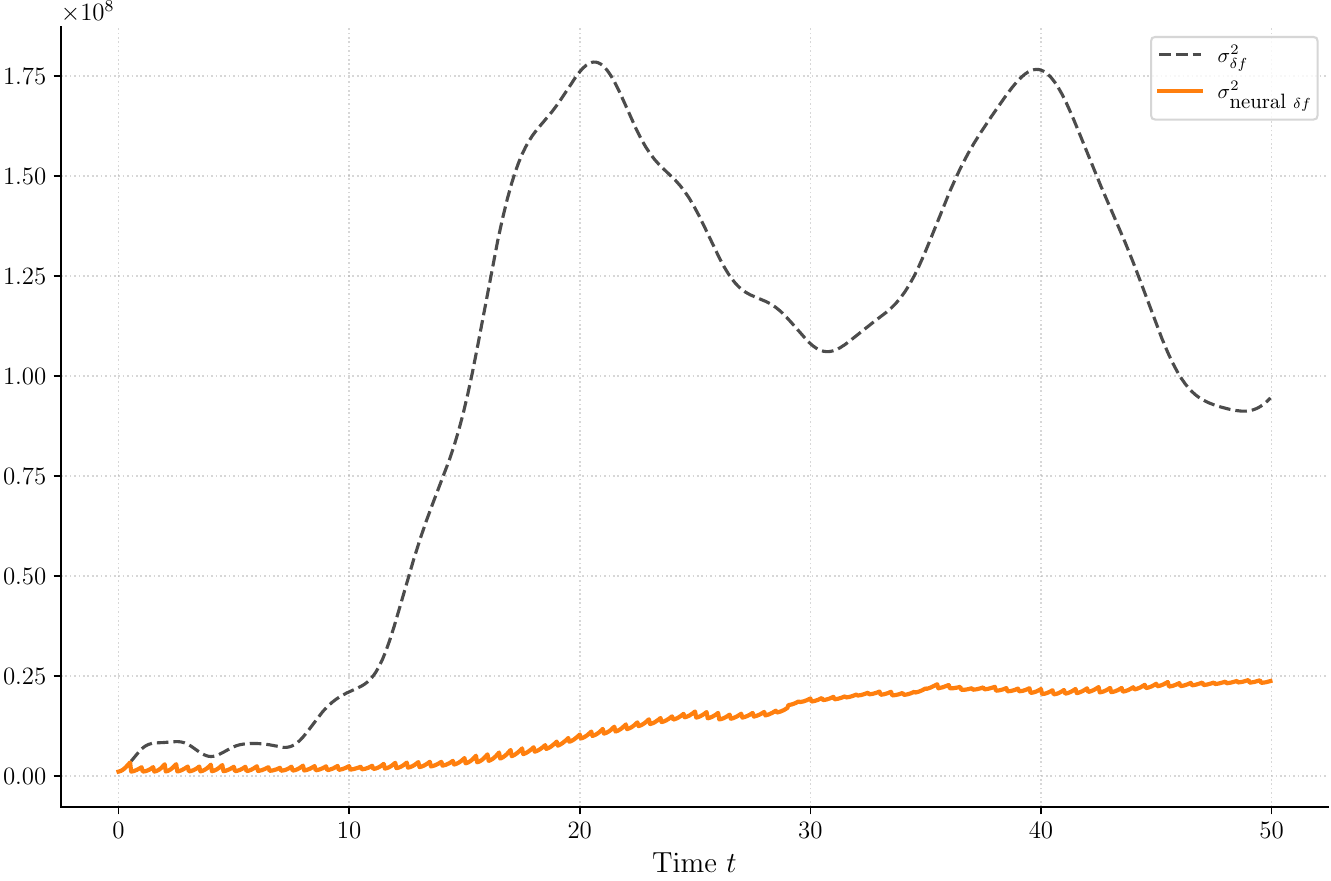}
        \caption{Evolution of the weight variance $\sigma_{\df}^2$.}
        \label{fig:SSI3D_variance}
    \end{subfigure}

    \caption{3D3V six-stream instability from \cref{sec:3D3V_six_stream}: Evolution of the electric energy and weight variance.}
    \label{fig:SSI3D}
\end{figure}

\Cref{fig:SSI3D} shows the evolution of the electric energy and of the empirical weight variance $\sigma_{\df}^2$,
both for the Neural $\df$-PIC method and for standard $\df$-PIC method. The two methods are
in qualitative agreement on the electric energy. The weight variance
for the Neural $\df$-PIC method is reduced by a factor of approximately 10.
The semi-Lagrangian scheme used a grid of size $64\times 64 \times 64 \times 63 \times 63 \times 63$. This most demanding case required $22$ networks over the run.

\subsubsection{3D3V bump-on-tail instability}
\label{sec:3D3V_bot}

\begin{figure}[!ht]
    \centering

    \begin{subfigure}[b]{\textwidth}
        \centering
        \includegraphics[width=\textwidth]{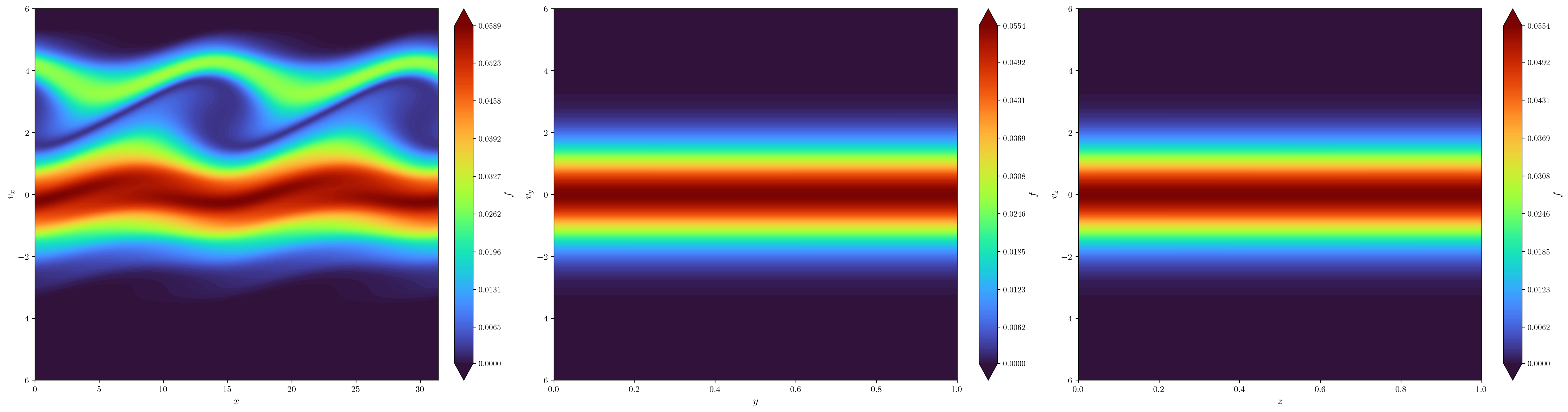}
        \caption{Cross sections of the density at $t=15$}
        \label{fig:BOT3D_1}
    \end{subfigure}

    \vspace{0.5cm}
    \begin{subfigure}[b]{\textwidth}
        \centering
        \includegraphics[width=\textwidth]{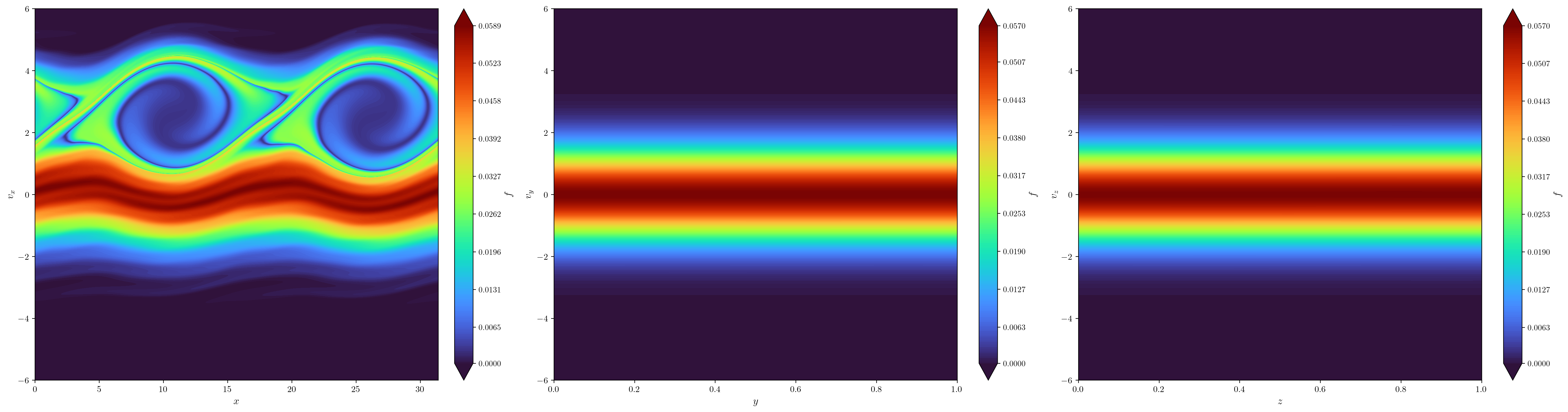}
        \caption{Cross sections of the density at $t=32$}
        \label{fig:BOT3D_2}
    \end{subfigure}

    \vspace{0.5cm}

    \begin{subfigure}[b]{\textwidth}
        \centering
        \includegraphics[width=\textwidth]{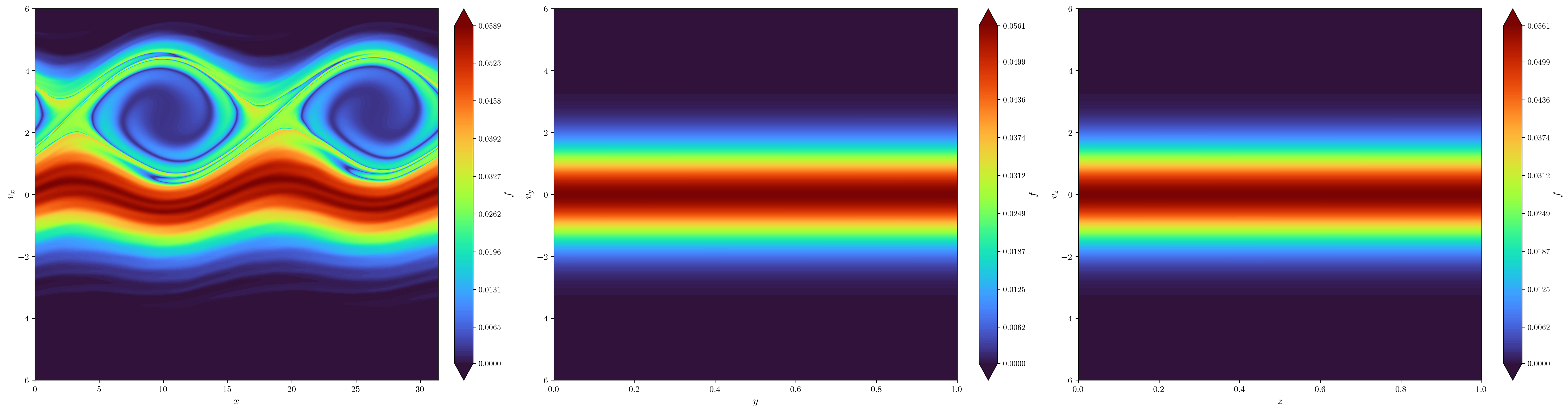}
        \caption{Cross sections of the density at $t=38$}
        \label{fig:BOT3D_3}
    \end{subfigure}

    \caption{3D3V bump-on-tail instability from \cref{sec:bot}: Cross sections of the phase-space density $f(x,y=0,z=0,v_x,v_y = 0,v_z=0)$, $f(x=0,y,z=0,v_x=0,v_y,v_z=0)$ and $f(x=0,y=0,z,v_x=0,v_y=0,v_z)$ at $t = 15$, $t=32$ and $t=38$.}
    \label{fig:BOT3D_density}
\end{figure}

\begin{figure}[!ht]
    \centering

    \begin{subfigure}[b]{0.48\textwidth}
        \centering
        \includegraphics[width=\textwidth]{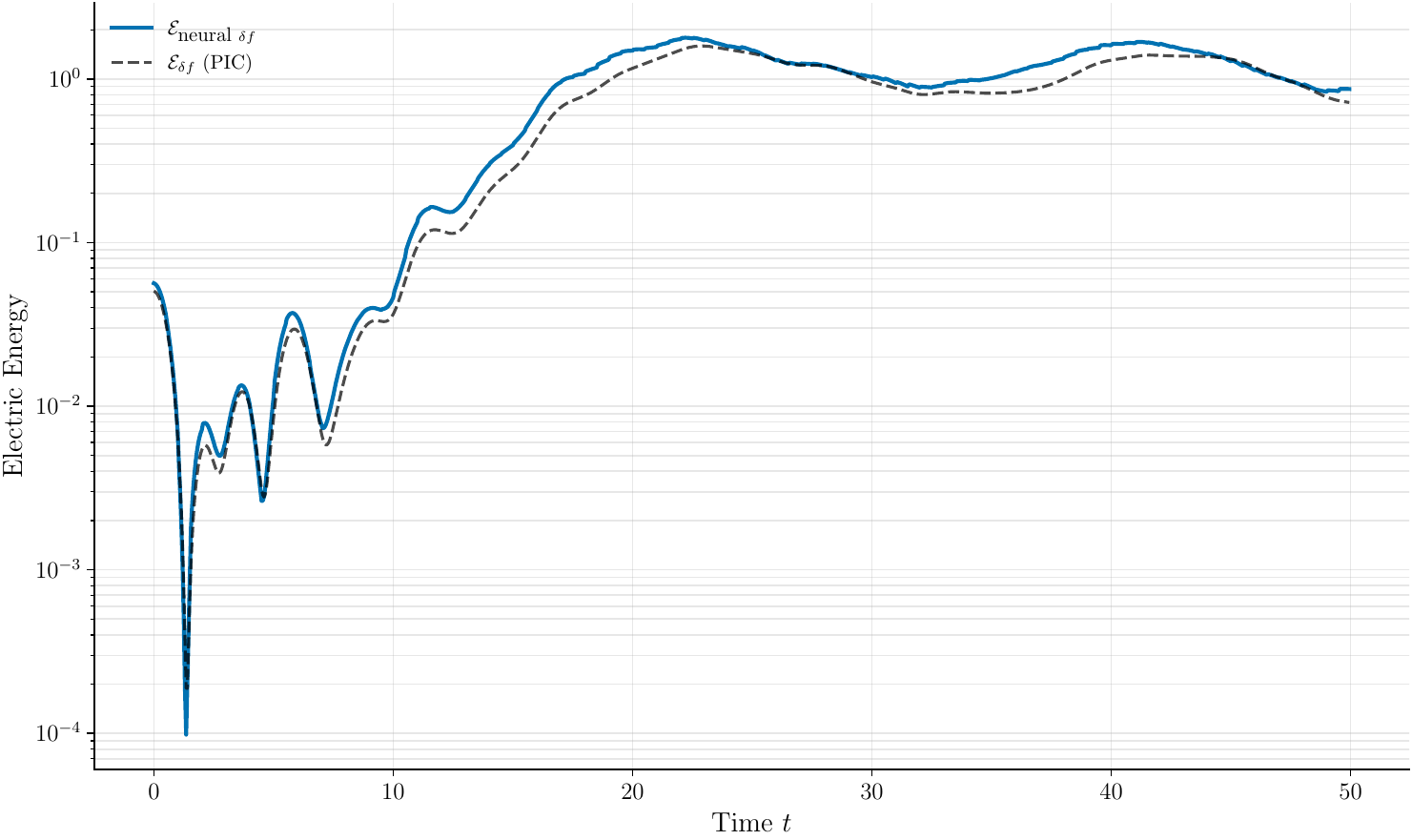}
        \caption{Evolution of the electric energy $\mathcal{E}$.}
        \label{fig:BOT3D_energy}
    \end{subfigure}
    \begin{subfigure}[b]{0.48\textwidth}
        \centering
        \includegraphics[width=\textwidth]{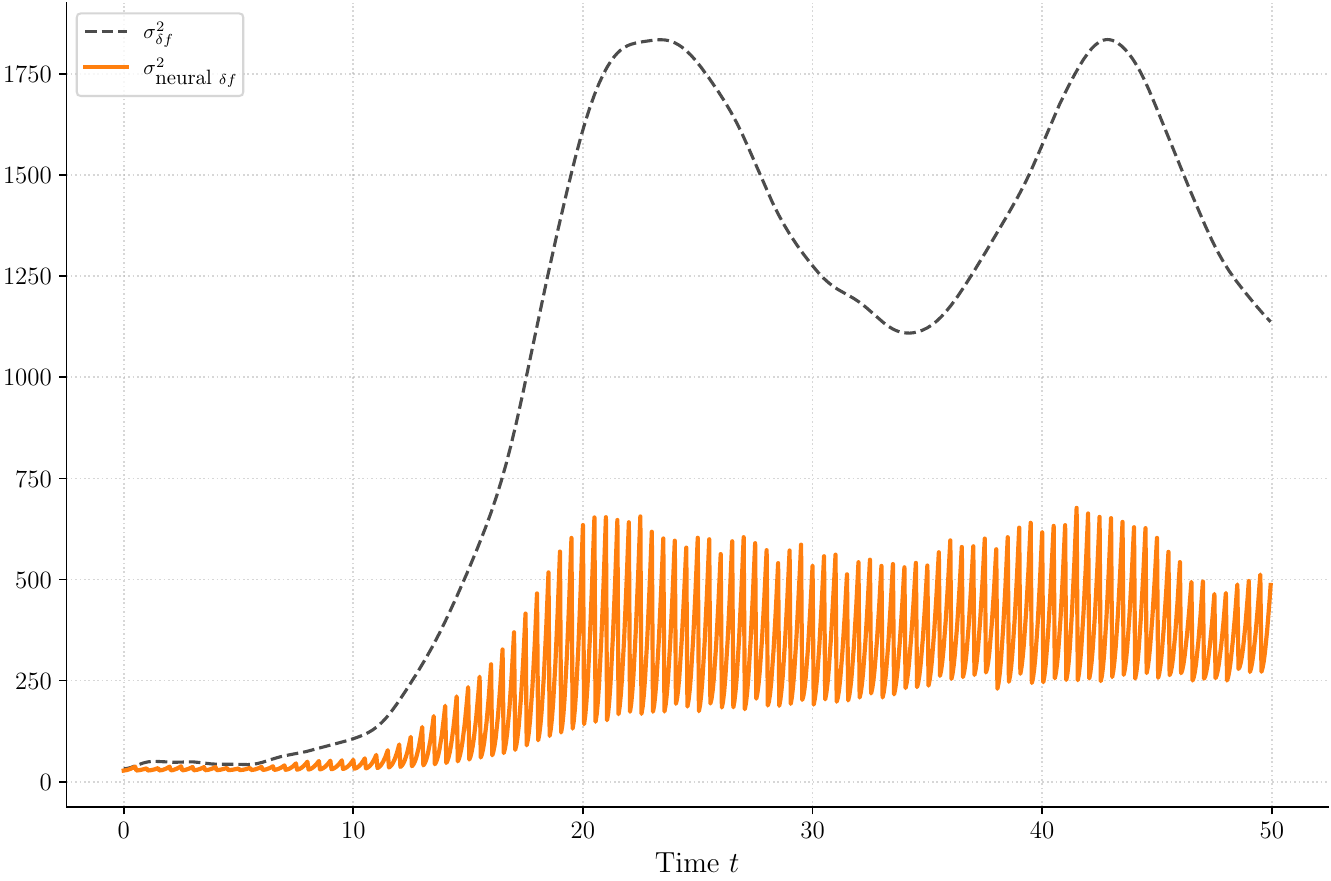}
        \caption{Evolution of the weight variance $\sigma_{\df}^2$.}
        \label{fig:BOT3D_variance}
    \end{subfigure}

    \caption{3D3V bump-on-tail instability from \cref{sec:3D3V_bot}: Evolution of the electric energy and weight variance.}

    \label{fig:BOT_3D}
\end{figure}

The 3D3V bump-on-tail initial condition is
\begin{equation}
    \finit(x,y,z,v_x,v_y,v_z)
    = \left(
    \frac{0.9}{\sqrt{2\pi}}\,e^{-v_x^2/2}
    + \frac{0.2}{\sqrt{10\pi}}\,e^{-(v_x-3.8)^2/10}
    \right)
    \frac{e^{-(v_y^2+v_z^2)/2}}{2\pi}
    \bigl(1 + 0.03\cos(0.4x)\bigr).
    \label{eq:bot_3d_init}
\end{equation}
The computational domain is $[0,10\pi] \times [0,1]^2 \times[-6,6]^3$.
\Cref{fig:BOT3D_density} shows cross sections of the phase space density at different times. The density is well reconstructed.
\Cref{fig:BOT_3D} shows the evolution of the electric energy and of the empirical weight variance $\sigma_{\df}^2$,
both for the Neural $\df$-PIC method and for standard $\df$-PIC method. The two methods are
in qualitative agreement on the electric energy. The weight variance \eqref{eq:weight_variance}
for the Neural $\df$-PIC method is reduced by a factor of approximately 10. Over the run, $15$ networks were trained.

\section{Conclusion}
\label{sec:conclusion}

We have presented the Neural $\df$-PIC method, a new approach for the kinetic
simulation of plasmas in which the bulk density, acting as a control
variate, is evolved using a sequence of symplectic neural networks trained on the
particle trajectories. The bulk is then reprojected on a coarse spline grid, which
makes the evaluation of the weights and of the velocity integrals relatively cheap.
The use of SympNets guarantees that the reconstructed backward flow is symplectic by
construction.
We also introduced a periodic variant of the SympNet architecture that natively
encodes the spatial periodicity of the problem, avoiding any penalisation term.

We first verified, on a controlled flow-learning experiment, that periodic SympNets
can accurately approximate the characteristic flow of the Vlasov-Poisson system from
particle data, and that an incremental training strategy
(akin to a preconditioning or a curriculum learning) is both more
stable and less prone to overfitting than a direct training of the flow over the
whole time interval. On the full scheme, numerical experiments in 1D1V and 3D3V show
that the dynamically evolved bulk keeps the particle weights small, reducing the
empirical weight variance by a factor of about $5$ to $10$ compared to a static
$\df$-PIC scheme, and reducing the error on the electric field accordingly. The
incremental strategy further keeps the number of stored networks well below the
number of bulk updates.
Nevertheless, several limitations remain. The present study is a proof of concept on the
Vlasov-Poisson system: in this setting the method is significantly more expensive than
a standard $\df$-PIC scheme, and its benefit is to be sought in regimes where a static bulk is inadequate. The evaluation cost grows linearly with the
number of composed networks, and the composition of many approximate flows, together
with the spline interpolation, induces a slow growth of the particle weights over time;
controlling this growth, for instance by periodically remapping the density onto a
single network or spline representation, is an important direction for long-time
simulations. Finally, the 3D3V results are exploratory: a fully converged reference is
out of reach in six dimensions, and the simulations are limited by a non-parallel,
single-GPU implementation.

Future work will focus on a parallel implementation, on adaptive strategies for the
bulk-update period $N_\Psi$ and for the network resets, on a control of the weight
growth through remapping, and on the application to gyrokinetic models, which is an
important objective of this approach.

\section*{Acknowledgement}
The authors would like to thank Nils Schild for kindly providing the data of the electric energy for the 3D3V tests from the BSL6D code.

\bibliographystyle{abbrvnat}
\bibliography{biblio}

\end{document}